\documentclass[a4paper,11pt]{article}
\pdfoutput=1
\usepackage{jheppub}

\usepackage{pstricks}
\usepackage{amsmath}
\usepackage{graphicx}
\usepackage{booktabs}
\usepackage{hyperref}
\usepackage{color}
\usepackage{feynmp}

\definecolor{lgris}{rgb}{0.95,0.95,0.95}

\def\be{\begin{equation}}
\def\ee{\end{equation}}
\def\bea{\begin{eqnarray}}
\def\eea{\end{eqnarray}}
\def\nnb{\nonumber}
\def\dps{\displaystyle}
\def\lk{\left(}
\def\rk{\right)}

\def\bbuildrel#1_#2^#3{\mathrel{\mathop{\kern 0pt#1}\limits_{#2}^{#3}}}
\def\slash#1{\setbox0=\hbox{$#1$}#1\hskip-\wd0\dimen0=5pt\advance
       \dimen0 by-\ht0\advance\dimen0 by\dp0\lower0.5\dimen0\hbox
         to\wd0{\hss\sl/\/\hss}}
\def\gev{{\rm GeV}}
\def\mev{{\rm MeV}}

\newcommand{\gae}{\lower 2pt \hbox{$\, \buildrel {\scriptstyle >}\over
    {\scriptstyle \sim}\,$}}
\newcommand{\lae}{\lower 2pt \hbox{$\, \buildrel {\scriptstyle <}\over
    {\scriptstyle \sim}\,$}}

\newcommand{\f}{\frac}
\newcommand{\fm}[2]{{\textstyle \frac{#1}{#2}}}

\newcommand{\aem}{\widetilde{\alpha}_{\mathrm e}}
\newcommand{\as}{\widetilde{\alpha}_{\mathrm s}}

\newcommand{\s}{\hat{s}}

\newcommand{\spp}{\vphantom{\bigg(}}

\newcommand{\bsll}{\bar{B} \to X_s \ell^+ \ell^-}

\newcommand{\skipp}[1]{}

\renewcommand\thefootnote{\fnsymbol{footnote}}

\title{Inclusive $\bsll$: Complete angular analysis \\ and a thorough study of collinear photons} 
\author{Tobias Huber$^a$,} 
\author{Tobias Hurth$^b$,} 
\author{and Enrico Lunghi$^c$} 
\affiliation[a]{Theoretische Physik 1, Naturwissenschaftlich-Technische Fakult\"at, \\
Universit\"at Siegen,  D-57068 Siegen, Germany} 
\affiliation[b]{PRISMA Cluster of Excellence and  Institute for Physics (THEP)\\
Johannes Gutenberg University, D-55099 Mainz, Germany}
\affiliation[c]{Physics Department, Indiana University, Bloomington, IN 47405, USA}
\emailAdd{huber@physik.uni-siegen.de} 
\emailAdd{hurth@uni-mainz.de} 
\emailAdd{elunghi@indiana.edu}

\abstract{We investigate logarithmically enhanced electromagnetic corrections of all angular observables in inclusive $\bsll$. We present analytical results, which are  supplemented by a dedicated Monte Carlo study on the treatment of collinear photons in order to determine the size of the electromagnetic logarithms. We then give the Standard Model predictions of all observables, considering all available NNLO QCD, NLO QED and power corrections, and investigate their sensitivity to New Physics. Since the structure of the double differential decay rate is modified in the presence of QED corrections, we also propose new observables which vanish if only QCD corrections are taken into account. Moreover, we study the experimental sensitivity to these new observables at Belle~II.
}
\preprint{
\begin{minipage}{3cm}
\small
\flushright
QFET-2015-13\\
SI-HEP-2015-12\\
MITP/15-001\\
IU-HET-593
\end{minipage}}

\begin{document}

\maketitle

\renewcommand{\thefootnote}{\arabic{footnote}}
\setcounter{footnote}{0}


\section{Introduction} 
\label{sec:introduction}
By now the LHC experiment has not discovered any new degrees of freedom beyond the Standard Model (SM). In particular, the measurements of the LHCb experiment and the $B$-physics experiments of ATLAS and CMS  have confirmed the simple Cabibbo-Kobayashi-Maskawa (CKM) theory of the SM~\cite{LHCb,ATLASb,CMSb}. This corresponds to  the general result of the $B$-factories~\cite{BABAR,BELLE} and of the Tevatron $B$-physics experiments~\cite{TEVATRONb} which have not indicated any sizable discrepancy from SM predictions in the $B$-meson sector (for reviews see refs.~\cite{Isidori:2010kg, Hurth:2010tk, Hurth:2012vp}).

However, recently the first measurement of new angular observables in the exclusive decay $B \to K^* \mu^+\mu^-$ has shown  a kind of anomaly~\cite{Aaij:2013qta}.  Due to the large hadronic uncertainties  it is not clear if this anomaly is a first sign for new physics beyond the SM, or a consequence of hadronic power corrections; but of course, it could turn out to just be a statistical fluctuation (see e.g.\ refs.~\cite{Descotes-Genon:2013wba,Altmannshofer:2013foa,Beaujean:2013soa,Hambrock:2013zya,Gauld:2013qba,Buras:2013dea,Hurth:2013ssa,Mahmoudi:2014mja,Lyon:2014hpa}. The LHCb analysis based on the 3 fb$^{-1}$ dataset is  eagerly awaited to clarify the situation. More recently, another slight discrepancy occurred. The ratio $R_K = {\rm BR}(B^+ \to K^+ \mu^+ \mu^-) / {\rm BR}(B^+ \to K^+ e^+ e^-)$ in the low-$q^2$ region ($q^2$ being the di-lepton invariant mass) has been measured by LHCb showing a $2.6\sigma$ deviation from the SM prediction~\cite{Aaij:2014ora}. In contrast to the anomaly in the rare decay $B  \rightarrow K^{*} \mu^+\mu^-$ which is affected by unknown power corrections, the ratio  $R_K$ is theoretically rather clean. This might be a sign for lepton non-universality (see e.g.\ refs.~\cite{Alonso:2014csa,Hiller:2014yaa,Ghosh:2014awa,Biswas:2014gga,Davidtalk:2014,Hurth:2014vma,Glashow:2014iga,Altmannshofer:2014rta,Hiller:2014ula,Bhattacharya:2014wla}).

The inclusive decay mode $\bar  B \to X_s \ell^+\ell^-$ is one of the most important, theoretically clean modes of the indirect search for new physics via flavour observables (for a review and updates see refs.~\cite{Misiak1,Misiak2,Hurth:2003vb}); especially it allows for a nontrivial crosscheck of the recent LHCb data on the exclusive mode~\cite{Hurth:2013ssa,Hurth:2014zja}.

The observables within this inclusive mode are dominated by perturbative  contributions if the  $c \bar c$ resonances that show up as large peaks in the dilepton invariant mass spectrum are removed by appropriate  kinematic cuts -- leading to so-called `perturbative di-lepton invariant mass windows', namely  the low di-lepton mass region $1\,{\rm GeV}^2 < s = q^2 = m_{\ell\ell}^2   < 6\,{\rm GeV}^2$, and also the high dilepton mass region with $q^2 > 14.4\,{\rm GeV}^2$ (or $q^2 > 14.2\,{\rm GeV}^2$). In these regions a theoretical precision of order $10\%$ is in principle possible.  

By now the branching fraction has been measured by Belle and BaBar using the sum-of-exclusive technique only. The latest published measurement of Belle~\cite{Iwasaki:2005sy} is based on a sample of $152 \times 10^6$ $B \bar B$ events only, which corresponds to less than $30\%$ of the dataset available at the end of the Belle experiment. Babar has just recently presented an analysis based on the whole dataset of Babar using a sample of $471 \times 10^6$ $B \bar B$ events~\cite{Lees:2013nxa} which updated the former analysis of 2004~\cite{Aubert:2004it}.

In the low- and high-dilepton invariant mass region the weighted averages of the experimental results read
\begin{align}
{\cal B} (\bar B\to X_s \ell^+\ell^-)_{\rm low}^{\rm exp} &= 
\left( 1.58 \pm 0.37 \right) \times 10^{-6}  \; ,\\
{\cal B} (\bar B\to X_s \ell^+\ell^-)_{\rm high}^{\rm exp} &=
\left( 0.48 \pm 0.10 \right) \times 10^{-6} \; .
\label{eq:expWA}
\end{align}
All the measurements are  still dominated  by the statistical error. The expectation  is that the final word of the present $B$ factories leads to an experimental accuracy of $15-20\%$. 

In addition, Belle has presented a first measurement of the forward-backward asymmetry~\cite{Sato:2014pjr}  and Babar a measurement of the CP violation in this channel~\cite{Lees:2013nxa}. 

The super flavour factory Belle~II at KEK will accumulate two orders of magnitude larger data samples~\cite{Belle2}. Such data will  push experimental precision to its limit. This is the main motivation for the present study to decrease the theoretical uncertainties accordingly.\\

The theoretical precision has already reached a highly sophisticated level. Let us briefly review the previous analyses. 
\begin{itemize}
\item 
Within the inclusive decay mode $\bar B \to X_s \ell^+\ell^-$ the  dominating perturbative QCD contributions  are calculated up to NNLL precision. The complete NLL QCD contributions have been  presented~\cite{Misiak:1992bc,Buras:1994dj}. For the NNLL calculation,  many components  were taken over from the NLL calculation of the $\bar B \to X_s \gamma$ mode.  The additional components  for the NNLL QCD precision have been  calculated in refs.~\cite{Bobeth:1999mk,Gambino:2003zm,Gorbahn:2004my,Asatryan:2001zw,Asatrian:2001de,Asatryan:2002iy,Ghinculov:2002pe,Asatrian:2002va,Asatrian:2003yk,Ghinculov:2003bx,Ghinculov:2003qd,Greub:2008cy,Bobeth:2003at}.

\item
If only the leading operator of the electroweak hamiltonian is considered, one is led to a local operator product expansion (OPE). In this case, the leading hadronic power corrections in the decay {\bf $\bar B \to X_s \ell^+ \ell^-$}  scale  with $1/m_b^2$ and $1/m_b^3$ only and have already been analysed~\cite{Falk:1993dh, Ali:1996bm, Chen:1997dj, Buchalla:1998mt, Bauer:1999kf, Ligeti:2007sn}. Power correction that scale with $1/m_c^2$~\cite{Buchalla:1997ky} have also been considered.  They can be calculated quite analogously to those in the decay $\bar B \rightarrow X_s \gamma$.  A systematic and careful analysis of hadronic power corrections including all relevant operators has been  performed in the case of the decay $\bar B \rightarrow X_s \gamma$~\cite{Benzke:2010js}.  Such analysis goes beyond the local OPE.  An additional uncertainty of $\pm 5\%$ has been identified. The analysis in the case of $\bar B \rightarrow X_s \ell^+\ell^-$ is fully analogous and work in progress. There is no reason to expect any large deviation from the $\bar B \to X_s \gamma$ result.

In the high-$q^2$ region, one encounters the breakdown of the heavy-mass expansion (HME) at the end point of the dilepton mass spectrum: Whereas the partonic contribution vanishes, the $1/m_b^2$ and $1/m_b^3$ corrections tend towards a finite, non-zero value. Contrary to the end-point region of the photon-energy spectrum in the $\bar B \rightarrow X_s \gamma$ decay, no partial all-order resummation into a shape-function is possible. However, for the {\it integrated}\/ high-$q^2$ spectrum an effective expansion is found in inverse powers of $m_b^{\rm eff} = m_b \times (1 - \sqrt{s_{\rm min}})$ instead of $m_b$~\cite{Neubert:2000ch,Bauer:2001rc}. The expansion converges less rapidly, and the convergence behaviour depends on the lower dilepton-mass cut $s_{\rm min} = q^2_{\rm min} / m_b^2$~\cite{Ghinculov:2003qd}.

The large theoretical uncertainties could be significantly reduced by normalizing the $\bar B \rightarrow X_s \ell^+ \ell^-$ decay rate to the semileptonic $\bar B \rightarrow X_u \ell \bar\nu$ decay rate with the same $s$ cut~\cite{Ligeti:2007sn}:
\begin{equation}
\label{eq:zoltanR}
{\cal R}(s_0) =  
\int_{\hat s_0}^1 {\rm d} \hat s \, {{\rm d} {\Gamma} (\bar B \to X_s \ell^+\ell^-) \over {\rm d} \hat s}\,  /\,   
\int_{\hat s_0}^1 {\rm d} \hat s \, {{\rm d} {\Gamma} (\bar B^0 \to X_u \ell \nu) \over {\rm d} \hat s}\,.
\end{equation}
For example, the uncertainty due to the dominating $1/m_b^3$ term could be reduced from $19\%$ to $9\%$~\cite{Huber:2007vv}.

\item 
In the inclusive decay {\bf $\bar B \to X_s \ell^+ \ell^-$}, 
the hadronic and dilepton invariant masses are independent
kinematical quantities.  A hadronic invariant-mass cut is imposed in the
experiments. The high-dilepton-mass region is not affected by this cut, but in the
low-dilepton mass region the kinematics with a jet-like $X_s$ and $m_X^2
\leq m_b \Lambda$ implies the relevance of the shape function. A recent
analysis in soft-collinear effective theory (SCET)  shows that by using the universality of the shape
function, a  $10-30\%$ reduction in  the dilepton-mass spectrum can be
accurately computed. Nevertheless, the  effects of subleading shape functions
lead to an additional uncertainty of $5\%$~\cite{Lee:2005pk,Lee:2005pw}.  
A more recent analysis~\cite{Lee:2008xc}
estimates the uncertainties due to subleading shape functions more
conservatively. By scanning over a range of models of these functions,
one finds corrections in the rates relative to the leading-order result
 to be between $-10\%$ to $+10\%$ with equally large
uncertainties.  In the future it may be possible to
decrease such uncertainties significantly by constraining both the
leading and subleading shape functions using the combined $\bar B \to
X_s\gamma$, $\bar B \to X_u\ell \bar \nu$ and $\bar B \to X_s \ell^+\ell^-$
data~\cite{Lee:2008xc,Bernlochner:2011di}. In~\cite{Bell:2010mg}, $\bsll$ in the
presence of a cut on $m_{X_s}$ was analysed by performing the matching from QCD
onto SCET at NNLO, and a prediction of the zero of the forward-backward asymmetry in this semi-inclusive channel was provided.

\item
As already  discussed, the $c\bar c$  resonances
can be removed by making appropriate kinematic cuts in the invariant mass
spectrum.  However, nonperturbative contributions away from the
resonances within the perturbative windows are also important. In the
KS approach~\cite{Kruger:1996cv,Kruger:1996dt}
one absorbs factorizable long-distance charm rescattering effects (in which the
$\bar B \to X_s c\bar cr$ transition can be factorized into the product of
$\bar s  b$ and $c\bar c$ color-singlet currents) into the matrix element
of the leading semileptonic operator ${\cal O}_9$.  Following the inclusion of 
nonperturbative corrections scaling with $1/m_c^2$, the KS approach
avoids double-counting.  For the integrated branching fractions one
finds an increase of $(1-2)\%$ in the low-$q^2$ region due to the
KS effect, whereas  in the high-$q^2$ region there is a decrease of $\sim 10\%$, which is still below the
uncertainty due to the $1/m_b$ corrections.
  
\item  The integrated branching fraction is dominated
by this resonance background which exceeds the nonresonant charm-loop
contribution by two orders of magnitude.  
This feature should not be
misinterpreted as a striking failure of global parton-hadron duality~\cite{Beneke:2009az},
which postulates that the sum over the hadronic final states, including
resonances, should be well approximated by a quark-level
calculation~\cite{Poggio:1975af}.  Crucially, the
charm-resonance contributions to the decay $\bar B \to X_s \ell^+\ell^-$
are expressed in terms of a phase-space integral over the absolute
square of a correlator.  For such a quantity global quark-hadron
duality is not expected to hold.  Nevertheless,
local quark-hadron duality (which, of course, also implies global duality) 
  may be reestablished by resumming Coulomb-like
interactions~\cite{Beneke:2009az}.

\item 
Also electromagnetic perturbative  corrections were calculated: NLL quantum
electrodynamics (QED) two-loop corrections to the
Wilson coefficients are of ${\cal O}(2\%)$~\cite{Bobeth:2003at}. 
In the QED one-loop corrections to matrix elements, large
collinear logarithms of the form $\log(m_b^2/m^2_\ell)$ survive
integration over phase space if only a restricted part of the dilepton mass spectrum is
considered. These collinear logarithms add another contribution of order $+2\%$ in the low-$q^2$ region of the dilepton mass spectrum in  $\bar B \to X_s \mu^+\mu^-$~\cite{Huber:2005ig}. For  the high-$q^2$ region,  one finds   $-8\%$~\cite{Huber:2007vv}.

\end{itemize}
Based on all these  scientific efforts of various groups, the latest theoretical predictions have been presented in ref.~\cite{Huber:2007vv}.\\

In the present manuscript, we make the effort to provide all missing relevant perturbative contributions to all  independent  observables in the decay $\bar B \rightarrow X_s \ell^+\ell^-$. 
As it is well-known, the angular decomposition of this inclusive decay rate provides three independent observables, $H_T$, $H_A$,
$H_L$ from which one can extract the short-distance electroweak Wilson coefficients that test for possible new physics~\cite{Lee:2006gs}:
\begin{eqnarray}\label{eq:d3Gamma}
\frac{d^2\Gamma}{d q^2\,  d z}
&=& \frac{3}{8} \bigl[(1 + z^2) H_T(q^2)
+ 2(1 - z^2) H_L(q^2)
+  2 z H_A(q^2)
\bigr]
\,.
\label{diffwidth}
\end{eqnarray}
Here, $z=\cos\theta$, where $\theta$ is the angle between the $\ell^{+}$ and $B$ meson three 
momenta in the di-lepton rest frame, $H_A$ is equivalent to the forward-backward asymmetry~\cite{Ali:1991is}, and the $q^2$ 
spectrum is given by $H_T + H_L$.  The observables  dominantly depend on the effective Wilson coefficients  
corresponding to the operators  ${\cal O}_7, {\cal O}_9$, and ${\cal O}_{10}$.

The paper is organized as follows. In section~\ref{sec:observables} we define the observables which we consider in the present analysis. In section~\ref{sec:QEDdoublediff} the derivation of the log-enhanced terms is presented. Master formulae for our observables are given in section~\ref{sec:masterformulas}, our phenomenological results in section~\ref{sec:results}.  We briefly discuss the new physics sensitivity of our observables in section~\ref{sec:NP}. Finally we explore the precise connection between experimental and theoretical quantities using Monte Carlo techniques in section~\ref{sec:PHOTOS}. The latter analysis updates, and in parts supersedes, our previous statements in ref.~\cite{Huber:2008ak}. We conclude in section~\ref{sec:conclusions}. In the appendices we collect various functions that arise in the computation of QED and QCD corrections to the observables (appendix~\ref{app:functions}), as well as formulas that parametrise the observables in terms of ratios of high-scale Wilson coefficients (appendix~\ref{app:NPformulae}).


\section{Definition of the observables}
\label{sec:observables}
The $z$ dependence of the double differential decay distribution presented in eq.~(\ref{eq:d3Gamma}) is exact to all orders in QCD because it is controlled by the square of the leptonic current. The inclusion of QED bremsstrahlung modifies the simple second order polynomial structure and replaces it with a complicated analytical $z$ dependence (see eqs.~(\ref{eq:QEDdoublediff})-(\ref{eq:xi910})). In particular this implies that, as long as QED effects are observably large, a simple fit to a quadratic polynomial will introduce non-negligible distorsions in the comparison between theory and experiment. In this section we explain the procedure that we adopt to construct various $q^2$ differential distributions and suggest that experimental analyses follow the same prescriptions.

The extraction of multiple differential distributions from eq.~(\ref{eq:d3Gamma}) is phenomenologically important because the various observables have different functional dependence on the Wilson coefficients. For instance, at next-to-leading order in QCD and without including any QED effect, the three $H_I$ defined in eq.~(\ref{eq:d3Gamma}) are given by~\cite{Lee:2006gs}:
\allowdisplaybreaks{
\begin{align}
 H_T(q^2)= \;& \frac{G_F^2 m_b^5\left| V_{ts}^* V_{tb} \right|^2}{48 \pi^3}  \, 2 \s (1-\s)^2 \Big[\big|C_9+\frac{2}{\s} \, C_7\big|^2+|C_{10}|^2\Big] \, , \label{eq:HT-nlo}\\
 H_L(q^2)= \;& \frac{G_F^2 m_b^5\left| V_{ts}^* V_{tb} \right|^2}{48 \pi^3}  \,  (1-\s)^2 \Big[\big|C_9+2 \, C_7\big|^2+|C_{10}|^2\Big] \, , \label{eq:HL-nlo}\\
 H_A(q^2)= \;& \frac{G_F^2 m_b^5\left| V_{ts}^* V_{tb} \right|^2}{48 \pi^3}  \,  (-4\s) \, (1-\s)^2 \; {\rm{Re}}\!\left[C_{10} \big(C_9+\frac{2}{\s} \, C_7\big) \right]\, .
\label{eq:HA-nlo}
\end{align}}

We decided to preserve the natural definitions of the differential decay width $d\Gamma/dq^2$ and of the forward--backward asymmetry $d{\cal A}_{\rm FB}/dq^2$:
\begin{align}
\frac{d\Gamma}{dq^2} &\equiv \int_{-1}^{+1} \frac{d^2\Gamma}{dq^2 dz} dz \; ,\\
\frac{dA_{\rm FB}}{dq^2} &\equiv \int_{-1}^{+1} \frac{d^2\Gamma}{dq^2 dz} {\rm sign}(z) dz \;,
\label{eq:fba}
\end{align}
with the understanding that $A_{\rm FB}$ does not coincide with the coefficient of the linear term in $z$ in the Taylor expansion of $d^2\Gamma/dq^2dz$. 

We extract other single-differential distributions by projecting the double-differential rate onto various Legendre polynomials, $P_n(z)$. These polynomials are orthogonal in the $[-1,1]$ interval and are, therefore, ideally suited as angular projectors. In order to make connection with the existing literature we choose the first two projections in such a way to reproduce $H_T$ and $H_L$ in the limit of no QED radiation. For the higher order terms we simply adopt the corresponding Legendre polynomials. The observables are defined as
\begin{align}
H_I (q^2) &= \int_{-1}^{+1} \frac{d^2\Gamma}{dq^2dz} W_I(z) dz \label{eq:projectionHI}
\end{align}
and the weights we use are:
\begin{equation}
\begin{aligned}
W_T &= \frac{2}{3} \; P_0 (z) + \frac{10}{3} \; P_2 (z) \, , &\quad\quad & W_3 = P_3 (z) \, , \\
W_L &= \frac{1}{3} \; P_0 (z) - \frac{10}{3} \; P_2 (z) \, , &\quad\quad & W_4 = P_4 (z) \, , \\
W_A &= \frac{4}{3} {\rm sign} (z) \; .
\label{weights}
\end{aligned}
\end{equation}
Note that $W_T + W_L = P_0(z) = 1$ implying that the relation $d\Gamma/dq^2 = H_T + H_L$ is exact. 

The unnormalized (defined in eq.~(\ref{eq:fba})) forward--backward asymmetry receives contributions form all odd powers of $z$ in the Taylor expansion of the double differential rate and is given by
\begin{align}
\frac{dA_{\rm FB}}{dq^2} &= \frac{3}{4} \; H_A (q^2) \; .
\end{align}
In the literature the normalized differential and integrated forward--backward asymmetries are often considered (see for instance ref.~\cite{Huber:2007vv}):
\begin{align}
\frac{d \overline A_{\rm FB}}{dq^2} &\equiv 
\frac{\displaystyle\int_{-1}^{+1} dz \frac{d^2\Gamma}{dq^2 dz} {\rm sign}(z) }{\displaystyle
\int_{-1}^{+1} dz\frac{d^2\Gamma}{dq^2 dz} } = \frac{3}{4} \frac{H_A (q^2)}{H_T(q^2) + H_L(q^2)} \, ,\\
\overline A_{\rm FB}[q_m^2,q_M^2] &\equiv 
\frac{\displaystyle\int_{q^2_m}^{q^2_M}dq^2 \int_{-1}^{+1}dz \frac{d^2\Gamma}{dq^2 dz} {\rm sign}(z)  }{\displaystyle
\int_{q^2_m}^{q^2_M} dq^2 \int_{-1}^{+1} dz \frac{d^2\Gamma}{dq^2 dz} }
 = \frac{3}{4} \frac{\displaystyle \int_{q^2_m}^{q^2_M} dq^2 H_A (q^2)}{
\displaystyle \int_{q^2_m}^{q^2_M} dq^2 \left[ H_T (q^2) + H_L(q^2) \right]} \, .
\label{eq:normFBA}
\end{align}

 The new observables $H_3$ and $H_4$ (obtained by employing the weights $W_3$ and $W_4$) vanish exactly in the limit of no QED radiation but are still potentially important for phenomenology because of their non trivial dependence on the Wilson coefficients. We find that projections with even higher Legendre polynomials are suppressed and will not be considered further.

Note that the expected statistical experimental uncertainties (at a given luminosity) are well understood in the total width ($H_T + H_L$) and forward-backward asymmetry ($3/4 \; H_A$) cases. On the other hand, $H_T$, $H_L$, $H_3$ and $H_4$ are obtained by projecting the double differential rate with weights that (especially for $W_3$ and $W_4$) are essentially arbitrary. As a consequence a simple rescaling of these weights implies a corresponding rescaling of the central values we find. In section~\ref{sec:NP} we show how to use the squared weights ($W_I^2$) to assess the expected Belle~II reach for each of these observables.

The experimental procedure that we recommend is to use the weights $W_I$ to extract single-differential distributions and to refrain from attempting polynomial fits to the data.


\section{Log-enhanced QED corrections to the double differential decay rate}
\label{sec:QEDdoublediff}

\begin{figure}[t]
\begin{center}
\includegraphics[width=0.23 \linewidth]{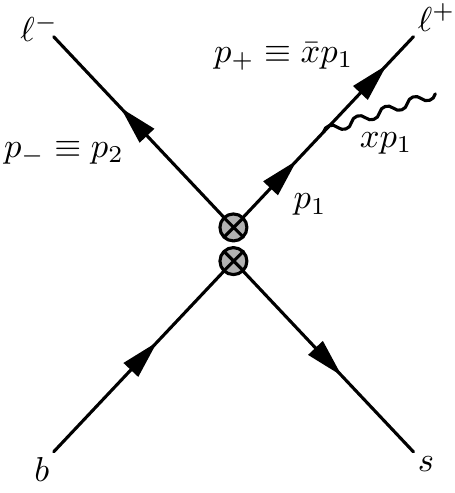}
\quad\quad\quad
\includegraphics[width=0.23 \linewidth]{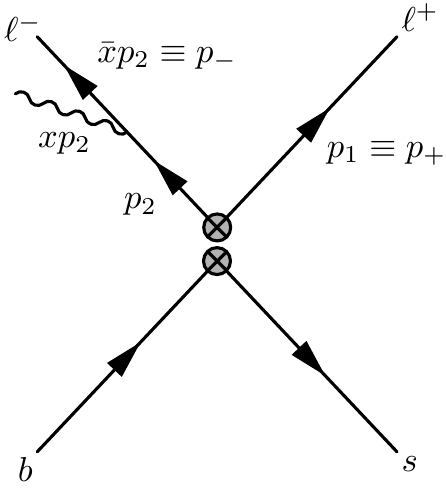}
\vspace*{10pt}
\caption{Kinematics of collinear photon radiation. The collinear photon is
radiated off $\ell^+$ (left panel) and $\ell^-$ (right panel), respectively.
The crossed grey circles denote an operator insertion from the effective weak Hamiltonian.
The arrows indicate momentum rather than fermion flow. $x$ denotes the momentum fraction
of the collinear photon.
\label{fig:splitting}}
\end{center}
\end{figure}

In this section we work out  the formulas for the logarithmically enhanced electromagnetic corrections of  the double differential decay
rate $d^2\Gamma/(dq^2 \, dz)$. The operators and Wilson coefficients of the effective weak Hamiltonian are the same as
in~\cite{Huber:2005ig,Huber:2007vv}. The kinematics can be inferred from figure~\ref{fig:splitting}.

Let us first consider the case without photon radiation. The momenta of the quarks are labelled $p_b$ and $p_s$, respectively. The
momenta of $\ell^+$ and $\ell^-$ are denoted by $p_1$ and $p_2$, respectively. From momentum conservation we arrive at  $p_b = p_1 + p_2 +
p_s$. We define the invariants
\be\label{eq:sij}
s_{ij} \equiv \frac{2p_i \, p_j}{m_b^2} \, , \qquad i \in \{1,2,s,b\} \, .
\ee
Moreover, we define
\be\label{eq:y12}
y_{1} \equiv \frac{2 \, E_{1}}{m_b} \, , \qquad y_{2} \equiv \frac{2 \, E_{2}}{m_b} \, , 
\ee
where $E_i$ ($i=1,2$) is the zero-component of $p_i$ when evaluated in the rest-frame of the decaying $b$-quark. From momentum
conservation and by treating all final-state particles as massless, we obtain the relation $y_1 + y_2 = 1 + s_{12}$. This relation  also implies
\be
\begin{array}{*{3}{l}}
s_{1s} = 1-y_2 \, , \qquad & s_{1b} = y_1 \, , \qquad & s_{sb} = 1-s_{12} \, ,\\
s_{2s} = 1-y_1 \, , \qquad & s_{2b} = y_2 \, . \qquad&  \\
\end{array}
\ee
For the double differential decay rate we also need $z \equiv \cos\theta$, where $\theta$ is the angle between the $b$-quark and the
positively charged lepton in the centre-of-mass system (c.m.s.) of the final-state lepton pair. Hence
\be
z = \cos\theta = \frac{\vec p_1^{\; \prime} \cdot \vec p_b^{\; \prime}}{\left|\vec p_1^{\; \prime}\right| \,  \left|\vec p_b^{\; \prime}\right|} \; ,
\ee
where all primed momenta are taken in the c.m.s.\ of the final-state lepton pair. It turns out that $z$ is simply given
by~\cite{Ali:1996bm}
\be\label{eq:z}
z = \frac{y_2-y_1}{1-s_{12}} \, .
\ee
At this point we stress that the LHS of this equation is evaluated in the lepton c.m.s., whereas its RHS is evaluated in the rest-frame of the
decaying $b$-quark. The connection between the angle $\theta$ and the leponic energy asymmetry has already been emphasized in~\cite{Ali:1996bm}.

We now switch on QED and consider the radiation of a collinear photon off a lepton leg as shown in figure~\ref{fig:splitting}.
The momentum of the positively (negatively) charged lepton is denoted by $p_1$ ($p_2$) before radiation and by $p_+$ ($p_-$) thereafter.
If the positively charged lepton radiates the photon (left panel of figure~\ref{fig:splitting}), its momentum $p_+$ after radiation is
given by $p_+  = \bar x p_1$, where $x$ denotes the momentum fraction of the collinear photon and $\bar x \equiv 1-x$. In this case, the
momentum of the negatively charged lepton remains unchanged and hence we have $p_-  = p_2$. If the negatively charged lepton radiates the
photon (right panel of figure~\ref{fig:splitting}), we obviously have $p_-  = \bar x p_2$ and $p_+  = p_1$. In analogy to
eq.~(\ref{eq:y12}), we define
\be
y_\pm \equiv \frac{2 \, E_\pm}{m_b} \, ,
\ee
where $E_\pm$ is the zero-component of $p_\pm$, again evaluated in the rest-frame of the decaying $b$-quark. We will also need the
definition
\be
s_{+-} \equiv \bar x \, s_{12} \; .
\ee

As already discussed in refs.~\cite{Huber:2005ig,Huber:2007vv}, the logarithmically enhanced contributions stemming  from collinear photon
radiation are evaluated by 
\be\label{eq:doubleminustriple}
\dps \frac{d^2{\Gamma}_{{\rm coll}}}{ds \, dz} = \frac{d^2{\Gamma}_{{\rm coll},2}}{ds \, dz} - \frac{d^2{\Gamma}_{{\rm coll},3}}{ds \, dz} \; ,
\ee
where we stay differential in the double invariant $s_{+-} = (p_{+} + p_{-})^2 = \bar x \, s_{12}$ and the triple invariant
$s_{12} = (p_{+} + p_{-} + p_\gamma)^2 = (p_1 +p_2)^2$, respectively. We first consider the case of the triple invariant, where the
formulae look exactly the same as in the case without QED, since we can lump the lepton and the collinear photon. We therefore arrive at 
\bea
d\Gamma_{{\rm coll},3} &=& PF \, ds_{12} \, dy_1 \, dy_2 \, dx \, \delta(1+s_{12}-y_1-y_2) \, f^{(m)}_{\gamma}(x)
\, \left[\left|{\cal A}\right|^2(s_{12},y_1,y_2)\right] \nnb \\
&& \times \theta(y_1) \, \theta(1-y_1) \, \theta(y_2) \, \theta(1-y_2) \, \theta(s_{12}) \, \theta(1-s_{12})
\, \theta(x) \, \theta(1-x) \, . \label{eq:gammacoll3}
\eea
Here $\theta$ denotes the heaviside step-function, $PF$ is the pre-factor 
\be
\dps PF = \frac{G_F^2 m_b |V_{tb} V_{ts}^{\ast}|^2}{32 \pi^3} \; ,
\ee
and $f^{(m)}_{\gamma}(x)$ is the mass-regularised splitting function for collinear photon radiation of which we only keep the
logarithmically enhanced part ($\aem = \alpha_e/(4\pi)$),
\be
f^{(m)}_{\gamma}(x) = 4 \, \aem \, \frac{[ 1+(1-x)^2]}{x} \, \ln\!\lk\frac{m_b}{m_{\ell}}\rk \; .
\ee
The squared matrix elements $\left|{\cal A}\right|^2$ for the different operators read
\allowdisplaybreaks{
\bea
\left|{\cal A}\right|^2_{77} (s_{12},y_1,y_2) &=&  \frac{8 m_b^4}{s_{12}} \, \left[(1-y_2) \, y_1 + (1-y_1) \, y_2\right]\, ,\nnb \\
\left|{\cal A}\right|^2_{79} (s_{12},y_1,y_2) &=&  4 m_b^4 \, (1-s_{12})\, ,\nnb \\
\left|{\cal A}\right|^2_{99} (s_{12},y_1,y_2) &=&  4 m_b^4 (1-y_1) (1-y_2) + 2 m_b^4 \, s_{12} \, (1-s_{12})\, ,\nnb \\
\left|{\cal A}\right|^2_{710} (s_{12},y_1,y_2) &=& 4 m_b^4 \, (y_1 - y_2) \, ,\nnb \\
\left|{\cal A}\right|^2_{910} (s_{12},y_1,y_2) &=& 2 m_b^4 \, s_{12} \, (y_1 - y_2) \, ,\nnb \\
\left|{\cal A}\right|^2_{29} (s_{12},y_1,y_2) &=& \aem \, f_2(s_{12}) \, \left|{\cal A}\right|^2_{99} (s_{12},y_1,y_2) \, ,\nnb \\
\left|{\cal A}\right|^2_{27} (s_{12},y_1,y_2) &=& \aem \, f_2(s_{12}) \, \left|{\cal A}\right|^2_{79} (s_{12},y_1,y_2) \, ,\nnb \\
\left|{\cal A}\right|^2_{22} (s_{12},y_1,y_2) &=& \aem^2 \, |f_2(s_{12})|^2 \, \left|{\cal A}\right|^2_{99} (s_{12},y_1,y_2) \, ,\nnb \\
\left|{\cal A}\right|^2_{210} (s_{12},y_1,y_2) &=& \aem \, f_2(s_{12}) \, \left|{\cal A}\right|^2_{910} (s_{12},y_1,y_2) \, . \label{eq:MEsquared}
\eea}
The function $f_2(s_{12})$ denotes the one-loop matrix element of $P_2$ and is given by
\bea
f_2(s_{12}) & = &\f{8}{9} \ln\left(\!\frac{\mu}{m_c}\!\right) +
\frac{8}{27} + \f{4}{9} y_c - \f{2}{9}(2+y_c) \sqrt{|1-y_c|} \left\{ \begin{array}{ll}
\ln \left|\f{1+\sqrt{1-y_c}}{1-\sqrt{1-y_c}}\right| - i \pi, & {\rm when}~ y_c < 1 \; ,\\[2mm]
2 \arctan \f{1}{\sqrt{y_c-1}},                     & {\rm when}~ y_c \ge 1 \;,
\end{array} \right. \nnb \\
\eea
with $y_c=4 m_c^2/(m_b^2 s_{12})$. $f_2(s_{12})$ is a complex function and therefore the $\left|{\cal A}\right|^2_{2j}$ with $j \neq 2$ are complex. However, after taking into account the Wilson coefficients and adding the appropriate complex conjugate expression, the double differential rate turns out to be real, see eq.~(\ref{eq:QEDdoublediff}).

Let us now come back to the evaluation of~(\ref{eq:gammacoll3}). After integrating over the $\delta$-function and changing variables according to eq.~(\ref{eq:z}) we arrive at
\bea\label{eq:triple}
\frac{d^2\Gamma_{{\rm coll},3}}{ds_{12} \, dz} &=& 2 \, PF \, \int\limits_{0}^{1} \!\! dx \, f^{(m)}_{\gamma}(x)
\, \left[\left|{\cal A}\right|^2\left(s_{12},\frac{1+s_{12}}{2}-\frac{1-s_{12}}{2} \,z ,\frac{1+s_{12}}{2}+\frac{1-s_{12}}{2}\,z\right)\right] \nnb \\
&& \times \, \frac{1-s_{12}}{2} \,\, \theta(1-z) \, \theta(1+z) \, \theta(s_{12}) \, \theta(1-s_{12}) \, .
\eea
The factor of two stems from the fact that both diagrams in figure~\ref{fig:splitting}  are relevant. Note that the integral in
eq.~(\ref{eq:triple}) is divergent at $x=0$. However, eq.~(\ref{eq:doubleminustriple}) is well-behaved once all expressions on its
RHS are plugged in.

\begin{figure}[t]
\begin{center}
 \includegraphics[scale = 1.5]{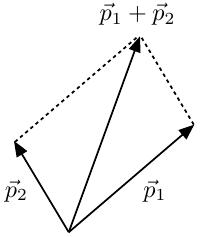}
\quad\quad\quad
 \includegraphics[scale = 1.5]{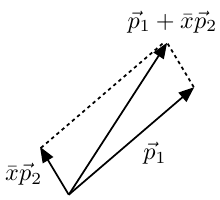}
\quad\quad\quad
 \includegraphics[scale = 1.5]{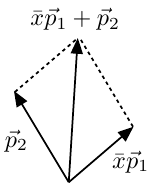}
\caption{\label{fig:boosts} Boosts that have to be performed in order to translate the $b$ rest-frame into the leptonic c.m.s. Left panel: without QED. Middle panel: Radiation off $\ell^-$. Right panel: Radiation off $\ell^+$.}
\end{center}
\end{figure}

We now turn our attention to the more complicated case of the double invariant $d^2{\Gamma}_{{\rm coll},2}/(ds \, dz)$, and first address
radiation off $\ell^-$. As can be seen from the middle panel of figure~\ref{fig:boosts}, the boost $\Lambda$ from the $b$-quark rest-frame
into the leptonic c.m.s.\ is determined by
\be
\vec p_1 + \bar x \vec p_2 \stackrel{\Lambda}{\longrightarrow} 0.
\ee
After the boost, we compute $z = \cos\theta$ via
\be\label{eq:zlminus}
z = \frac{\vec p_+^{\; \prime} \cdot \vec p_b^{\; \prime}}{\left|\vec p_+^{\; \prime}\right| \,  \left|\vec p_b^{\; \prime}\right|}
= \frac{\bar x y_2 - y_1}{\sqrt{\left(y_1+\bar x y_2\right)^2-4\bar x s_{12}}}\; ,
\ee
Again, the primed momenta are evaluated in the lepton c.m.s., whereas the RHS of the equation is evaluated in the rest-frame of the
$b$-quark. The differential decay width reads
\bea
d\Gamma_{{\rm coll},2}^{\ell^-} &=& PF \, ds_{12} \, dy_1 \, dy_2 \, dx \, \delta(1+s_{12}-y_1-y_2) \, f^{(m)}_{\gamma}(x)
\, \left[\left|{\cal A}\right|^2(s_{12},y_1,y_2)\right] \nnb \\
&& \times \theta(y_1) \, \theta(1-y_1) \, \theta(y_2) \, \theta(1-y_2) \, \theta(s_{12}) \, \theta(1-s_{12})
\, \theta(x) \, \theta(1-x) \, .
\eea
We first eliminate $y_1$ by integrating over the $\delta$-function. Subsequently, we eliminate $y_2$ in favour of  $z$ according to
eq.~(\ref{eq:zlminus}). This transformation reads
\be\label{eq:y2vonz}
y^{(\pm)}_2(z) = \frac{(1+s_{12}) (2-x-xz^2) \pm 2z\sqrt{1-x}\sqrt{(1-s_{12})^2\bar x-s_{12}x^2(1-z^2)}}{x^2(1-z^2)+4 \bar x} \, .
\ee
It turns out that this in an injective mapping only for $s_{12} < \bar x$. For $s_{12} > \bar x$ we have to subdivide the $y_2$-interval
into two pieces, so that we get a total of three contributions. After the additional variable substitution $s_{12} = s_{+-}/\bar x$ they
read
\allowdisplaybreaks{
\bea
\frac{d^2\Gamma_{{\rm coll},2; \; 1}^{\ell^-}}{ds_{+-} \, dz} &=& PF \, \int\limits_0^{1-\sqrt{s_{+-}}} \!\!\!\!\! dx \;\;
\frac{f^{(m)}_{\gamma}(x)}{\bar x} \, \left[\frac{\partial}{\partial z} y^{(+)}_2(z)\right]
\,\left[\left|{\cal A}\right|^2(s_{12},1+s_{12}-y_2,y_2)\right]
 {}_{\Bigg| \begin{array}{l} y_2 = y_2^{(+)}(z) \\ s_{12} = s_{+-}/\bar x\end{array}} \nnb \\
&& \times \theta(1-z) \, \theta(1+z) \, \theta(s_{+-}) \, \theta(1-s_{+-}) \; , \\
&&\nnb \\
&&\nnb \\
\frac{d^2\Gamma_{{\rm coll},2; \; 2/3}^{\ell^-}}{ds_{+-} \, dz} &=& \pm PF \, \int\limits_{1-\sqrt{s_{+-}}}^{x_{-}} \!\!\!\!\! dx \;\;
\frac{f^{(m)}_{\gamma}(x)}{\bar x} \, \left[\frac{\partial}{\partial z} y^{(\pm)}_2(z)\right]
\,\left[\left|{\cal A}\right|^2(s_{12},1+s_{12}-y_2,y_2)\right]
 {}_{\Bigg| \begin{array}{l} y_2 = y_2^{(\pm)}(z) \\ s_{12} = s_{+-}/\bar x\end{array}} \nnb \\
&& \times \theta(-z) \, \theta(1+z) \, \theta(s_{+-}) \, \theta(1-s_{+-}) \; ,
\eea}
where
\be
x_\pm = \frac{1-s_{+-}}{1\mp \sqrt{(1-z^2) s_{+-}}} \; .
\ee

Once the photon is radiated off $\ell^+$,  we apply very similar steps. As can be seen from the right panel of figure~\ref{fig:boosts}, the
boost $\Lambda$, from the $b$-quark rest-frame into the leptonic c.m.s.,  is determined by
\be
\bar x \vec p_1 + \vec p_2 \stackrel{\Lambda}{\longrightarrow} 0.
\ee
After the boost, we compute $z = \cos\theta$ by 
\be\label{eq:zlplus}
z = \frac{\vec p_+^{\; \prime} \cdot \vec p_b^{\; \prime}}{\left|\vec p_+^{\; \prime}\right| \,  \left|\vec p_b^{\; \prime}\right|}
= \frac{ y_2 - \bar x y_1}{\sqrt{\left(\bar x y_1+y_2\right)^2-4\bar x s_{12}}}\; ,
\ee
We now eliminate $y_2$ by integrating over the $\delta$-function. Subsequently, we eliminate $y_1$ in favour of  $z$ according to
eq.~(\ref{eq:zlplus}). This transformation reads
\be\label{eq:y1vonz}
y^{(\pm)}_1(z) = \frac{(1+s_{12}) (2-x-xz^2) \pm 2z\sqrt{1-x}\sqrt{(1-s_{12})^2\bar x-s_{12}x^2(1-z^2)}}{x^2(1-z^2)+4 \bar x} \, .
\ee
As mentioned before, this is an injective mapping only for $s_{12} < \bar x$. For $s_{12} > \bar x$ we have to subdivide the $y_1$-interval
into two pieces, so that in this case we also get a total of three contributions. After the additional variable substitution $s_{12} =
s_{+-}/\bar x$ they read
\allowdisplaybreaks{
\bea
\frac{d^2\Gamma_{{\rm coll},2; \; 1}^{\ell^+}}{ds_{+-} \, dz} &=& -PF \, \int\limits_0^{1-\sqrt{s_{+-}}} \!\!\!\!\! dx \;\;
\frac{f^{(m)}_{\gamma}(x)}{\bar x} \, \left[\frac{\partial}{\partial z} y^{(-)}_1(z)\right]
\,\left[\left|{\cal A}\right|^2(s_{12},y_1,1+s_{12}-y_1)\right]
 {}_{\Bigg| \begin{array}{l} y_1 = y_1^{(-)}(z) \\ s_{12} = s_{+-}/\bar x\end{array}} \nnb \\
&& \times \theta(1-z) \, \theta(1+z) \, \theta(s_{+-}) \, \theta(1-s_{+-}) \; , \\
&&\nnb \\
&&\nnb \\
\frac{d^2\Gamma_{{\rm coll},2; \; 2/3}^{\ell^+}}{ds_{+-} \, dz} &=& \mp PF \, \int\limits_{1-\sqrt{s_{+-}}}^{x_{-}} \!\!\!\!\! dx \;\;
\frac{f^{(m)}_{\gamma}(x)}{\bar x} \, \left[\frac{\partial}{\partial z} y^{(\mp)}_1(z)\right]
\,\left[\left|{\cal A}\right|^2(s_{12},y_1,1+s_{12}-y_1)\right]
 {}_{\Bigg| \begin{array}{l} y_1 = y_1^{(\mp)}(z) \\ s_{12} = s_{+-}/\bar x\end{array}} \nnb \\
&& \times \theta(z) \, \theta(1-z) \, \theta(s_{+-}) \, \theta(1-s_{+-}) \; .
\eea}
The total contribution in case of the double invariant is now obtained by
\be\label{eq:doubletotal}
\frac{d^2\Gamma_{{\rm coll},2}}{ds_{+-} \, dz} = \sum\limits_{i=1}^{3} \, \left[\frac{d^2\Gamma_{{\rm coll},2; \; i}^{\ell^+}}{ds_{+-} \, dz}
+ \frac{d^2\Gamma_{{\rm coll},2; \; i}^{\ell^-}}{ds_{+-} \, dz} \right] \; .
\ee
We finally identify $s_{12} \equiv s$ in eq.~(\ref{eq:triple}) and $s_{+-} \equiv s$ in eq.~(\ref{eq:doubletotal}) and plug everything
into eq.~(\ref{eq:doubleminustriple}). This leads us to the following expression for the logarithmically enhanced collinear decay width
\bea
\dps \frac{d^2{\Gamma}_{{\rm coll}}}{ds \, dz} &=& \frac{G_F^2 m_b^5 |V_{tb} V_{ts}^{\ast}|^2}{32 \pi^3} \; \aem \,
\ln\!\left(\frac{m_b^2}{m_\ell^2}\right) \Bigg\{ |C_9|^2 \;  \xi_{99}^{\rm (em)}(s,z) + |C_{10}|^2 \; \xi_{99}^{\rm (em)} (s,z) \nnb \\
&& + \aem^2 \; |C_{7}^{{\rm eff}}|^2 \; \xi_{77}^{\rm (em)} (s,z)
+ \aem \; {\rm Re} \left[C_7^{{\rm eff}} C_9^*\right] \; \xi_{79}^{\rm (em)}(s,z)
+ \aem \;  {\rm Re} \left[C_7^{{\rm eff}} C_{10}^*\right] \;  \xi_{710}^{\rm (em)}(s,z) \nnb \\
&& + {\rm Re} \left[C_9 C_{10}^*\right] \;  \xi_{910}^{\rm (em)}(s,z)
+ \aem^2 \; {\rm Re} \left[(C_2 + C_F C_1) \, C_7^{{\rm eff} \, *} \;  \xi_{27}^{\rm (em)}(s,z)\right] \nnb \\
&& + \aem \; {\rm Re} \left[(C_2 + C_F C_1) C_9^* \;  \xi_{29}^{\rm (em)}(s,z) \right]
+ \aem \; {\rm Re} \left[(C_2 + C_F C_1) C_{10}^* \;  \xi_{210}^{\rm (em)}(s,z) \right] \nnb \\
&& + \aem^2 \; (C_2 + C_F C_1)^2 \;  \xi_{22}^{\rm (em)} (s,z) \Bigg\} \; \; , \label{eq:QEDdoublediff}
\eea
where we assumed that the Wilson coefficients $C_1$ and $C_2$ are real, and we neglected contributions from the penguin operators
$P_{3-6}$ due to their small Wilson coefficients. The functions $\xi^{\rm (em)}_{ij}(s,z)$ are given by
\allowdisplaybreaks
\bea
\xi^{\rm{(em)}}_{77}(s,z) &=& -\frac{64 \, p_1(s,z) \, \sqrt{s} \, \ln\!
   \left(\sqrt{\frac{s}{1-z^2}}-\sqrt{\frac{s}{1-z^2}-1}\right)}{\left(
   z^2-1\right)^3 \sqrt{s+z^2-1}}
   +\frac{64 \, z \, p_2(s,z) \, \ln\! \left(\frac{1-z}{z+1}\right)}{s (z^2-1)^3} \nnb \\
   &&+\frac{64 \, p_3(s,z) \, \ln\! \left(\sqrt{\frac{1}{s (1-z^2)}}-\sqrt{\frac{1}{s (1-z^2)}-1}\right)}{s (z^2-1)^3 \left(s
   \left(z^2-1\right)+1\right)^{3/2}}
   +\frac{16 \, p_4(s,z) \,\ln(s)}{s (z^2-1)^3} \nnb \\
   &&+\frac{4 \, p_5(s,z)}{3 s \left(z^2-1\right)^2 \left(s
   \left(z^2-1\right)+1\right)}-\frac{16 (s-1)^2 \, p_6(s,z) \, \ln\! \left(\frac{2
   (1-s)}{\sqrt{1-z^2}}\right)}{s} \; , \\
\xi^{\rm{(em)}}_{99}(s,z) &=& -\frac{16 \, s \, z \, p_7(s,z) \, \ln\!
   \left(\frac{1-z}{z+1}\right)}{\left(z^2-1\right)^4}
   +\frac{4 \, s \, p_8(s,z) \, \ln(s)}{\left(z^2-1\right)^4} \nnb \\
   &&+\frac{8\, s^{3/2} \, p_9(s,z) \, \ln\!\left(\sqrt{\frac{s}{1-z^2}}-\sqrt{\frac{s}{1-z^2}-1}\right)}{\left(
   z^2-1\right)^4 \left(s+z^2-1\right)^{5/2}}
   +\frac{p_{10}(s,z)}{3(z^2-1)^3 \left(s+z^2-1\right)^2}\nnb \\
   &&+4 (s-1)^2 \left(s z^2+s-z^2+1\right) \ln\! \left(\frac{2 (1-s)}{\sqrt{1-z^2}}\right)\; , \\
\xi^{\rm{(em)}}_{79}(s,z) &=&-\frac{64 \, z \, p_{11}(s,z)\, \ln\!\left(\frac{1-z}{z+1}\right)}{(z^2-1)^3}
   -\frac{32 \, p_{12}(s,z) \, \ln(s)}{(z^2-1)^3} \nnb \\
   &&-\frac{8 \, p_{13}(s,z)}{\left(z^2-1\right)^2\left(s+z^2-1\right)}
   +\frac{64 \sqrt{s} \, p_{14}(s,z) \, \ln\! \left(\sqrt{\frac{s}{1-z^2}}-\sqrt{\frac{s}{1-z^2}-1}\right)}{\left(
   z^2-1\right)^3 \left(s+z^2-1\right)^{3/2}} \nnb \\
   &&+32 \, (s-1)^2 \ln\!\left(\frac{2 (1-s)}{\sqrt{1-z^2}}\right)
   +\frac{64 \, p_{15}(s,z) \, \ln\! \left(\sqrt{\frac{1}{s (1- z^2)}}-\sqrt{\frac{1}{s (1-z^2)}-1}\right)}
   {\left(z^2-1\right)^2 \sqrt{s\left(z^2-1\right)+1}} , \\
\xi^{\rm{(em)}}_{710}(s,z) &=& -\frac{64 \, p_{15}(s,z) \; {\rm{sign}}(z) \ln\!
   \left(\frac{-\sqrt{s} \left(z^2-1\right)-\sqrt{z^2} \sqrt{s
   \left(z^2-1\right)+1}+1}{\left(\sqrt{s}+1\right)
   \sqrt{1-z^2}}\right)}{\left(z^2-1\right)^2 \sqrt{s
   \left(z^2-1\right)+1}} \nnb \\
   &&-\frac{64 \, p_{16}(s,z) \, \sqrt{s} \; {\rm{sign}}(z) \ln\! \left(\frac{-\sqrt{z^2}
   \sqrt{s+z^2-1}+\sqrt{s}-z^2+1}{\left(\sqrt{s}+1\right)\sqrt{1-z^2}}\right)}{(z^2-1)^3\left(s+z^2-1\right)^{3/2}} \nnb \\
   &&+\frac{32 \, z \, p_{17}(s,z) \, \ln\! \left(\frac{1}{2}
   \left(\sqrt{s}+1\right) \sqrt{1-z^2}\right)}{(z^2-1)^3}
   +\frac{8\left(\sqrt{s}-1\right)^2 z \,p_{18}(s,z)}{\left(z^2-1\right)^2\left(s+z^2-1\right)} \nnb \\
   &&-\frac{64 \, s \, z \, \left(9 s z^2+7 s+4 z^2-4\right)\ln(s)}{(z^2-1)^3}-32 (s-1)^2 \, z \, \ln\! \left(1-\sqrt{s}\right)\; , \\
\xi^{\rm{(em)}}_{910}(s,z) &=&-\frac{32 \, s \, z \, p_{19}(s,z) \, \ln(s)}{\left(z^2-1\right)^4}
   +\frac{16 \, s \, z \, p_{20}(s,z) \, \ln\! \left(\frac{1}{2} \left(\sqrt{s}+1\right)
   \sqrt{1-z^2}\right)}{\left(z^2-1\right)^4} \nnb \\
   && +\frac{4 s \left(\sqrt{s}-1\right)^2 z \, p_{21}(s,z)}{(z^2-1)^3 \left(s+z^2-1\right)^2}
   -\frac{16 \, p_{22}(s,z) \, s^{3/2} \; {\rm{sign}}(z) \ln\! \left(\frac{-\sqrt{z^2}
   \sqrt{s+z^2-1}+\sqrt{s}-z^2+1}{\left(\sqrt{s}+1\right)
   \sqrt{1-z^2}}\right)}{\left(z^2-1\right)^4
   \left(s+z^2-1\right)^{5/2}} \nnb \\
   &&-16 (s-1)^2 \, s \, z \, \ln\!\left(1-\sqrt{s}\right) \; .
\label{eq:xi910}
\eea
The $p_i(s,z)$ are polynomials in $s$ and $z$ and are given in appendix~\ref{app:functions}.
In case of negative or complex arguments, the logarithms and square-roots are defined as
\allowdisplaybreaks{
\bea
\sqrt{z} &=& \sqrt{|z|} \, e^{i/2 \, \arg(z)}\; , \nnb \\
\ln(z) &=& \ln\! |z| \, + \, i \, \arg(z) \; , \nnb \\
\arg(z) &\in& (-\pi,\pi] \; .
\eea}

The functions $\xi^{\rm{(em)}}_{2x}(s,z)$ with $x=2,7,9,10$ cannot be computed analytically since the squared matrix elements (see eq.~(\ref{eq:MEsquared})) are complicated functions of $s_{12}$. We therefore refrain from presenting their explicit expressions. They can easily be computed numerically by applying the steps outlined above.

A strong cross-check is done if we weight  the $\xi^{\rm{(em)}}_{ij}(s,z)$ by unity  or by ${\rm{sign}}(z)$ and subsequently integrating
over $z$. After proper normalisation one obtains the functions $\omega^{\rm{(em)}}_{ij}(s)$ from~\cite{Huber:2005ig,Huber:2007vv}. Note
that this cross-check is non-trivial due to the fact that in our former work we computed the
$\omega^{\rm{(em)}}_{ij}(s)$ in a different way: Since there was no need to introduce the variable $z$  we performed the
calculation entirely in terms of the rescaled energies $y_i$. Moreover, there was more freedom in choosing the order of integrations
since we were not forced to perform the $x$-integration immediately after that over the $\delta$-function. These two simplifications led
to significantly simpler variable substitutions and shorter expressions. With the ability to reproduce them by the more
complicated calculation can therefore be regarded as a non-trivial cross-check.


\section{Master formulas for the observables}
\label{sec:masterformulas}

We start again from the double differential decay width
\begin{align}
\frac{d^2\Gamma}{dz\, d q^2}
&= \frac{3}{8} \bigl[(1 + z^2) H_T(q^2)+  2 z H_A(q^2) + 2(1 - z^2) H_L(q^2) \bigr] \, ,
\end{align}
where $z=\cos\theta$ and $\theta$ is the angle between the $\ell^{+}$ and the $B$ meson three 
momenta in the di-lepton rest frame. This formula is  modified once QED corrections are taken into account (see sections~\ref{sec:observables} and~\ref{sec:QEDdoublediff}) due to the appearance of higher powers of $z$. As stated in section~\ref{sec:observables},  we project out the $H_I$ ($I=T,A,L$) by eq.~(\ref{eq:projectionHI}) in this case. Then the $H_I$ are functions of the dilepton-invariant mass $q^2=m^2_{\ell\ell}$, but obviously not of $z$. $H_A$ is proportional to the lepton forward-backward asymmetry; the $q^2$-spectrum is given by $H_T + H_L$,
\begin{align}
\frac{d\Gamma}{d q^2} =& \int_{-1}^{1} \! dz \, \frac{d^2\Gamma}{dz\, d q^2}
= H_T(q^2) + H_L(q^2) \; , \label{eq:GammaHTHL}\\ 
\frac{d{\cal A}_{FB}}{d q^2}=& \int_{-1}^{1} \! dz \, \frac{d^2\Gamma}{dz\, d q^2} \, {\rm{sign}}(z)
= \frac{3}{4} \, H_A(q^2) \; .\label{eq:AFBHA}
\end{align}
Each of the $H_I$ can be expressed as follows ($\s = q^2/m_{b,{\rm pole}}^2$):
\bea
H_I(q^2)
& = &
\frac{G_F^2 m_{b,{\rm pole}}^5}{48 \pi^3} \left| V_{ts}^* V_{tb} \right|^2 \; \Phi^I_{\ell\ell}(\s),
\eea
where the dimensionless functions $\Phi^I_{\ell\ell}(\s)$ include both perturbative and non-perturbative contributions. Moreover, we normalise the observables to the inclusive semi-leptonic $b \to X_c e \bar{\nu}$ decay. However, the normalisation proceeds in such a way that we insert the perturbative expansion of the inclusive semi-leptonic $b \to X_u e \bar{\nu}$ decay (including power-corrections), and also use the ratio~\cite{Bobeth:2003at,Gambino:2001ew}
\be
C = \left| \frac{V_{ub}}{V_{cb}} \right|^2 
       \frac{\Gamma (\bar B\to X_c e\bar\nu)}{\Gamma (\bar B\to X_u e\bar\nu)} \;,
\label{eq:C}
\ee
which was recently reanalysed in~\cite{Gambino:2013rza}. We therefore use $C = 0.574 \pm 0.019$  (see also table~\ref{tab:inputs}). Consequently, our expression of the normalised angular observables ${\cal H}_I$ reads
\bea
{{\cal H}_I} & = &
{\cal B} (B\to X_c e \bar\nu)_{\rm exp} \; 
\left| \frac{V_{ts}^* V_{tb}}{V_{cb}} \right|^2 \; 
\frac{4}{C} \; \frac{\Phi^I_{\ell\ell}(\s)}{\Phi_u} \;, 
\label{eq:calHi}
\eea
where $\Phi_u$ is defined by~\cite{Huber:2005ig}
\be \label{eq:busl}
\Gamma (B\to X_u e\bar\nu) =
\frac{G_F^2 m_{b,{\rm pole}}^5}{192 \pi^3} \left| V_{ub}\right|^2 \; \Phi_u.
\ee
The expansion of $\Phi_u$ is given by
\begin{align}
\Phi_u &= 1+ \as \varphi^{(1)} + \kappa \left[\frac{12}{23}\left(1-\eta^{-1}\right)\right] + \as^2 \left[\varphi^{(2)} + 2 \beta_0^{(5)} \varphi^{(1)} \ln\left(\frac{\mu_b}{m_b}\right)\right]
+ \frac{\lambda_1}{2 m_b^2}- \frac{9\lambda_2}{2 m_b^2} \nnb \\
& + {\cal O}(\as^3,\kappa^2,\as \kappa,\as\Lambda^2/m_b^2,\Lambda^3/m_b^3)\; ,\nnb \\[0.5em]
\varphi^{(1)} &= \frac{50}{3}-\frac{8\pi^2}{3} \; , \nnb \\[0.5em]
\varphi^{(2)} &= n_h \left(-\frac{2048 \zeta_3}{9}+\frac{16987}{54}-\frac{340 \pi^2}{81}\right)+n_l
   \left(\frac{256 \zeta_3}{9}-\frac{1009}{27}+\frac{308 \pi^2}{81}\right) \nnb \\
   &-\frac{41848 \zeta_3}{81}+\frac{578 \pi^4}{81}-\frac{104480 \pi^2}{729}+\frac{1571095}{1458}-\frac{848}{27} \pi^2 \ln(2) \; .
\end{align}

As explained in detail in~\cite{Huber:2005ig}, a consistent perturbative expansion in inclusive $\bsll$ in the presence of QED corrections is done in $\as = \alpha_s(\mu_b)/(4\pi)$ and $\kappa = \alpha_e(\mu_b)/\alpha_s(\mu_b)$. We will also briefly sketch the structure of this expansion later below.

In the above equation,  the ${\cal O}(\kappa)$ is taken from~\cite{Sirlin:1981ie}, with $\eta = \alpha_s(\mu_0)/\alpha_s(\mu_b)$. There also exist  QED corrections at ${\cal O}(\as\kappa)$ which could be computed in principle. However, they are not logarithmically enhanced since the fully integrated $\bar B \to X_u e \bar\nu$ rate is an infrared safe observable with respect to collinear photon radiation. We therefore neglect this contribution, but include it lateron in the quantity ${\cal R}(s_0)$, where QED logs will be present in the normalisation.

The two-loop correction of ${\cal O}(\as^2)$ was taken from~\cite{vanRitbergen:1999gs}. Here, $n_h$ and $n_l$ are the numbers of heavy and light quark flavours, respectively, and $\beta_0^{(5)} = 23/3$ is the one-loop QCD $\beta$-function for five active flavours. The coefficients $\lambda_{1}$ and $\lambda_{2}$ in the power-suppressed terms represent the matrix element of the kinetic energy and magnetic moment operator, respectively, and are defined as
\begin{align}
\lambda_1 & = \langle B | \bar h (iD)^2 h | B\rangle /(2 M_B) \, , \nnb \\
\lambda_2 & = - \langle B | \bar h i\sigma^{\mu\nu} G_{\mu\nu} h | B\rangle /(12 M_B) \approx \frac{1}{4} (M_{B^\ast}^2-M_B^2)\, .
\end{align}

As far as the quantity $\Phi^I_{\ell\ell}(\s)$ is concerned, we expand it in the following way in terms of products of the low-scale Wilson coefficients and various functions arising  from the matrix elements,
\bea \label{masterphii}
\Phi^I_{\ell\ell}(\s) &=& \sum_{i\leq j} {\rm Re} \left[ C_i^{\rm eff} (\mu_b) \; C_j^{{\rm eff}*} (\mu_b)  
                \; H^I_{ij} (\mu_b,\s) \right] \;,
\eea
where $C_i^{\rm eff} (\mu_b) \neq C_i (\mu_b)$ only for $i=7,8$. Here $i$ and $j$ run over all operators of  eqs.~(15) and~(16)  in~\cite{Huber:2005ig}. Their low-scale Wilson coefficients are also given explicitly (analytically and numerically) in that paper. For $I=T,L$ the functions $H^I_{ij}(\mu_b,\s)$ are given by
 \be 
H^I_{ij} = \left\{ \begin{array}{ll}
{}~\sum~~ |M_i^N|^2 \;S^I_{NN} + \;{\rm Re} (M_i^7 M_i^{9*}) \;S^I_{79} +\Delta H^I_{ii}\;,
& \mbox{for~} i=j \, , \\[-2mm]
{\!\!\!\scriptscriptstyle N=7,9,10}\\[2mm]
{}~\sum~~ 2 M_i^N M_j^{N*} \; S^I_{NN} 
+ \; \left(M_i^7 M_j^{9*} + M_i^9 M_j^{7*} \right) \; S^I_{79} +\Delta H^I_{ij}\;,
& \mbox{for~} i\neq j \, . \\[-2mm]
{\!\!\!\scriptscriptstyle N=7,9,10}
\end{array}\right.\label{eq:hijI}
\ee 
For $I=A$ the formula is simpler,
 \be 
H^A_{ij} = \left\{ \begin{array}{ll}
 \;\; 0\; , & \mbox{for~} i=j \; , \\
{}~\sum~~ \; \left(M_i^N M_j^{10*} + M_i^{10} M_j^{N*} \right) \; S^A_{N10} +\Delta H^A_{ij}\;,
& \mbox{for~} i\neq j \; .\\[-2mm]
{\,\scriptscriptstyle N=7,9}
\end{array}\right.\label{eq:hijA}
\ee 
The coefficients $M_i^A$ are listed in table~6 of~\cite{Huber:2005ig}. The building blocks 
$S_{NM}^I$ have the following structure,
\begin{align}
{\rm S}^I_{NM} & = \sigma_{NM}^I(\s) \left\{ 
                   1 + 8 \, \as \, \omega_{NM,I}^{(1)} (\s) + 16 \, \as^2 \, \omega_{NM,I}^{(2)} (\s) \right\} 
		   + \frac{\lambda_1}{m_b^2} \; \chi_{1,NM}^I(\s) + \frac{\lambda_2}{m_b^2} \; \chi_{2,NM}^I(\s)  \, .
\end{align}
From~(\ref{eq:hijI}) and~(\ref{eq:hijA}) we see that the possible combinations of indices are $NM=77$,~$79$,~$99$,~$1010$ for $I=T,L$ and $NM=710$,~$910$ for $I=A$. Moreover, we have $S_{99}^I = S_{1010}^I$ for $I=T,L$. Explicitly, the phase-space factors $\sigma_{NM}^I(\s)$ read
\begin{align}
\sigma_{77}^T(\s) &= 8 (1-\s)^2/\s \, , & \sigma_{77}^L(\s) &= 4 (1-\s)^2 \, , & \sigma_{710}^A(\s) &= -8 (1-\s)^2 \, , \nnb \\
\sigma_{79}^T(\s) &= 8 (1-\s)^2    \, , & \sigma_{79}^L(\s) &= 4 (1-\s)^2 \, , & \sigma_{910}^A(\s) &= -4 \s (1-\s)^2 \, , \nnb \\
\sigma_{99}^T(\s) &= 2 \s (1-\s)^2    \, , & \sigma_{99}^L(\s) &= (1-\s)^2 \, . \label{eq:phasespace}
\end{align}

The one-loop QCD functions $\omega_{NM,I}^{(1)} (\s)$ can be extracted from~\cite{Asatrian:2002va} and have  already been given in~\cite{Lee:2006gs}. The two-loop functions $\omega_{NM,I}^{(2)} (\s)$ have  so far only been available for the $q^2$-spectrum~\cite{Chetyrkin:1999ju,Czarnecki:2001cz,Blokland:2004ye,Blokland:2005vq}, but not for the double differential rate. Due to a recent calculation of the double differential rate of the inclusive semi-leptonic $b \to X_u \ell \bar{\nu_\ell}$ decay at two loops in QCD~\cite{Brucherseifer:2013cu}, they can be extracted for $NM=99$,~$1010$ and $I=T,L$ as well as for $NM=910$ and $I=A$. The data to extract these functions was kindly provided by the authors of~\cite{Brucherseifer:2013iv,Brucherseifer:2013cu} and we can therefore present them here for the first time. All functions $\omega_{NM,I}^{(1,2)} (\s)$ are rather lengthy and we therefore relegate their explicit expressions to appendix~\ref{app:functions}.

The functions $\chi_{i,NM}^I(\s)$ ($i=1,2$) that accompany the non-perturbative ${\cal O}(\Lambda_{\rm QCD}^2/m_b^2)$ corrections can be obtained from~\cite{Ali:1996bm} (see also~\cite{Falk:1993dh,Buchalla:1998mt}) and were previously computed in~\cite{Lee:2006gs}. We confirm their expressions,
\begin{align}
\chi_{1,77}^T(\s) &= \frac{4}{3\s} \, (1-\s) (5 \s+3) \, , 
& \chi_{1,77}^L(\s) &= \frac{2}{3} (\s-1) (3\s+13) \, , \nnb \\
\chi_{1,79}^T(\s) &= 4 (1-\s)^2\, , 
& \chi_{1,79}^L(\s) &= 2 (1-\s)^2 \, , \nnb \\
\chi_{1,99}^T(\s) &= -\frac{\s}{3} (1-\s) (3 \s+5) \, , 
& \chi_{1,99}^L(\s) &= \frac{1}{6} (1-\s) (13\s+3) \, , \nnb \\
\chi_{1,710}^A(\s) &= -\frac{4}{3} \left(3 \s^2+2 \s+3\right) \, ,
& \chi_{1,910}^A(\s) &= -\frac{2}{3} \s \left(3 \s^2+2 \s+3\right) \, , \label{eq:chi1}
\end{align}
\begin{align}
\chi_{2,77}^T(\s) &= \frac{4}{\s} \left(3 \s^2+2 \s-9\right) \, , 
& \chi_{2,77}^L(\s) &= 2 \left(15 \s^2-6 \s-13\right) \, , \nnb \\
\chi_{2,79}^T(\s) &= 4 \left(9 \s^2-6 \s-7\right) \, , 
& \chi_{2,79}^L(\s) &= 2 \left(3 \s^2-6 \s-1\right) \, , \nnb \\
\chi_{2,99}^T(\s) &= \s \left(15 \s^2-14 \s-5\right) \, , 
& \chi_{2,99}^L(\s) &= \frac{1}{2} \left(-17 \s^2+10 \s+3\right) \, , \nnb \\
\chi_{2,710}^A(\s) &= -4 \left(9 \s^2-10 \s-7\right) \, ,
& \chi_{2,910}^A(\s) &= -2 \s \left(15 \s^2-14 \s-9\right) \, . \label{eq:chi2}
\end{align}

The quantities $\Delta H^I_{ij}$ can be further decomposed into
\be
\Delta H^I_{ij} = b^I_{ij} + c^I_{ij} + e^I_{ij} \; . 
\ee
Here the contributions $b^I_{ij}$ represent finite bremsstrahlung corrections that appear at
NNLO. They are known for the $q^2$-spectrum (i.e.\ $I=T+L$)~\cite{Asatryan:2002iy} and the forward-backward asymmetry (equivalent to $I=A$)~\cite{Asatrian:2003yk}, but not for the double differential rate. Hence we only include them for these two cases, but not for $H_T$ and $H_L$ separately. This is still an excellent approximation since the effect of finite bremsstrahlung corrections is very small  anyway. The explicit formulas can be found in~\cite{Asatryan:2002iy,Asatrian:2003yk} and will therefore not be repeated. 

The coefficients $c^I_{ij}$ comprise non-perturbative ${\cal O}(\Lambda_{\rm QCD}^2/m_c^2)$ contributions and were calculated in
ref.~\cite{Buchalla:1997ky} for~$I=T+L$ and~$I=A$. Moreover, the coefficients of the double differential rate can be inferred from that paper. One obtains
\bea
c^T_{2j} &=& - \as \kappa \frac{8\lambda_2}{9 m_c^2} \, (1-\s)^2 (1+3\s) \, F(r) \left[\frac{1}{s} \, M_j^{7\ast} + \frac{1}{2} \, M_j^{9\ast}\right] \;, \hskip 1.5cm {\rm for \; j \neq 1,2} \;,\nnb\\
c^T_{1j} &=& -\fm{1}{6}  \; c^T_{2j}\;, \hskip 1.5cm {\rm for \; j \neq 1,2} \;,\nnb\\
c^T_{22} &=& - \as \kappa \frac{8\lambda_2}{9 m_c^2} \, (1-\s)^2 (1+3\s) \, F(r) \left[\frac{1}{s} \, M_2^{7\ast} + \frac{1}{2} \, M_2^{9\ast}\right] \;, \nnb\\
c^T_{11} &=& + \as \kappa \frac{4\lambda_2}{27 m_c^2} \, (1-\s)^2 (1+3\s) \, F(r) \left[\frac{1}{s} \, M_1^{7\ast} + \frac{1}{2} \, M_1^{9\ast}\right] \;, \nnb\\
c^T_{12} &=& - \as \kappa \frac{8\lambda_2}{9 m_c^2} \, (1-\s)^2 (1+3\s) \! \left[F^\ast(r) \left(\frac{1}{s} \, M_1^{7} + \frac{1}{2} \, M_1^{9}\right)-\frac{1}{6} \, F(r) \left(\frac{1}{s} \, M_2^{7\ast} + \frac{1}{2} \, M_2^{9\ast}\right)\right] \;, \nnb\\[1.5em]
c^L_{2j} &=& - \as \kappa  \frac{8\lambda_2}{9 m_c^2} \, (1-\s)^2 (3-\s) \, F(r) \left[ M_j^{7\ast} + \frac{1}{2} \, M_j^{9\ast}\right] \;, \hskip 1.5cm {\rm for \; j \neq 1,2} \;,\nnb\\
c^L_{1j} &=& -\fm{1}{6}  \; c^L_{2j}\;, \hskip 1.5cm {\rm for \; j \neq 1,2} \;,\nnb\\
c^L_{22} &=& - \as \kappa \frac{8\lambda_2}{9 m_c^2} \, (1-\s)^2 (3-\s) \, F(r) \left[ M_2^{7\ast} + \frac{1}{2} \, M_2^{9\ast}\right] \;, \nnb\\
c^L_{11} &=& + \as \kappa \frac{4\lambda_2}{27 m_c^2} \, (1-\s)^2 (3-\s) \, F(r) \left[ M_1^{7\ast} + \frac{1}{2} \, M_1^{9\ast}\right] \;, \nnb\\
c^L_{12} &=&  - \as \kappa \frac{8\lambda_2}{9 m_c^2} \, (1-\s)^2 (3-\s) \! \left[F^\ast(r) \left( M_1^{7} + \frac{1}{2} \, M_1^{9}\right)-\frac{1}{6} \, F(r) \left( M_2^{7\ast} + \frac{1}{2} \, M_2^{9\ast}\right)\right] \;, \nnb\\[1.5em]
c^A_{210} &=& +\as \kappa  \frac{4\lambda_2}{9 m_c^2} (1-\s)^2
                  (1+3\s)\,
		  F(r)  \;,\nnb\\
c^A_{110} & = &  -\fm{1}{6}  \; c^A_{210} \; , 
\eea
where $r = q^2/(4 m_c^2)$. The function  $F(r)$ can be found in the appendix of~\cite{Huber:2005ig}. Moreover, we also include factorisable non-perturbative charm contributions which we implement by means of  the Kr\"uger-Sehgal approach~\cite{Kruger:1996cv,Kruger:1996dt}. We elaborated extensively on this approach and also  the formulas by means of which these corrections are taken into account in ref.~\cite{Huber:2007vv}. Given their length we do not repeat these formulas here but refer the inclined reader to refs.~\cite{Kruger:1996cv,Kruger:1996dt,Huber:2007vv} for all necessary details.

Finally, the coefficients $e^I_{ij}$ collect the $\ln (m_b^2/m_\ell^2)$-enhanced electromagnetic corrections which we calculated in section~\ref{sec:QEDdoublediff} for the double differential rate. Their contribution to the $H_I$ can be derived from~(\ref{eq:QEDdoublediff}) by applying the projections given in section~\ref{sec:observables}. One finds
\begin{align}
e^I_{77} & = 8 \, \as^3 \kappa^3 \, \sigma_{77}^I(\s) \, \omega_{77,I}^{\rm (em)}(\s) \; , &
e^I_{29} & = 8 \, \as^2 \kappa^2 \, \sigma_{99}^I(\s) \,\omega_{29,I}^{\rm (em)}(\s) \, , \nnb\\
e^I_{79} & = 8 \, \as^2 \kappa^2 \, \sigma_{79}^I(\s) \, \omega_{79,I}^{\rm (em)}(\s) \; , &
e^I_{22} & = 8\, \as^3 \kappa^3\, \sigma_{99}^I(\s) \, \omega_{22,I}^{\rm (em)}(\s) \, , \nnb\\
e^I_{99} & = 8 \, \as \kappa \, \sigma_{99}^I(\s) \, \omega_{99,I}^{\rm (em)}(\s) \; , &
e^I_{11} & =   \fm{16}{9}  \; e^I_{22} \, ,\nnb\\
e^I_{1010} & = e^I_{99}\; , &
e^I_{12} & =  \fm{8}{3}   \; e^I_{22} \, ,\nnb\\
e^I_{27} & = 8 \,\as^3 \kappa^3 \, \sigma_{79}^I(\s) \, \omega_{27,I}^{\rm (em)}(\s) \, , &
e^I_{1j} & =  \fm{4}{3}  e^I_{2j}, \hskip 1.5cm {\rm{ for}} \;\, j=7,9 \, ,
\end{align}
for $I=T,L$, while for $I=A$ one gets
\begin{align}
e^A_{710} & = 8 \, \as^2 \kappa^2 \, \sigma_{710}^A(\s) \, \omega_{710,A}^{\rm (em)}(\s) \; , &
e^A_{210} & = 8 \, \as^2 \kappa^2 \, \sigma_{910}^A(\s) \, \omega_{210,A}^{\rm (em)}(\s) \; , \nnb\\
e^A_{910} & = 8 \, \as \kappa \, \sigma_{910}^A(\s) \, \omega_{910,A}^{\rm (em)}(\s) \; , &
e^A_{110} & =  \fm{4}{3}  e^A_{210}.
\end{align}
The functions $\omega_{ij,I}^{\rm (em)}(\s)$ have again been moved to appendix~\ref{app:functions}. 

We consider the observables $H_I$ (or equivalently ${\cal H}_I$) in the low-$q^2$ region only, because their sensitivity to New Physics is highest in this region~\cite{Lee:2006gs}. Besides, there are two more observables which we compute in the low-$\s$ region. First, there is the zero crossing $q_0^2$ of the forward-backward asymmetry, which we extract numerically from ${\cal H}_A$ by means of the formulas given above. Moreover, there is the branching ratio. In principle, it can be obtained by taking the sum of ${\cal H}_T$ and ${\cal H}_L$. Its master formula has  already been given in~\cite{Huber:2005ig}. We therefore only highlight two small pieces which are available for the branching ratio only, but not for ${\cal H}_T$ and ${\cal H}_L$ individually. These are only the finite bremsstrahlung contributions from~\cite{Asatryan:2002iy} and the non-log enhanced terms of $\omega_{99}^{({\rm em})}(\s)$ (see eq.~(94) of ref.~\cite{Huber:2005ig}).

In the high-$q^2$ region we consider two observables. The first one is the branching ratio, where we include the same terms as in the low-$q^2$ region. As far as QED corrections are concerned, the functions $\omega_{99}^{\rm (em)}(\s)$, $\omega_{1010}^{\rm (em)}(\s)$, $\omega_{77}^{\rm (em)}(\s)$, and $\omega_{79}^{\rm (em)}(\s)$ (see eqs.~(94) and (100) --~(102) of~\cite{Huber:2005ig}) are valid in the entire $q^2$-region, while the functions $\omega_{2x}^{\rm (em)}(\s)$ are again obtained from a numerical fit. To take into account our most recent input parameters (see table~\ref{tab:inputs}), we re-did the fits and collected the results in appendix~\ref{app:functions}. In addition, the two-loop QCD matrix element functions $F_{1,2}^7(\s)$ and $F_{1,2}^9(\s)$, which were originally computed in~\cite{Ghinculov:2003qd}, were given explicitly only in~\cite{Greub:2008cy}. We implement these formulas in our numerical code. Moreover, non-perturbative $1/m_b^3$ corrections become sizable in the high-$\s$ region. They were originally computed in~\cite{Bauer:1999kf} and we implement the formulas of refs.~\cite{Bauer:1999kf,Ligeti:2007sn}. The second observable is the ratio ${\cal R}(s_0)$ which we have already mentioned in the introduction. It was proposed in~\cite{Ligeti:2007sn}\footnote{Note that we use a different pre-factor here.} and is obtained by normalizing the $\bar B \rightarrow X_s \ell^+ \ell^-$ decay rate to the semileptonic $\bar B^0 \rightarrow X_u \ell \bar\nu$ rate \emph{with the same cut in $q^2$}. In this way, large theoretical uncertainties that stem from poorly known parameters in the $1/m_b^2$ and $1/m_b^3$ power-corrections can be significantly reduced, as we will see in our numerical analysis in section~\ref{sec:results}. In terms of our perturbative quantities, it reads
\bea\label{eq:R0}
{\cal R}(s_0) & = & \frac{\displaystyle
\int_{\hat s_0}^1 {\rm d} \hat s \, {{\rm d} {\Gamma} (\bar B\to X_s \ell^+\ell^-) \over {\rm d} \hat s}
}{\displaystyle
\int_{\hat s_0}^1 {\rm d} \hat s \, {{\rm d} {\Gamma} (\bar B^0\to X_u \ell \nu) \over {\rm d} \hat s}
} 
=
4 \left| V_{ts}^\ast V_{tb} \over V_{ub} \right|^2 \frac{
\int_{\hat s_0}^1 {\rm d} \hat s \, \Phi_{\ell\ell} (\hat s)}{
\int_{\hat s_0}^1 {\rm d} \hat s \, \Phi_{u} (\hat s)
}  \; .
\eea
The quantity $\Phi_{\ell\ell}(\s)$ is known from the branching ratio. The differential $\Phi_u (\s)$ is given by
\bea
\frac{{\rm d} \Gamma(\bar B^0\to X_u \ell \nu)}{{\rm d} \s} = 
\frac{G_F^2 \left| V_{ub} \right|^2 m_{b,{\rm pole}}^5}{192 \pi^3} \; \Phi_u (\s) \; .
\eea
We elaborated extensively in ref.~\cite{Huber:2007vv} about  how to obtain the ${\cal O}(1,\as,\as^2,1/m_b^2,1/m_b^3)$ contributions to $\Phi_u(\s)$, and will therefore  not repeat these formulas. We would rather like to describe the ${\cal O}(\as \kappa)$ contribution to $\Phi_u(\s)$, which we include in the present work and which was absent in~\cite{Huber:2007vv}. Once the integration over $\s$ is restricted to the high-$q^2$ region, the corrections of ${\cal O}(\as \kappa)$ to $\Phi_u(\s)$ contain residual terms logarithmically enhanced by $\ln(m_b^2/m_\ell^2)$. These must be proportional to $\omega_{99}^{\rm (em)}(\s)$. We take into account that we only have one charged lepton in the final state, and that the leptonic current is $V-A$. Moreover, we average over $e$ and $\mu$, and arrive at
\bea
\Phi_u (\s)_{\big|\as \kappa} &=& 8 \, \as \kappa \, (1-\s)^2 \, (1+2\s) \, \omega_{99}^{\rm (em)}(\s)_{\big|\ln\left(\frac{m_b^2}{m_\ell^2}\right) \, \longrightarrow \, \ln\left(\frac{m_b^2}{m_e m_\mu}\right)}\; .
\eea
As in our previous analysis~\cite{Huber:2007vv} we do not include electromagnetic corrections of order ${\cal O}(\kappa)$ to $\Phi_u(\s)$ because they are unknown.

Let us conclude this section by a few remarks on the renormalisation schemes for the quark masses, as well as on the expansion in $\as$ and $\kappa$. The pole masses of the $b$ and $c$ quark that are present in the definition of $\s$ and in several loop functions suffer from renormalon ambiguities~\cite{Beneke:1998ui,Hoang:1998hm}. We therefore convert them analytically to short-distance schemes (1S and $\overline{\rm MS}$, respectively) before any numerical evaluation of the observables is carried out. In our numerical analysis we use the conversion formulas up to order ${\cal O}(\as^2)$~\cite{Hoang:2000fm}. As far as the mass of the top quark is concerned we take the pole mass as input and convert it to the $\overline{\rm MS}$ scheme at order ${\cal O}(\as^3)$ using {\tt RunDec}~\cite{Chetyrkin:2000yt}. 
We also take into account electroweak corrections presented in eq.~(31) of ref.~\cite{Buchalla:1997kz}, consistently to the other contributions. 
Turning our attention to the expansion in $\as$ and $\kappa$, we observe that the amplitude has the structure
\bea
{\cal A} & = & \kappa \left[ {\cal A}_{LO}  + \alpha_s  \; {\cal A}_{NLO}+  
               \alpha_s^2 \; {\cal A}_{NNLO} + {\cal O}(\alpha_s^3) \right]
               \nonumber\\
         & + & \kappa^2 \left[
               {\cal A}_{LO}^{\mathrm em} + \alpha_s \; {\cal A}_{NLO}^{\mathrm em}  + 
               \alpha_s^2  \; {\cal A}_{NNLO}^{\mathrm em} + {\cal O}(\alpha_s^3) \right] 
\; + {\cal O}(\kappa^3) \; , \label{eq:schematic}
\eea 
and that the ratio $\Phi^I_{\ell\ell}(\s)/\Phi_u$ in~(\ref{eq:calHi}) has a similar structure to that of the squared amplitude (up to bremsstrahlung and non-perturbative corrections),
\bea
{\cal A}^2 
& = & 
\kappa^2 \Big[ {\cal A}_{LO}^2 
+  \alpha_s \; 2 {\cal A}_{LO} {\cal A}_{NLO} 
+ \alpha_s^2 \; ({\cal A}_{NLO}^2 + 2  {\cal A}_{LO} {\cal A}_{NNLO}  ) 
\nonumber\\
& & \hskip 0.5 cm 
+  \alpha_s^3 \; 2 ({\cal A}_{NLO} {\cal A}_{NNLO} + \ldots)  + {\cal O}(\alpha_s^4)
\Big]   \nonumber \\
&+ &
\kappa^3 \Big[
2 {\cal A}_{LO}^{} {\cal A}_{LO}^{\mathrm em} + 
\alpha_s \; 2 ({\cal A}_{NLO}^{} {\cal A}_{LO}^{\mathrm em} + {\cal A}_{LO}^{} {\cal A}_{NLO}^{\mathrm em}) 
\nonumber\\
& &
\hskip 0.5 cm 
+\alpha_s^2 \; 2 ( {\cal A}_{NLO}^{} {\cal A}_{NLO}^{\mathrm em}
+ {\cal A}_{NNLO}^{} {\cal A}_{LO}^{\mathrm em} 
+ {\cal A}_{LO}^{} {\cal A}_{NNLO}^{\mathrm em}) \nonumber\\
&& \hskip 0.5 cm
+ \alpha_s^3 \; 2 ({\cal A}_{NLO}^{} {\cal A}_{NNLO}^{\mathrm em} + {\cal A}_{NNLO}^{} {\cal A}_{NLO}^{\mathrm em}
+ \ldots)
+ {\cal O}(\alpha_s^4)\Big] +{\cal O}( \kappa^4)\; .
\label{eq:schematic2}
\eea
We already argued in refs.~\cite{Huber:2005ig,Huber:2007vv} that an expansion of this kind up to and including ${\cal O}(\as^3\kappa^3)$ also captures the dominant N$^3$LO QCD corrections, since the missing terms ${\cal A}_{LO} {\cal A}_{N^3LO}$, ${\cal A}_{LO} {\cal A}^{\mathrm em}_{N^3LO}$, and ${\cal A}_{LO}^{\mathrm em} {\cal A}_{N^3LO}$ (represented by the dots) are small. It is therefore justified to refer to the accuracy of our calculations as {\it improved} NNLO. Hence we expand all products in eq.~(\ref{eq:calHi}) (and in all other observables) in $\as$ and $\kappa$ up to the aforementioned order, and neglect all higher terms. The observables are also expanded in the power-correction parameters $\lambda_{1,2}, \rho_1, f_u^{0,\pm}, f_s$ up to linear terms. Higher powers as well as products of these parameters are dropped.


\section{Phenomenological results}
\label{sec:results}
\begin{table}[t]
\begin{center}
\begin{displaymath}
\begin{tabular}{|l|l|}
\hline\spp 
$\alpha_s (M_z) = 0.1184 \pm 0.0007$ &  
  $m_e = 0.51099892 \;\mev $ \\ \spp 
$\alpha_e (M_z) =  1/127.918 $ & 
  $m_\mu = 105.658369 \;\mev$ \\ \spp 
$s_W^2 \equiv \sin^2\theta_W = 0.2312$ & 
  $m_\tau = 1.77699 \;\gev$ \\ \spp 
$|V_{ts}^* V_{tb}/V_{cb}|^2 = 0.9621 \pm 0.0027$~\cite{Charles:2004jd} &
  $m_c(m_c) = (1.275 \pm 0.025)\;\gev$ \\\spp 
$|V_{ts}^* V_{tb}/V_{ub}|^2 = 130.5 \pm 11.6 $~\cite{Charles:2004jd} &
  $m_b^{1S} = (4.691 \pm 0.037)\;\gev$~\cite{Amhis:2012bh,Schwanda:2013bg} \\ \spp 
$BR(B\to X_c e \bar\nu)_{\rm exp}=0.1051 \pm 0.0013$~\cite{Amhis:2012bh} & 
  $m_{t,{\rm pole}}= (173.5 \pm 1.0) \;\gev$ \\ \spp 
$M_Z = 91.1876\;\gev$ & 
  $m_B = 5.2794\;\gev$ \\ \spp 
$M_W = 80.385\;\gev$ & 
  $C = 0.574 \pm 0.019$~\cite{Gambino:2013rza} \\ \spp 
 $\mu_b = 5^{+5}_{-2.5}\;\gev$ & $\mu_0 = 120^{+120}_{-60}\;\gev$ \\ \spp
$\lambda_2^{\rm eff} = (0.12 \pm 0.02)\;\gev^2$ & 
  $\rho_1 = (0.06 \pm 0.06)\;\gev^3$~\cite{Bauer:2004ve} \\ \spp
$\lambda_1^{\rm eff} = (-0.362 \pm 0.067)\;\gev^2$~\cite{Amhis:2012bh,Schwanda:2013bg} & 
  $f_u^0+f_s = (0 \pm 0.2)\;\gev^3$~\cite{Ligeti:2007sn} \\ \spp
$f_u^0-f_s = (0 \pm 0.04)\;\gev^3$~\cite{Ligeti:2007sn} &
$f_u^\pm = (0 \pm 0.4)\;\gev^3$~\cite{Ligeti:2007sn} \\ \hline
\end{tabular}
\end{displaymath}
\caption{Numerical inputs used in the phenomenological analysis. Unless specified otherwise, they are taken from PDG~\cite{Agashe:2014kda}.}
\label{tab:inputs}
\end{center}
\end{table}
In this section we give the numerical results of our phenomenological analysis. We use the input parameters as given in table~\ref{tab:inputs}. For each variable we give the integral over bin~1 ($1 \; {\rm GeV}^2 < q^2 < 3.5 \; {\rm GeV}^2$), bin~2 ($3.5 \; {\rm GeV}^2 < q^2 < 6 \; {\rm GeV}^2$), and the entire low-$q^2$ region ($1 \; {\rm GeV}^2 < q^2 < 6 \; {\rm GeV}^2$). In the high-$q^2$ region we integrate over all $q^2>14.4 \; {\rm GeV}^2$. The respective $q^2$-interval is indicated by the argument of the observables. We give the numbers for electron and muon final state separately, and remind the reader that, depending on the channel and the experimental setup, our numbers have to be modified according to our Monte Carlo study in section~\ref{sec:PHOTOS}.

The quoted uncertainties are the parametric and perturbative ones only. Additional uncertainties from subleading non-perturbative corrections are not included. In particular, the ${\cal O}(\alpha_s(\mu_b) \Lambda_{\rm QCD}/m_{c,b})$ non-perturbative corrections are estimated to be around $\sim 5\% $ in the low-$q^2$ region. The individual error bars are obtained by varying the parameters in the range indicated in table~\ref{tab:inputs}, where we assume the errors on $C$ and $m_c$ to be fully correlated. The total error is obtained by adding the individual ones in quadrature. By default we give two decimal digits. In case this leads to $0.00$ we give the number up to the first significant digit.

Before presenting our actual results, we would like to comment on the size of QED corrections. In table \ref{table:QEDexact} the first two columns in each of the three sections are, respectively, the observable at NNLL and its QED correction expressed as a percentage of the branching ratio integrated in the whole low-$q^2$ region (${\cal B}[1,6]_{ee}$). The third column is the relative size of the QED correction with respect to each NNLL observable.

\begin{table}
\begin{center}
\begin{tabular}{|c|lll|lll|lll|}
\hline
& \multicolumn{3}{c|}{$q^2 \in [1,6] \; {\rm GeV}^2$} & \multicolumn{3}{c|}{$q^2 \in [1,3.5] \; {\rm GeV}^2$} & \multicolumn{3}{c|}{$q^2 \in [3.5,6] \; {\rm GeV}^2$} \cr
& $\frac{O_{[1,6]}}{{\cal B}_{[1,6]}}$ & $ \frac{\Delta O_{[1,6]}}{{\cal B}_{[1,6]}}$ & $\frac{\Delta O_{[1,6]}}{O_{[1,6]}}$ 
& $\frac{O_{[1,3.5]}}{{\cal B}_{[1,6]}}$ & $ \frac{\Delta O_{[1,3.5]}}{{\cal B}_{[1,6]}}$ & $\frac{\Delta O_{\rm [1,3.5]}}{O_{\rm [1,3.5]}}$
& $\frac{O_{[3.5,6]}}{{\cal B}_{[1,6]}}$ & $ \frac{\Delta O_{[3.5,6]}}{{\cal B}_{[1,6]}}$ & $\frac{\Delta O_{\rm [3.5,6]}}{O_{\rm [3.5,6]}}$ \cr \hline
$\cal B$ & 100 & 5.1 & 5.1 & 54.6 & 3.7 & 6.8 & 45.4 & 1.4 & 3.1\cr
${\cal H}_T$ & 19.5 & 14.1 & 72.5 & 9.5 & 8.8 & 92.1 & 10.0 & 5.4 & 53.6 \cr
${\cal H}_L$ & 80.0 & -8.7 & -10.9 & 44.7 & -4.7 & -10.6 & 35.3 & -4.0 & -11.3\cr
${\cal H}_A$ & -3.3 & 1.4 & -43.6 & -7.2 & 0.8 & -10.7 & 4.0 & 0.6 & 16.2 \cr
\hline
\end{tabular}
\caption{Relative size of QED effects on $b\to s e^+ e^-$ at low-$q^2$ (the muon case can be easily obtained by rescaling). All entries are given in percent. For each of the three bins the first two columns are the integrated observable and its QED correction normalized to the total low-$q^2$ branching ratio, respectively ($\int_{s_1}^{s_2} O/\int_1^6 {\cal B}$ and $\int_{s_1}^{s_2} \Delta O/\int_1^6 {\cal B}$). The third column is the relative size of the QED correction ($\int_{s_1}^{s_2} \Delta O/\int_{s_1}^{s_2} O$). The sum of ${\cal H}_T$ and ${\cal H}_L$ does not exactly reproduce the branching ratio because in the latter we include finite bremsstrahlung and non-log enhanced QED corrections that are not available for the first two. \label{table:QEDexact}}
\end{center}
\end{table}

One can see immediately that the relative size of QED corrections to ${\cal H}_T$ is large, see third column in each section in table~\ref{table:QEDexact}. Therefore, a few remarks on this observable are in order. It turns out that ${\cal H}_T$ is suppressed in the low-$q^2$ region. To see this, let us look at the tree--level dependence of 
$H_T$ and $H_L$ on the Wilson coefficients presented in eqs.~(\ref{eq:HT-nlo}) and (\ref{eq:HL-nlo}). The phase space corresponding to $|C_9|^2$ is suppressed in $H_T$ compared to $H_L$, whereas that associated to $|C_7|^2$ is enhanced. Surprisingly, this leads to a two-fold suppression of $H_T$. First, there is an additional factor of $2\s$ in the overall phase space w.r.t.\ $H_L$. Second, the factor $|C_9+2/\s \, C_7|^2$ is small in the low-$q^2$ region, and even vanishes at the position of the zero of $H_A$.

The QED corrections to the $H_I$, however, do not follow this pattern of suppression. In fact, from the inspection of the second columns in each section in table~\ref{table:QEDexact} we see that the absolute values of these corrections are natural in size and that all entries in these columns have the same order of magnitude. In the case of $H_T$ the smallness of the NNLL QCD result implies that their relative size is anomalously large (see the third columns in table~\ref{table:QEDexact}). 
However, we emphasize here that this does {\emph{not}} indicate a breakdown of the perturbative series because the large relative size of QED corrections is almost entirely due to the suppression of the tree-level plus QCD contribution, and not due to a large absolute value of the QED corrections. To support our analytical findings, we investigated the situation in a Monte Carlo study (for details, see section~\ref{sec:PHOTOS}) and find exactly the same pattern once we use EVTGEN and PHOTOS, see figures~\ref{fig:bsll-spectrum} and~\ref{fig:bsll-QEDcorrectionTL} in section~\ref{sec:MontecarloHTHL}.

We can even turn the argument around and regard the relative size of QED corrections in ${\cal H}_T$ as a virtue rather than a drawback, because it offers a good opportunity to be sensitive to QED corrections even without the pure QED observables ${\cal H}_3$ and ${\cal H}_4$ defined in section~\ref{sec:observables}.

Finally, let us point out that similar large effects on ${\cal H}_A$ (or, equivalently, the forward--backward asymmetry) integrated in the whole low-$q^2$ region are simply a due to the large cancellation between the integrated asymmetry in the two bins. This cancellation originates from the presence of a zero in the differential ${\cal H}_A$ spectrum and is not reproduced in the pattern of QED corrections. As we see in table~\ref{table:QEDexact}, the latter imply a positive shift on ${\cal H}_A$ in both bins.

\begin{figure}
\begin{center}
\includegraphics[scale=0.8]{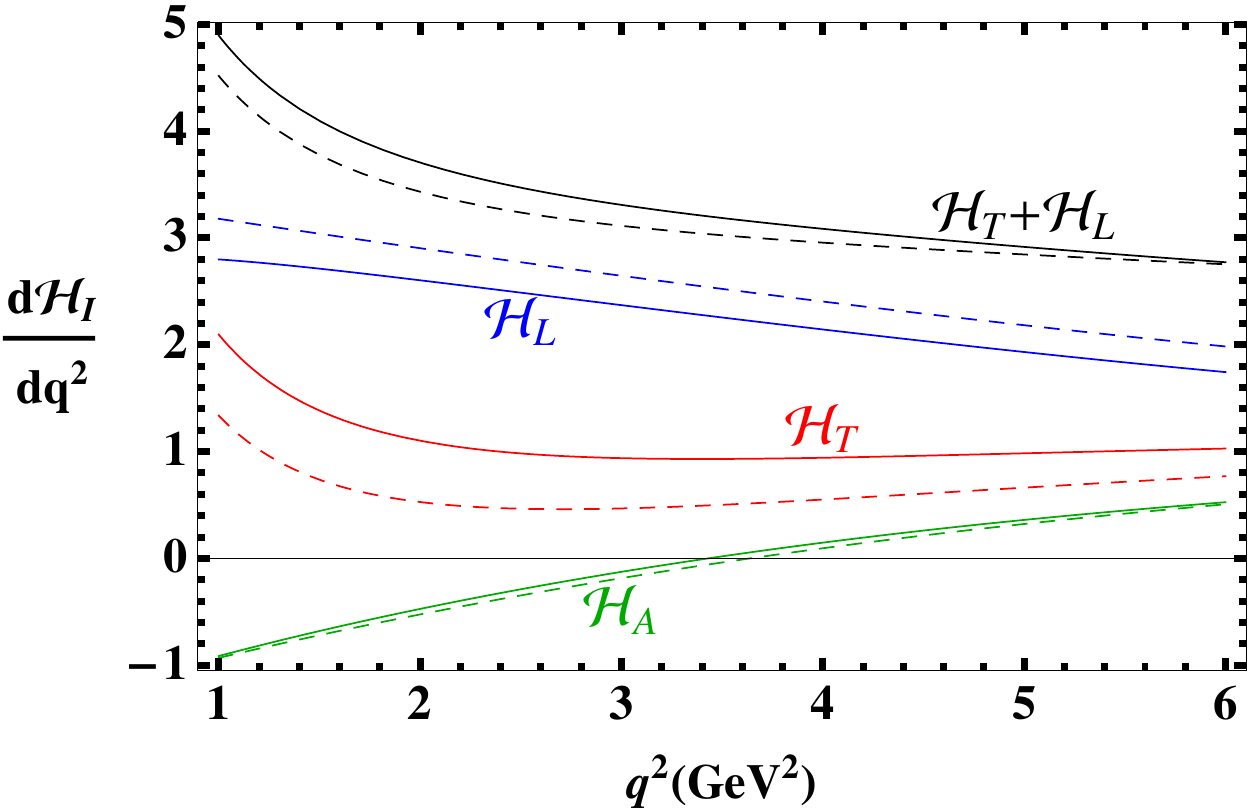}\\[1.5em]

\hspace*{-32pt}
\includegraphics[scale=0.87]{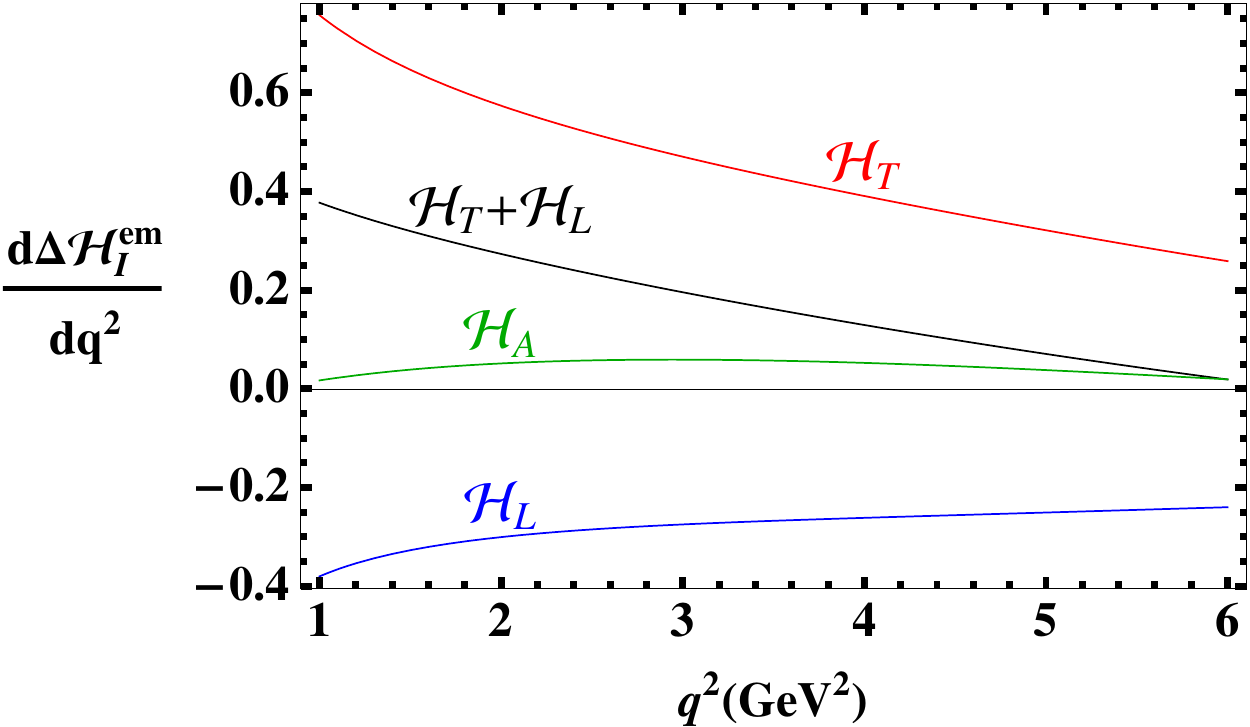}
\caption{Differential distributions for the various observables (upper panel) and their respective log--enhanced QED correction (lower panel) in units of $10^{-7}$. Dashed lines correspond to switching off all QED corrections. The integrals under the curves reproduce the results presented in section~\ref{sec:results} for the electron channel.
\label{fig:diff}}
\end{center}
\end{figure}

In the upper panel of figure~\ref{fig:diff} we show the differential distributions that we obtain for the various ${\cal H}_I$ in the electron channel; dashed lines are obtained by switching QED corrections off. In the lower panel of figure~\ref{fig:diff} we show the log--enhanced QED correction itself, i.e.\ the difference between solid and dashed lines in the upper panel.

\subsection{${\cal H}_T$ and   ${\cal H}_L$ }
\label{sec:HT}

For the quantities ${\cal H}_T$ and   ${\cal H}_L$
 we find theoretical uncertainties of $6$ to $9 \%$. In this sense the QED corrections listed in table~\ref{table:QEDexact} are really  significant. 
\begin{align}
\dps {\cal H}_T[1,3.5]_{ee} =&( 2.91 \pm 0.16_{\text{scale}} \pm 0.03_{m_t} \pm 0.08_{C,m_c} \pm 0.02_{m_b} \nonumber \\
                      &\hspace*{25pt} \pm  0.003_{\alpha_s} \pm 0.01_{\text{CKM}} \pm 0.04_{\text{BR}_{\text{sl}}}) \cdot 10^{-7} = ( 2.91 \pm 0.19 ) \cdot 10^{-7} \; ,  \nonumber\\[0.5em]
\dps {\cal H}_T[3.5,6]_{ee} =&( 2.43 \pm 0.16_{\text{scale}} \pm 0.04_{m_t} \pm 0.08_{C,m_c} \pm 0.05_{m_b} \nonumber \\
                      &\hspace*{25pt} \pm  0.01_{\alpha_s} \pm 0.01_{\text{CKM}} \pm 0.03_{\text{BR}_{\text{sl}}}) \cdot 10^{-7} = ( 2.43 \pm 0.20 ) \cdot 10^{-7} \; , \nonumber\\[0.5em]
\dps {\cal H}_T[1,6]_{ee} =&( 5.34 \pm 0.33_{\text{scale}} \pm 0.07_{m_t} \pm 0.16_{C,m_c} \pm 0.06_{m_b} \nonumber \\
                    &\hspace*{25pt} \pm    0.01_{\alpha_s} \pm 0.01_{\text{CKM}} \pm 0.06_{\text{BR}_{\text{sl}}}) \cdot 10^{-7}  = ( 5.34 \pm 0.38  ) \cdot 10^{-7} \; .
\end{align}
\begin{align}
\dps {\cal H}_T[1,3.5]_{\mu\mu} =&( 2.09 \pm 0.10_{\text{scale}} \pm 0.02_{m_t} \pm 0.06_{C,m_c} \pm 0.01_{m_b} \nonumber \\
                      &\hspace*{25pt} \pm  0.01_{\alpha_s} \pm 0.01_{\text{CKM}} \pm 0.03_{\text{BR}_{\text{sl}}}) \cdot 10^{-7} = ( 2.09 \pm 0.12 ) \cdot 10^{-7} \; ,  \nonumber\\[0.5em]
\dps {\cal H}_T[3.5,6]_{\mu\mu} =&( 1.94 \pm 0.13_{\text{scale}} \pm 0.03_{m_t} \pm 0.07_{C,m_c} \pm 0.05_{m_b} \nonumber \\
                      &\hspace*{25pt} \pm  0.01_{\alpha_s} \pm 0.01_{\text{CKM}} \pm 0.02_{\text{BR}_{\text{sl}}}) \cdot 10^{-7} = ( 1.94 \pm 0.16 ) \cdot 10^{-7} \; , \nonumber\\[0.5em]
\dps {\cal H}_T[1,6]_{\mu\mu} =&( 4.03 \pm 0.23_{\text{scale}} \pm 0.06_{m_t} \pm 0.12_{C,m_c} \pm 0.06_{m_b} \nonumber \\
                    &\hspace*{25pt} \pm    0.002_{\alpha_s} \pm 0.01_{\text{CKM}} \pm 0.05_{\text{BR}_{\text{sl}}}) \cdot 10^{-7}  = ( 4.03 \pm 0.28  ) \cdot 10^{-7} \; .
\end{align}

\begin{align}
\dps {\cal H}_L[1,3.5]_{ee} =&( 6.35 \pm 0.23_{\text{scale}} \pm 0.08_{m_t} \pm 0.22_{C,m_c} \pm 0.08_{m_b} \nonumber \\
                      &\hspace*{25pt} \pm  0.03_{\alpha_s} \pm 0.02_{\text{CKM}} \pm 0.08_{\text{BR}_{\text{sl}}}) \cdot 10^{-7} = ( 6.35 \pm 0.35 ) \cdot 10^{-7} \; ,  \nonumber\\[0.5em]
\dps {\cal H}_L[3.5,6]_{ee} =&( 4.97 \pm 0.22_{\text{scale}} \pm 0.06_{m_t} \pm 0.17_{C,m_c} \pm 0.04_{m_b} \nonumber \\
                      &\hspace*{25pt} \pm  0.02_{\alpha_s} \pm 0.01_{\text{CKM}} \pm 0.06_{\text{BR}_{\text{sl}}}) \cdot 10^{-7} = ( 4.97 \pm 0.29 ) \cdot 10^{-7} \; , \nonumber\\[0.5em]
\dps {\cal H}_L[1,6]_{ee} =&( 1.13 \pm 0.04_{\text{scale}} \pm 0.01_{m_t} \pm 0.04_{C,m_c} \pm 0.01_{m_b} \nonumber \\
                    &\hspace*{25pt} \pm    0.01_{\alpha_s} \pm 0.003_{\text{CKM}} \pm 0.01_{\text{BR}_{\text{sl}}}) \cdot 10^{-6}  = (1.13 \pm 0.06  ) \cdot 10^{-6} \; .
\end{align}
\begin{align}
\dps {\cal H}_L[1,3.5]_{\mu\mu} =&( 6.79 \pm 0.23_{\text{scale}} \pm 0.08_{m_t} \pm 0.23_{C,m_c} \pm 0.09_{m_b} \nonumber \\
                      &\hspace*{25pt} \pm  0.03_{\alpha_s} \pm 0.02_{\text{CKM}} \pm 0.08_{\text{BR}_{\text{sl}}}) \cdot 10^{-7} = ( 6.79 \pm 0.36 ) \cdot 10^{-7} \; ,  \nonumber\\[0.5em]
\dps {\cal H}_L[3.5,6]_{\mu\mu} =&( 5.34 \pm 0.23_{\text{scale}} \pm 0.06_{m_t} \pm 0.19_{C,m_c} \pm 0.04_{m_b} \nonumber \\
                      &\hspace*{25pt} \pm  0.03_{\alpha_s} \pm 0.01_{\text{CKM}} \pm 0.07_{\text{BR}_{\text{sl}}}) \cdot 10^{-7} = ( 5.34 \pm 0.32 ) \cdot 10^{-7} \; , \nonumber\\[0.5em]
\dps {\cal H}_L[1,6]_{\mu\mu} =&( 1.21 \pm 0.04_{\text{scale}} \pm 0.01_{m_t} \pm 0.04_{C,m_c} \pm 0.01_{m_b} \nonumber \\
                    &\hspace*{25pt} \pm    0.01_{\alpha_s} \pm 0.003_{\text{CKM}} \pm 0.02_{\text{BR}_{\text{sl}}}) \cdot 10^{-6}  = ( 1.21 \pm 0.07  ) \cdot 10^{-6} \; .
\end{align}

\subsection{${\cal H}_A$}

For the zero-crossing $q_0^2$ of ${\cal H}_A$, which is equivalent to the zero of the forward-backward asymmetry due to equation~(\ref{eq:AFBHA}), we find
\begin{align}
\dps (q_0^2)_{ee} =& (3.46 \pm 0.10_{\text{scale}} \pm 0.001_{m_t} \pm 0.02_{C,m_c} \pm 0.06_{m_b} \pm 0.02_{\alpha
   _s}) {\text{ GeV}}^2 \nonumber \\[0.5em]
                       =& (3.46 \pm 0.11) {\text{ GeV}}^2 \; , \\[1.0em]
\dps (q_0^2)_{\mu\mu} =&  (3.58 \pm 0.10_{\text{scale}} \pm 0.001_{m_t} \pm 0.02_{C,m_c} \pm 0.06_{m_b} \pm 0.02_{\alpha
   _s}) {\text{ GeV}}^2 \nonumber \\[0.5em]
                       =& (3.58 \pm 0.12) {\text{ GeV}}^2 \,\; .
\end{align}
We observe that the inclusive zero is in the same region as the semi-inclusive one obtained in the presence of a cut on $m_{X_s}$~\cite{Bell:2010mg}, but considerably lower than in the exclusive $\bar B \to K^{\ast}\ell^+\ell^-$ case~\cite{Beneke:2001at}. The integrated ${\cal H}_A$ reads
\begin{align}
\dps {\cal H}_A[1,3.5]_{ee} =&(-1.03 \pm 0.04_{\text{scale}} \pm 0.01_{m_t} \pm 0.02_{C,m_c} \pm 0.02_{m_b} \nonumber \\
                      &\hspace*{34pt} \pm  0.01_{\alpha_s} \pm 0.003_{\text{CKM}} \pm 0.01_{\text{BR}_{\text{sl}}}) \cdot 10^{-7} = ( -1.03 \pm 0.05 ) \cdot 10^{-7} \; ,  \nonumber\\[0.5em]
\dps {\cal H}_A[3.5,6]_{ee} =&(+0.73 \pm 0.11_{\text{scale}} \pm 0.01_{m_t} \pm 0.04_{C,m_c} \pm 0.05_{m_b} \nonumber \\
                      &\hspace*{34pt} \pm  0.02_{\alpha_s} \pm 0.002_{\text{CKM}} \pm 0.01_{\text{BR}_{\text{sl}}}) \cdot 10^{-7} = ( +0.73 \pm 0.12 ) \cdot 10^{-7} \; , \nonumber\\[0.5em]
\dps {\cal H}_A[1,6]_{ee} =&( -0.29 \pm 0.14_{\text{scale}} \pm 0.002_{m_t} \pm 0.02_{C,m_c} \pm 0.06_{m_b} \nonumber \\
                    &\hspace*{34pt} \pm    0.03_{\alpha_s} \pm 0.001_{\text{CKM}} \pm 0.004_{\text{BR}_{\text{sl}}}) \cdot 10^{-7}  = ( -0.29 \pm 0.16  ) \cdot 10^{-7} \; . \\[0.5em]
\dps {\cal H}_A[1,3.5]_{\mu\mu} =&(-1.10  \pm 0.03_{\text{scale}} \pm 0.01_{m_t} \pm 0.02_{C,m_c} \pm 0.02_{m_b} \nonumber \\
                      &\hspace*{34pt} \pm  0.01_{\alpha_s} \pm 0.003_{\text{CKM}} \pm 0.01_{\text{BR}_{\text{sl}}}) \cdot 10^{-7} = (-1.10  \pm 0.05 ) \cdot 10^{-7} \; ,  \nonumber\\[0.5em]
\dps {\cal H}_A[3.5,6]_{\mu\mu} =&(+0.67  \pm 0.11_{\text{scale}} \pm 0.01_{m_t} \pm 0.04_{C,m_c} \pm 0.05_{m_b} \nonumber \\
                      &\hspace*{34pt} \pm  0.02_{\alpha_s} \pm 0.002_{\text{CKM}} \pm 0.01_{\text{BR}_{\text{sl}}}) \cdot 10^{-7} = (+0.67  \pm 0.12 ) \cdot 10^{-7} \; , \nonumber\\[0.5em]
\dps {\cal H}_A[1,6]_{\mu\mu} =&(-0.42 \pm 0.14_{\text{scale}} \pm 0.003_{m_t} \pm 0.01_{C,m_c} \pm 0.06_{m_b} \nonumber \\
                    &\hspace*{34pt} \pm    0.03_{\alpha_s} \pm 0.001_{\text{CKM}} \pm 0.01_{\text{BR}_{\text{sl}}}) \cdot 10^{-7}  = (-0.42 \pm 0.16  ) \cdot 10^{-7} \; .
\end{align}
As far as the total error is concerned, the single bins are much better behaved than the entire low-$q^2$ region. This is due to the large cancellation of the central values of bin~1 and bin~2, which is owed to the presence of the zero. The value of the latter happens to be almost exactly at the position where we subdivide the low-$q^2$ region into bin~1 and bin~2.

\subsection{${\cal H}_3$ and ${\cal H}_4$}

For the observables    ${\cal H}_3$ and ${\cal H}_4$,           sensitive to QED corrections,  we find
\begin{align}
\dps {\cal H}_3[1,3.5]_{ee} =&(4.04 \pm 0.64_{\text{scale}} \pm 0.04_{m_t} \pm 0.13_{C,m_c} \pm 0.10_{m_b} \nonumber \\
                      &\hspace*{34pt} \pm  0.03_{\alpha_s} \pm 0.01_{\text{CKM}} \pm 0.05_{\text{BR}_{\text{sl}}}) \cdot 10^{-9} = ( 4.04 \pm 0.67 ) \cdot 10^{-9} \; ,  \nonumber\\[0.5em]
\dps {\cal H}_3[3.5,6]_{ee} =&(4.88 \pm 0.50_{\text{scale}} \pm 0.05_{m_t} \pm 0.16_{C,m_c} \pm 0.07_{m_b} \nonumber \\
                      &\hspace*{34pt} \pm  0.02_{\alpha_s} \pm 0.01_{\text{CKM}} \pm 0.06_{\text{BR}_{\text{sl}}}) \cdot 10^{-9} = ( 4.88 \pm 0.54 ) \cdot 10^{-9} \; , \nonumber\\[0.5em]
\dps {\cal H}_3[1,6]_{ee} =&( 8.92 \pm 1.14_{\text{scale}} \pm 0.10_{m_t} \pm 0.30_{C,m_c} \pm 0.16_{m_b} \nonumber \\
                    &\hspace*{34pt} \pm    0.06_{\alpha_s} \pm 0.03_{\text{CKM}} \pm 0.11_{\text{BR}_{\text{sl}}}) \cdot 10^{-9}  = ( 8.92 \pm 1.20  ) \cdot 10^{-9} \; .
\end{align}
\begin{align}
\dps {\cal H}_3[1,3.5]_{\mu\mu} =&(1.68  \pm 0.26_{\text{scale}} \pm 0.02_{m_t} \pm 0.06_{C,m_c} \pm 0.04_{m_b} \nonumber \\
                      &\hspace*{34pt} \pm  0.01_{\alpha_s} \pm 0.005_{\text{CKM}} \pm 0.02_{\text{BR}_{\text{sl}}}) \cdot 10^{-9} = (1.68  \pm 0.27 ) \cdot 10^{-9} \; ,  \nonumber\\[0.5em]
\dps {\cal H}_3[3.5,6]_{\mu\mu} =&(2.03  \pm 0.21_{\text{scale}} \pm 0.02_{m_t} \pm 0.07_{C,m_c} \pm 0.03_{m_b} \nonumber \\
                      &\hspace*{34pt} \pm  0.01_{\alpha_s} \pm 0.006_{\text{CKM}} \pm 0.03_{\text{BR}_{\text{sl}}}) \cdot 10^{-9} = (2.03  \pm 0.22 ) \cdot 10^{-9} \; , \nonumber\\[0.5em]
\dps {\cal H}_3[1,6]_{\mu\mu} =&(3.71 \pm 0.47_{\text{scale}} \pm 0.04_{m_t} \pm 0.12_{C,m_c} \pm 0.06_{m_b} \nonumber \\
                    &\hspace*{34pt} \pm    0.02_{\alpha_s} \pm 0.01_{\text{CKM}} \pm 0.05_{\text{BR}_{\text{sl}}}) \cdot 10^{-9}  = (3.71 \pm 0.50  ) \cdot 10^{-9} \; .
\end{align}
\begin{align}
\dps {\cal H}_4[1,3.5]_{ee} =&(6.23 \pm 0.55_{\text{scale}} \pm 0.07_{m_t} \pm 0.21_{C,m_c} \pm 0.01_{m_b} \nonumber \\
                      &\hspace*{34pt} \pm  0.02_{\alpha_s} \pm 0.02_{\text{CKM}} \pm 0.08_{\text{BR}_{\text{sl}}}) \cdot 10^{-9} = ( 6.23 \pm 0.60 ) \cdot 10^{-9} \; ,  \nonumber\\[0.5em]
\dps {\cal H}_4[3.5,6]_{ee} =&(2.19 \pm 0.16_{\text{scale}} \pm 0.03_{m_t} \pm 0.07_{C,m_c} \pm 0.02_{m_b} \nonumber \\
                      &\hspace*{34pt} \pm  0.006_{\alpha_s} \pm 0.006_{\text{CKM}} \pm 0.03_{\text{BR}_{\text{sl}}}) \cdot 10^{-9} = ( 2.19 \pm 0.18 ) \cdot 10^{-9} \; , \nonumber\\[0.5em]
\dps {\cal H}_4[1,6]_{ee} =&(8.41 \pm 0.71_{\text{scale}} \pm 0.10_{m_t} \pm 0.28_{C,m_c} \pm 0.02_{m_b} \nonumber \\
                    &\hspace*{34pt} \pm    0.02_{\alpha_s} \pm 0.02_{\text{CKM}} \pm 0.10_{\text{BR}_{\text{sl}}}) \cdot 10^{-9}  = ( 8.41 \pm 0.78  ) \cdot 10^{-9} \; .
\end{align}
\begin{align}
\dps {\cal H}_4[1,3.5]_{\mu\mu} =&(2.59  \pm 0.23_{\text{scale}} \pm 0.03_{m_t} \pm 0.09_{C,m_c} \pm 0.006_{m_b} \nonumber \\
                      &\hspace*{34pt} \pm  0.007_{\alpha_s} \pm 0.007_{\text{CKM}} \pm 0.03_{\text{BR}_{\text{sl}}}) \cdot 10^{-9} = (2.59  \pm 0.25 ) \cdot 10^{-9} \; ,  \nonumber\\[0.5em]
\dps {\cal H}_4[3.5,6]_{\mu\mu} =&(0.91  \pm 0.07_{\text{scale}} \pm 0.01_{m_t} \pm 0.03_{C,m_c} \pm 0.008_{m_b} \nonumber \\
                      &\hspace*{34pt} \pm  0.002_{\alpha_s} \pm 0.003_{\text{CKM}} \pm 0.01_{\text{BR}_{\text{sl}}}) \cdot 10^{-9} = (0.91  \pm 0.075 ) \cdot 10^{-9} \; , \nonumber\\[0.5em]
\dps {\cal H}_4[1,6]_{\mu\mu} =&(3.50 \pm 0.29_{\text{scale}} \pm 0.04_{m_t} \pm 0.12_{C,m_c} \pm 0.01_{m_b} \nonumber \\
                    &\hspace*{34pt} \pm    0.01_{\alpha_s} \pm 0.01_{\text{CKM}} \pm 0.04_{\text{BR}_{\text{sl}}}) \cdot 10^{-9}  = (3.50 \pm 0.32  ) \cdot 10^{-9} \; .
\end{align}
\subsection{Branching ratio, low-$q^2$ region}

The decay width is simply given by the sum of $H_T$ and $H_L$ and hence can in principle be derived by the numbers given in the previous subsections. However, we give the numbers explicitly here, for two reasons. First, the branching ratio is an important quantity, also experimentally. Second, there are two more contributions which are available only for the branching ratio, but not for ${\cal H}_T$ or ${\cal H}_L$ individually. These are the finite bremsstrahlung contributions from~\cite{Asatryan:2002iy} and the non-log enhanced terms of $\omega_{99}^{({\rm em})}(\s)$. Both give only a small correction, but we include them for the sake of completeness. This yields

\begin{align}
\dps {\cal B}[1,3.5]_{ee} =& ( 9.26 \pm 0.34_{\text{scale}} \pm 0.11_{m_t} \pm 0.30_{C,m_c} \pm 0.10_{m_b} \nonumber \\
 & \hspace*{25pt}\pm 0.02_{\alpha_s} \pm 0.03_{\text{CKM}} \pm 0.11_{\text{BR}_{\text{sl}}}) \cdot 10^{-7}
                       = ( 9.26 \pm 0.49 ) \cdot 10^{-7} \; , \nonumber\\[0.5em]
\dps {\cal B}[3.5,6]_{ee} =& ( 7.44 \pm 0.37_{\text{scale}} \pm 0.10_{m_t} \pm 0.26_{C,m_c} \pm 0.08_{m_b} \nonumber \\
 & \hspace*{25pt}\pm 0.03_{\alpha_s} \pm 0.02_{\text{CKM}} \pm 0.09_{\text{BR}_{\text{sl}}} ) \cdot 10^{-7}
                       = ( 7.44 \pm 0.48 ) \cdot 10^{-7} \; , \nonumber\\[0.5em]
\dps {\cal B}[1,6]_{ee} =& (1.67 \pm 0.07_{\text{scale}} \pm 0.02_{m_t} \pm 0.06_{C,m_c} \pm 0.02_{m_b} \nonumber \\
& \hspace*{25pt} \pm 0.01_{\alpha_s} \pm 0.005_{\text{CKM}} \pm 0.02_{\text{BR}_{\text{sl}}}) \cdot 10^{-6}  = (1.67 \pm 0.10) \cdot 10^{-6} \; .
\end{align}
\begin{align}
\dps {\cal B}[1,3.5]_{\mu\mu} =& ( 8.88 \pm 0.31_{\text{scale}} \pm 0.11_{m_t} \pm 0.29_{C,m_c} \pm 0.10_{m_b} \nonumber \\
& \hspace{25pt} \pm 0.02_{\alpha_s} \pm 0.02_{\text{CKM}} \pm 0.11_{\text{BR}_{\text{sl}}} ) \cdot 10^{-7}
                         = (8.88 \pm 0.46 ) \cdot 10^{-7}\;  , \nonumber\\[0.5em]
\dps {\cal B}[3.5,6]_{\mu\mu} =& ( 7.31 \pm 0.36_{\text{scale}} \pm 0.09_{m_t} \pm 0.25_{C,m_c} \pm 0.09_{m_b} \nonumber \\
&\hspace{25pt} \pm 0.03_{\alpha
   _s} \pm 0.02_{\text{CKM}} \pm 0.09_{\text{BR}_{\text{sl}}}) \cdot 10^{-7}
                       = (7.31 \pm 0.47 ) \cdot 10^{-7}\; ,\nonumber\\[0.5em]
\dps {\cal B}[1,6]_{\mu\mu} =& (1.62 \pm 0.07_{\text{scale}} \pm 0.02_{m_t} \pm 0.05_{C,m_c} \pm 0.02_{m_b} \nonumber \\
& \hspace*{25pt} \pm 0.01_{\alpha_s} \pm 0.005_{\text{CKM}} \pm 0.02_{\text{BR}_{\text{sl}}}) \cdot 10^{-6} 
 = (1.62 \pm 0.09) \cdot 10^{-6} \; .
\end{align}

The values are about 2\% larger compared to our previous analysis~\cite{Huber:2005ig}. This is due to updated input parameters and the inclusion of the Kr\"uger-Sehgal corrections~\cite{Kruger:1996cv,Kruger:1996dt}.

\subsection{Branching ratio, high-$q^2$ region}

The branching ratio in the high-$q^2$ region suffers from large uncertainties stemming from hadronic input parameters in the $1/m_b^{2,3}$ power-corrections, which results in total error bars of ${\cal O}(30\%)$,

\begin{align}
\dps {\cal B}[>14.4]_{ee} =& ( 2.20 \pm 0.30_{\text{scale}} \pm 0.03_{m_t} \pm 0.06_{C,m_c} \pm 0.16_{m_b} \pm 0.003_{\alpha_s} \pm 0.01_{\text{CKM}} \pm 0.03_{\text{BR}_{\text{sl}}} \nonumber \\[0.5em]
   & \hspace*{25pt} \pm 0.12_{\lambda _2} \pm 0.48_{\rho_1} \pm 0.36_{f_s} \pm 0.05_{f_u}) \cdot 10^{-7}\nonumber \\[0.5em]
                       =& ( 2.20 \pm 0.70 ) \cdot 10^{-7} \; , \nonumber\\[1.0em]
\dps {\cal B}[>14.4]_{\mu\mu} =& ( 2.53  \pm 0.29_{\text{scale}} \pm 0.03_{m_t} \pm 0.07_{C,m_c} \pm 0.18_{m_b} \pm 0.003_{\alpha_s} \pm 0.01_{\text{CKM}} \pm 0.03_{\text{BR}_{\text{sl}}} \nonumber \\[0.5em]
   & \hspace{25pt} \pm 0.12_{\lambda _2} \pm 0.48_{\rho_1} \pm 0.36_{f_s} \pm 0.05_{f_u}) \cdot 10^{-7}\nonumber \\[0.5em]
                      =  & ( 2.53 \pm 0.70 ) \cdot 10^{-7} \; .
\end{align}

Comparing these results to earlier analyses on the high-$q^2$ branching ratio shows that our numbers are considerably lower than the ones in~\cite{Ghinculov:2003qd,Greub:2008cy}. In the following, we show that this is the result of several effects which all give corrections in the same direction. Once we turn to the prescriptions given in~\cite{Ghinculov:2003qd,Greub:2008cy} we reproduce their results, as can be seen below.

We first perform the comparison to Greub et al.~\cite{Greub:2008cy}. We start with the above numbers and first switch off the $\ln(m_b^2/m_\ell^2)$-enhanced QED corrections, which also removes the difference between the muon and the electron channel, and yields 2.74 (all numbers that follow are in units of $10^{-7}$). Next, we turn off the finite bremsstrahlung contributions, which is only a minor effect and does not change the digits given before. Taking out the Kr\"uger-Sehgal corrections, on the other hand, is a rather large effect in the high-$q^2$ region and results in 3.05. We also have to remove the $1/m_b^{2,3}$ and $1/m_c^2$ non-factorisable power-corrections which further increases the result to 3.36. Switching furthermore off those QED corrections which are not $\ln(m_b^2/m_\ell^2)$-enhanced, we get 3.56. This shift is rather large, but we remind the reader that some of these terms are enhanced by $m_t^2/(M_W^2 \sin^2\theta_W)$. Changing from four- to two-loop running for $\alpha_s$ has again only a minor impact and gives 3.55. We now switch off the change in renormalisation scheme for the quark masses, i.e.\ we use the pole mass for charm and bottom. Furthermore, we use the input parameters from~\cite{Greub:2008cy}. Both effects taken together give 3.68. We now take into account that the integration interval in~\cite{Greub:2008cy} is given in the variable $\s=q^2/m_b^2$. Hence a change in the value for $m_b$ results in the modified lower integration limit $q^2_{\rm min}=13.824~{\rm GeV}^2$. This effect must not be underestimated because it brings the branching ratio up to 4.36. We now turn to the normalisation prescription given in~\cite{Greub:2008cy}, which instead of the factor $C$ from eq.~(\ref{eq:C}) and the perturbative expansion of $\Gamma(b \to u \, e \, \bar\nu)$ makes direct use of the perturbative expansion of $\Gamma(b \to c \, e \, \bar\nu)$, including charm-mass dependent phase-space factors and radiative corrections. This increases the branching ratio further to 4.57. Finally, we divide by the experimentally measured semileptonic $b \to c$ branching ratio (see table~\ref{tab:inputs}) and get 43, which is precisely the value of $R_{\rm high,~pert}$ in eq.~(48) of~\cite{Greub:2008cy}.

The comparison to Ghinculov et al.\ in~\cite{Ghinculov:2003qd} proceeds along the same lines. The differences to the analysis by Greub et al.\ are the Kr\"uger-Sehgal corrections and the $1/m_b^{2,3}$, $1/m_c^2$ power corrections, both are taken into account in~\cite{Ghinculov:2003qd}. Moreover, different input parameters are used and the lower integration limit is formulated in $q^2$ rather than in $\s$. To quantify these effects, we first switch off again $\ln(m_b^2/m_\ell^2)$-enhanced QED corrections and finite bremsstrahlung effects first and end up with 2.74. We then also remove those QED corrections that are not enhanced by $\ln(m_b^2/m_\ell^2)$, which gives 2.93. Changing from four- to two-loop running for $\alpha_s$ is again only a small effect and gives 2.92. The biggest effect comes from the change of input parameters and the removal of the renormalisation-scheme conversion for the quark masses, i.e.\ we now use the pole mass for charm and bottom. These two effects taken together result in 3.89. Finally, we switch to the normalisation that is used in~\cite{Ghinculov:2003qd} and get 4.02. This number coincides within a fraction of a percent with the value 4.04 from eq.~(6.36) in~\cite{Ghinculov:2003qd}. The obtained level of accuracy shall be sufficient for the present check.

\subsection{The ratio ${\cal R}(s_0)$}

\allowdisplaybreaks{
\begin{align}
\dps {\cal R}(14.4)_{ee} =& ( 2.25  \pm 0.12_{\text{scale}} \pm 0.03_{m_t} \pm 0.02_{C,m_c} \pm 0.01_{m_b} \pm 0.01_{\alpha
   _s} \pm 0.20_{\text{CKM}} \nonumber \\[0.5em]
   & \hspace*{25pt} \pm 0.02_{\lambda _2} \pm 0.14_{\rho_1} \pm 0.08_{f_u^0+f_s} \pm 0.12_{f_u^0-f_s}) \cdot 10^{-3}\nonumber \\[0.5em]
                       =& (2.25 \pm 0.31 ) \cdot 10^{-3} \; ,\nonumber\\[1.0em]
\dps {\cal R}(14.4)_{\mu\mu} =& ( 2.62 \pm 0.09_{\text{scale}} \pm 0.03_{m_t} \pm 0.01_{C,m_c} \pm 0.01_{m_b} \pm 0.01_{\alpha
   _s} \pm 0.23_{\text{CKM}} \nonumber \\[0.5em]
   & \hspace*{25pt} \pm 0.0002_{\lambda _2} \pm 0.09_{\rho_1} \pm 0.04_{f_u^0+f_s} \pm 0.12_{f_u^0-f_s}) \cdot 10^{-3}\nonumber \\[0.5em]
                       =& ( 2.62 \pm 0.30 ) \cdot 10^{-3} \; .
\end{align}
}
We clearly see a reduction of the total error bars from ${\cal O}(30\%)$ in the high-$q^2$ branching ratio to $14\%$ and $11\%$ in the electron and muon channel of ${\cal R}(s_0)$, respectively. Besides the uncertainties due to power corrections, also the scale uncertainty gets significantly reduced. The largest source of error are CKM elements (notably $V_{ub}$).


\section{New Physics sensitivities}
\label{sec:NP}
In this section we present the constraints on the most relevant Wilson coefficients ($C_9$ and $C_{10}$) that we obtain using the current experimental results, and investigate the reach of Belle~II with an expected final integrated luminosity of $50\; {\rm ab}^{-1}$.

Previous model-independent new physics analyses~\cite{AGHL,Lee:2006gs,Kim:1998hp,Gambino:2004mv}, as well as studies in specific models such as minimal-flavour-violation~\cite{Bobeth:2005ck,Hurth:2008jc,Hurth:2012jn}, two-Higgs doublet models~\cite{Schilling:2004gk,Xiao:2006dq}, and supersymmetry~\cite{Kim:1998hp,Bertolini:1990if,Cho:1996we,Goto:1996dh,Hewett:1996ct,Huang:1998vb,Lunghi:1999uk,Bobeth:2004jz} can  be found in the literature.

The weighted averages for the low- and high-$q^2$ branching fractions have been presented in eq.~(\ref{eq:expWA}). Here we need the results on the individual channels:
\bea
{\cal B} (\bar B\to X_s \ell^+\ell^-)_{\rm low}^{\rm exp} &=& 
\begin{cases}
\left( 1.493 \pm 0.504^{+0.411}_{-0.321} \right) \times 10^{-6} 
& ({\rm Belle,\; }\ell\ell) \cr 
 \left( 1.93^{+0.47 +0.21}_{-0.45 -0.16}  \pm 0.18 \right) \times 10^{-6} 
& ({\rm BaBar,\; }ee) \cr
 \left( 0.66^{+0.82 +0.30}_{-0.76 -0.24}  \pm 0.18 \right) \times 10^{-6} 
& ({\rm BaBar,\; }\mu\mu) \cr
 \left( 1.6^{+0.41 +0.17}_{-0.39 -0.13}  \pm 0.18 \right) \times 10^{-6} 
& ({\rm BaBar,\; }\ell\ell) \, ,\cr
\end{cases}
\label{eq:explow}
\eea
\bea
{\cal B} (\bar B\to X_s \ell^+\ell^-)_{\rm high}^{\rm exp} &=& 
\begin{cases}
\left( 0.418 \pm 0.117^{+0.061}_{-0.068} \right) \times 10^{-6} 
& ({\rm Belle,\;}\ell\ell) \cr 
\left( 0.56^{+0.19 +0.03}_{-0.18 -0.03}  \pm 0.00  \right) \times 10^{-6} 
& ({\rm BaBar, \; }ee) \cr
\left( 0.60^{+0.31 +0.05}_{-0.29 -0.04}  \pm 0.00  \right) \times 10^{-6} 
& ({\rm BaBar, \; }\mu\mu \; ) \cr
\left( 0.57^{+0.16 +0.03}_{-0.15 -0.02}  \pm 0.00  \right) \times 10^{-6} 
& ({\rm BaBar, \; }\ell\ell) \, .\cr
\end{cases}
\label{eq:exphigh}
\eea
In each result, the first error is statistical, the second systematics and the third model-depedent systematics which is included in case of Belle in the second error. Note that the high-$q^2$ region chosen by BaBar and Belle have a slightly different $q^2$ minimum (14.4 and 14.2 $\rm GeV^2$ for Belle and BaBar, respectively).

In ref.~\cite{Sato:2014pjr} Belle presented a measurement of the normalized forward--backward asymmetry defined in eq.~(\ref{eq:normFBA}) in the low- and high-$q^2$ regions. The binning chosen to present the measurement (bin1 = [0.2,4.3] $\rm GeV^2$ and bin2 = [4.3,7.3(8.1)]  $\rm GeV^2$ for electrons (muons)) differs from the one proposed in this work. In particular, the larger integration end-point in the second bin includes a region of the spectrum where sizable interference from the tail of the $J/\psi$ is present. From ref.~\cite{Sato:2014pjr} we read:
\begin{align}
{\overline A}_{\rm FB}^{\rm exp} (\bar B\to X_s \ell^+\ell^-) &= 
\begin{cases}
0.34 \pm 0.24 \pm 0.02 & {\rm Belle, bin1} \cr
0.04 \pm 0.31 \pm 0.05 & {\rm Belle, bin2.} \cr
\end{cases}
\label{eq:expFBA}
\end{align}
In order to preserve the cancellation of systematic uncertainties, Belle averaged the normalized asymmetries in the electron and muon channels; i.e.\
\begin{align}
{\overline A}_{\rm FB} (\bar B\to X_s \ell^+\ell^-) = \left( {\overline A}_{\rm FB}(\bar B\to X_s e^+e^-) +{\overline A}_{\rm FB}(\bar B\to X_s \mu^+\mu^-) \right)/2 \, .
\end{align}
We integrated our differential spectra in the above bins in order to investigate the impact that this measurement has on the Wilson coefficients, but we caution the reader that the uncertainties we quote could be underestimated. We find:
\begin{align}
{\overline A}_{\rm FB} (\bar B\to X_s \ell^+\ell^-) &= 
\begin{cases}
-0.0773\pm 0.0057  & {\rm bin1} \cr
+0.049 \pm  0.018 & {\rm bin2.} \cr
\end{cases}
\label{eq:thFBA}
\end{align}

We define the following ratios of high-scale Wilson coefficients (see~\cite{Huber:2005ig} for the precise definitions of the Wilson coefficients),
\bea
R_{7,8} = \frac{C_{7,8}^{(00){\rm eff}} (\mu_0)}{C_{7,8}^{(00){\rm eff,SM}}(\mu_0)}  
\hspace{1cm} {\rm and} \hspace{1cm}
R_{9,10} = \frac{C_{9,10}^{(11)} (\mu_0)}{C_{9,10}^{(11){\rm SM}}(\mu_0)} \; .
\eea
\begin{figure}
\begin{center}
\includegraphics[scale=0.5]{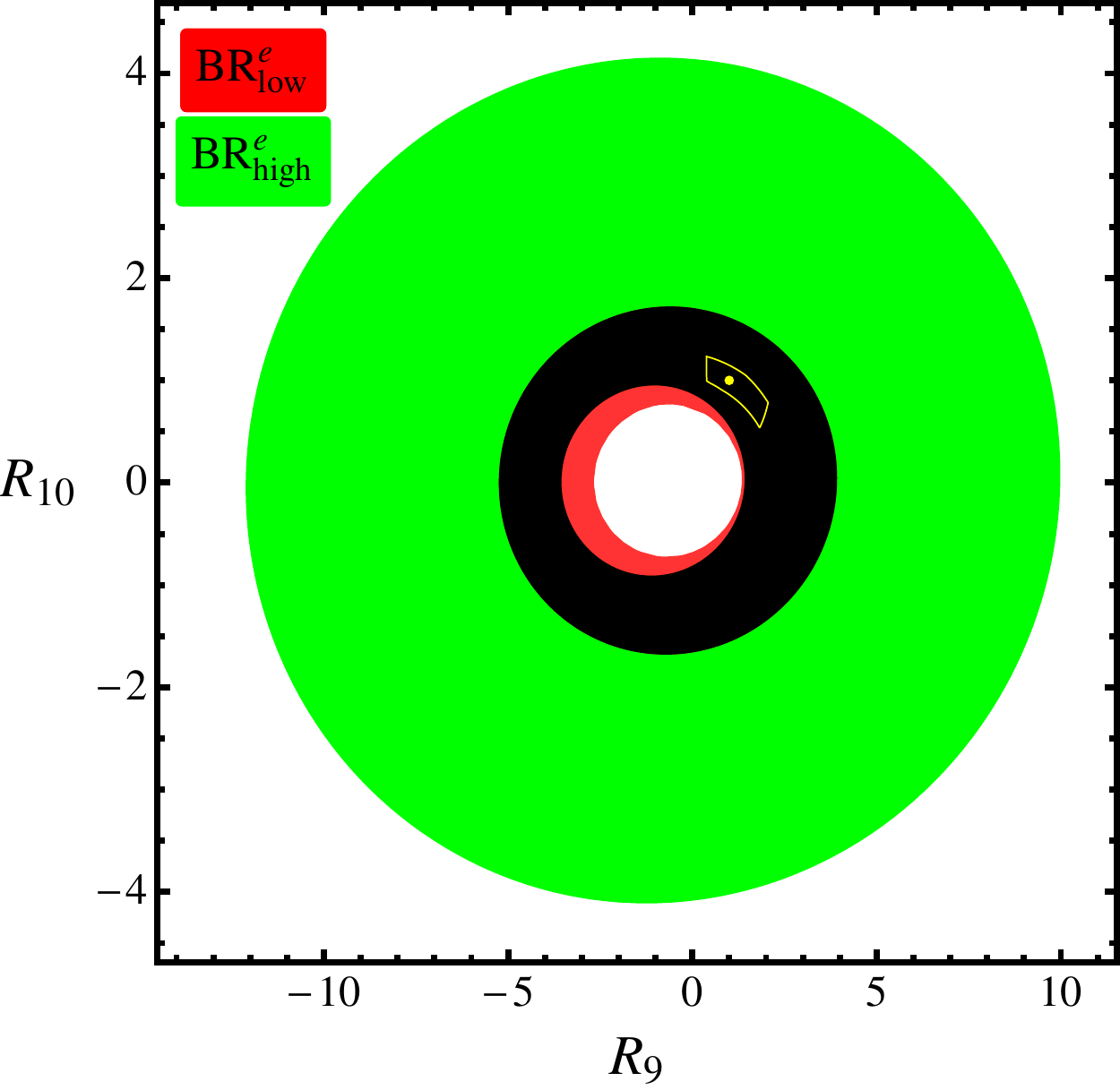}
\includegraphics[scale=0.5]{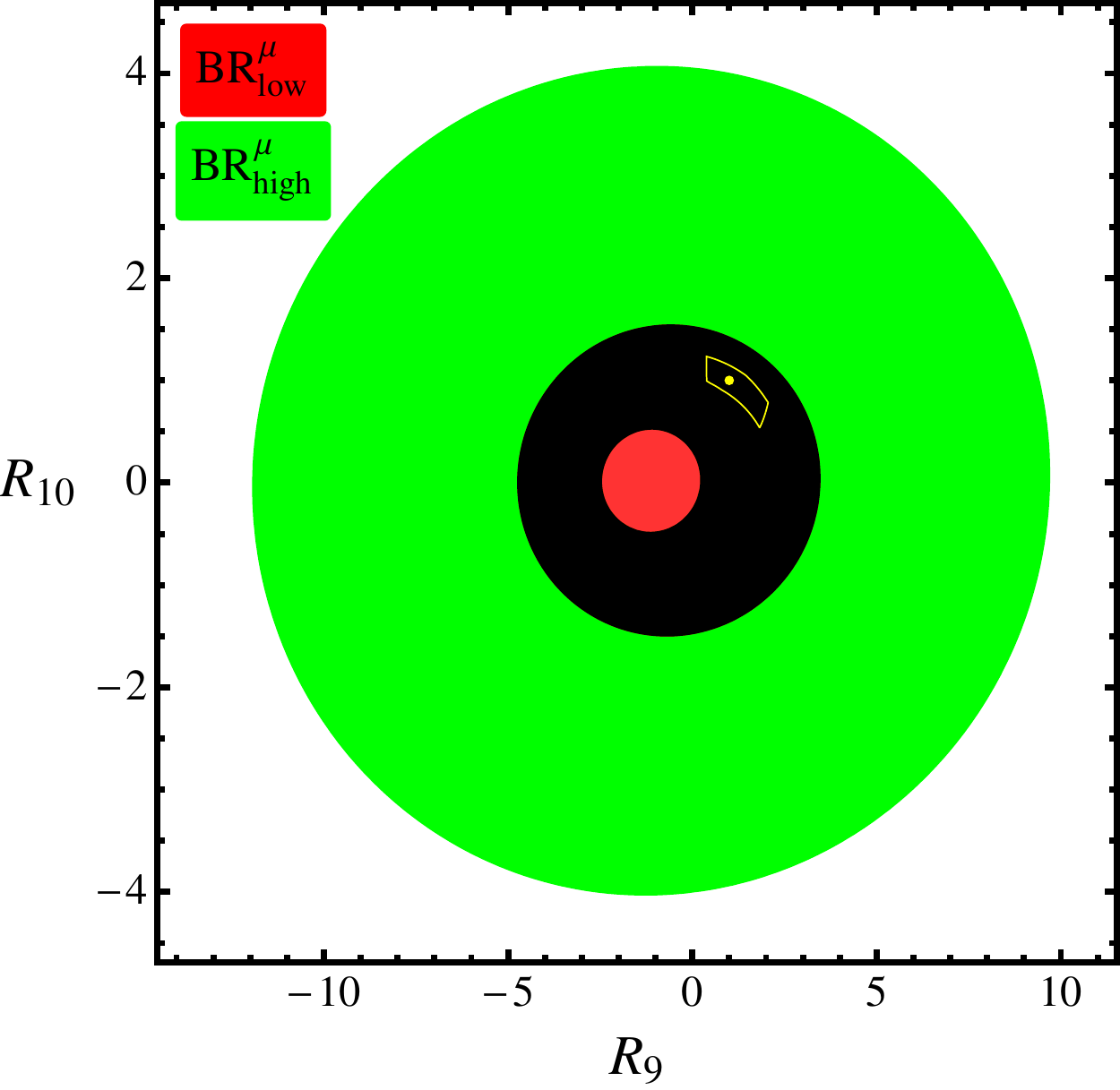}
\includegraphics[scale=0.5]{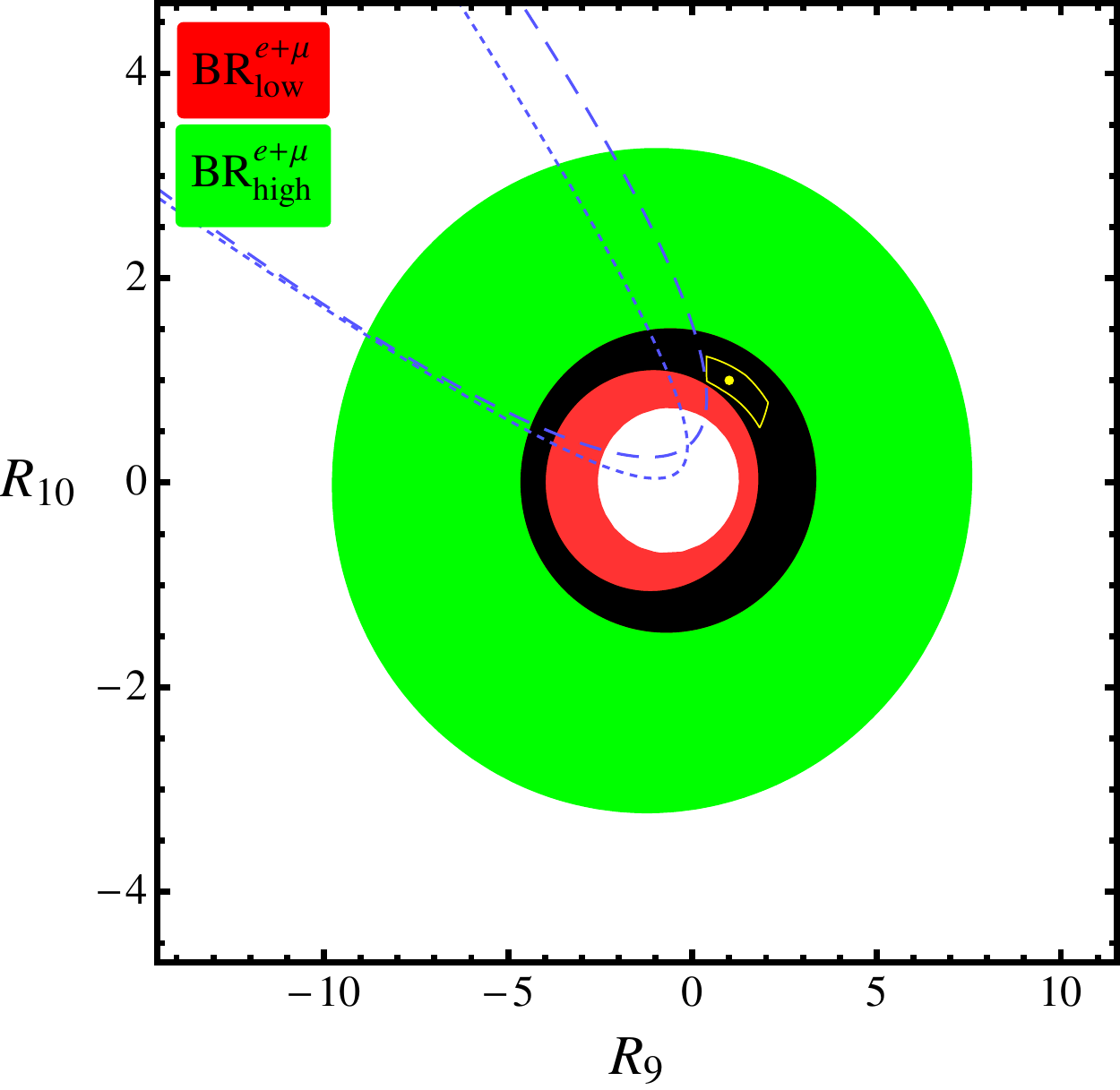}
\caption{Constraints on the high--scale Wilson coefficient ratios $[R_9,R_{10}]$ that we obtain at 95\% C.L.\ from the present BaBar and Belle experimental branching ratios measurements. In the upper left (upper right, lower) plot we show the constraints obtained from the measured branching ratios in the low-$q^2$ and high-$q^2$ region in the electron (muon, electron plus muon) channel. The red and green regions correspond to the low- and high-$q^2$ regions, respectively. The black region is the overlap of these two constraints. The dot is the SM expectation ($[R_9,R_{10}]=[1,1]$). The yellow contour is the Belle~II reach (see figures~\ref{fig:Ri-br-future}--\ref{fig:Ri-hah3h4-future}). The region outside the dashed (dotted) parabola shaped regions are allowed by the Belle measurement of the normalized forward--backward asymmetry in bin1 (bin2). \label{fig:Ri-present}}
\end{center}
\end{figure}
The numerical formulas for all observables in terms of the ratios $R_i$ can be found in appendix~\ref{app:NPformulae}. We assume that the relative theoretical uncertainty on a given observable ($\delta O/O$) is mostly independent of the precise values of Wilson coefficients and that it can be extracted from the SM predictions presented in section~\ref{sec:results}. 

We present the bounds on the ratios $R_9$ and $R_{10}$ under the assumption of no new physics contributions to the magnetic and chromo-magnetic dipole operators ($R_{7,8} = 1$) in figure~\ref{fig:Ri-present} (similar analyses were done, e.g., in~\cite{AGHL,Lee:2006gs}). The contours are the 95\% C.L.\ regions allowed by the experimental results in eqs.~(\ref{eq:explow}), (\ref{eq:exphigh}) and (\ref{eq:expWA}); two sigma theoretical uncertainties are added linearly. In each plot we show the impact of the branching ratio measurement in the low-$q^2$ (red regions) and high-$q^2$ (green regions) and their overlap (black regions). The SM corresponds to the point $[R_9, R_{10}] = [1,1]$. As we discuss below, the small yellow contours correspond to the Belle~II estimated reach, assuming that the observed central values coincide with our predictions. The top left, top right and lower plot consider the $B\to X_s e^+e^-$, $B\to X_s \mu^+\mu^-$  and $B\to X_s \ell^+\ell^-$ cases, respectively. In the lower plot in figure~\ref{fig:Ri-present} we include also the 95\% C.L.\ bounds from the Belle measurement of the normalized forward--backward asymmetry given in eq.~(\ref{eq:expFBA}); the region outside the dashed and dotted parabola shaped regions are allowed by the measurement in bin1 and bin2, respectively. The resulting picture is in overall agreement with the SM expectations at the 95\% C.L.; though we should note that at the one sigma level there are some statistically insignificant tensions driven by a disagreement between low- and high-$q^2$ measurements in the muon channel.

In order to study the expected Belle~II reach, we estimate the statistical uncertainties on the various observables using the squared weight method detailed in ref.~\cite{Cowan:2012}. Let us consider the following differential quantity:
\begin{align}
\frac{d^2{\cal N}}{d\s dz} &= \frac{{\cal L} \; \sigma_{\rm prod}}{\Gamma_{\rm tot}} \frac{d^2\Gamma}{d\s dz} 
\end{align}
where the ${\cal L}$ is the integrated luminosity, $\sigma_{\rm prod}$ is the production cross section for $e^+e^- \to B\bar B$ at the B-factories' center of mass energy, $\Gamma_{\rm tot}$ is the total $B$ decay width and $d^2\Gamma/d\s dz$ is the double differential $B\to X_s \ell^+\ell^-$ decay rate. The number of events that we expect to observe in a certain range of $\s$ and $z$ is
\begin{align}
{\cal N}_{\rm exp} &= \int \frac{d^2{\cal N}}{d\s dz} d\s \; dz  \; ,\\
\delta {\cal N}_{\rm exp} &= \sqrt{{\cal N}_{\rm exp}}
\end{align}
where $\delta {\cal N}_{\rm exp}$ is the expected statistical error. If instead of considering simple slices of the integration region we utilize a weight function $W[\s,z]$ to define an observable (that cannot be anymore interpreted in terms of ``number of events''), the above equations generalize to
\begin{align}
{\cal O}_{\rm exp} &= \int \frac{d^2{\cal N}}{d\s dz} \; W[\s,z]\;  d\s \; dz \; , \label{oexp}\\
\delta {\cal O}_{\rm exp} &= \left[ \int \frac{d^2{\cal N}}{d\s dz} \; W[\s,z]^2\;  d\s \; dz \right]^{\frac{1}{2}} \; .
\label{errw2}
\end{align}
Note that eq.~(\ref{errw2}) reproduces the correct uncertainties for the simple case in which the weight is a product of theta functions (i.e.\ the integral is restricted to a certain region of phase space) and that the relative uncertainty $\delta {\cal O}_{\rm exp}/{\cal O}_{\rm exp}$ is invariant under rescaling of the weight function.

\begin{table}
\begin{center}
\begin{tabular}{|c|c c c c|}
\hline
         & $[1,3.5]$  & $[3.5, 6]$ & $[1,6]$ & $> 14.4$ \\ \hline
$\cal B$ & 3.7 \% & 4.0 \% & 3.0 \% & 4.1\%\\
${\cal H}_T$    & 24 \% & 21 \% & 16 \% & - \\
${\cal H}_L$    & 5.8 \% & 6.8 \% & 4.6 \% & - \\
${\cal H}_A$    & 37 \% & 44 \% & 200 \% & - \\
${\cal H}_3$    & 240 \% & 180 \% & 150 \% & - \\
${\cal H}_4$    & 140 \% & 360 \% & 140 \% & - \\
\hline
\end{tabular}
\caption{Statistical uncertainties that we expect at Belle~II with $50\; {\rm ab}^{-1}$ of integrated luminosity. The first row gives the considered $q^2$ bin in ${\rm GeV}^2$. \label{table:future-errors}}
\end{center}
\end{table}

\begin{figure}
\begin{center}
\includegraphics[width=0.47 \linewidth]{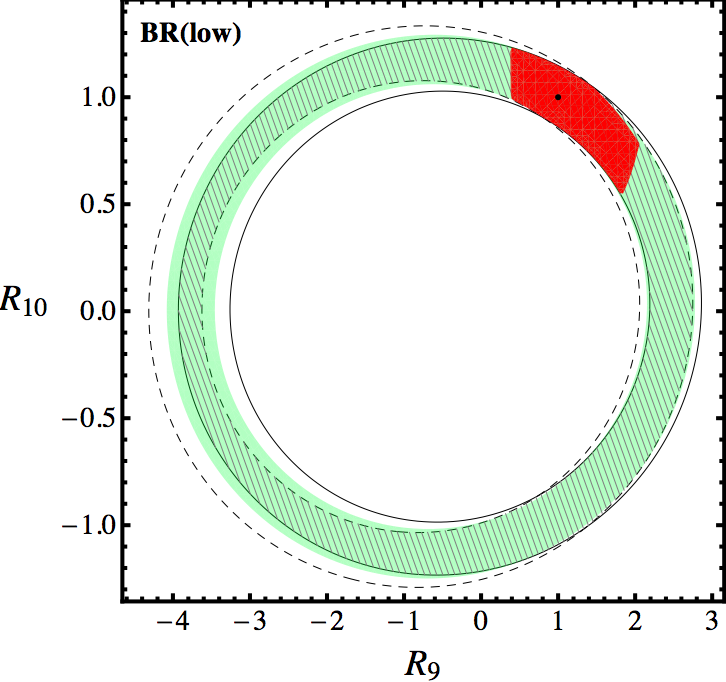}\hspace*{15pt}
\includegraphics[width=0.47 \linewidth]{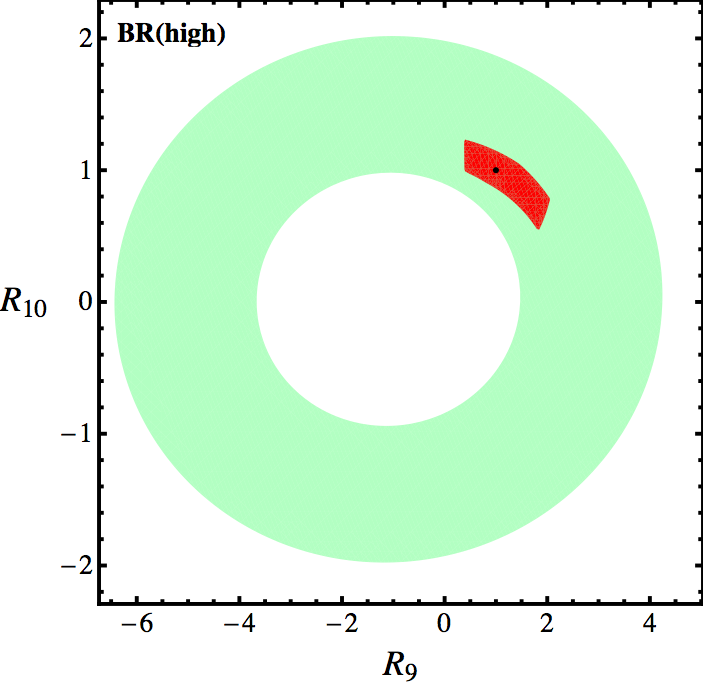}
\caption{Constraints on $[R_9,R_{10}]$ that we expect at 95\% C.L.\ from Belle~II measurements of the branching ratio in the low-$q^2$ (left plot) and high-$q^2$ (right plot) regions with $50\; {\rm ab}^{-1}$ of integrated luminosity. For the low-$q^2$ case, the solid and dashed contours correspond to the branching ratio restricted to the low ($[1,3.5]\; {\rm GeV}^2$) and high ($[3.5,6]\; {\rm GeV}^2$) bin, respectively. The hashed region is the overlap of the expected constraints from these two bins. The shaded region is the constraint we obtain by considering the branching ratio integrated in the whole low-$q^2$ region. The black dot is the SM expectation. The solid red area is the overlap of all constraints we consider (it corresponds to the yellow contour in figure~\ref{fig:Ri-present}). \label{fig:Ri-br-future}}
\end{center}
\end{figure}

\begin{figure}
\begin{center}
\includegraphics[width=0.47 \linewidth]{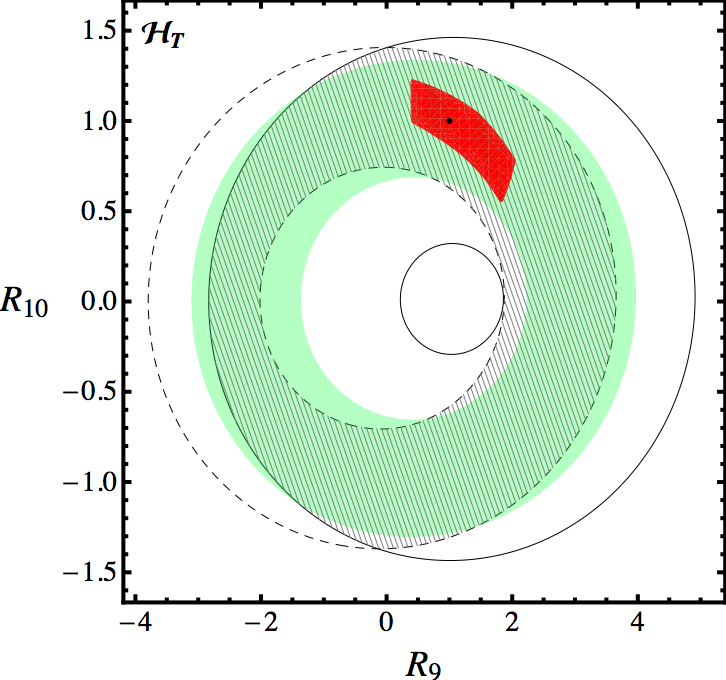}\hspace*{15pt}
\includegraphics[width=0.47 \linewidth]{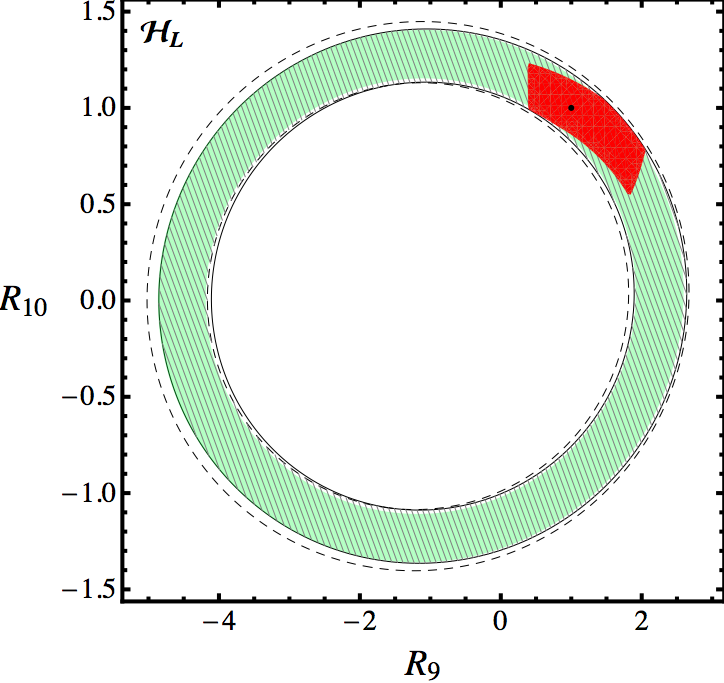}
\caption{Constraints on $[R_9,R_{10}]$ that we expect at 95\% C.L.\ from Belle~II measurements of ${\cal H}_T$ (left plot) and ${\cal H}_L$ (right plot) in the low-$q^2$ region with $50\; {\rm ab}^{-1}$ of integrated luminosity. See figure~\ref{fig:Ri-br-future} for further details.
\label{fig:Ri-hthl-future}}
\end{center}
\end{figure}

\begin{figure}
\begin{center}
\includegraphics[width=0.47 \linewidth]{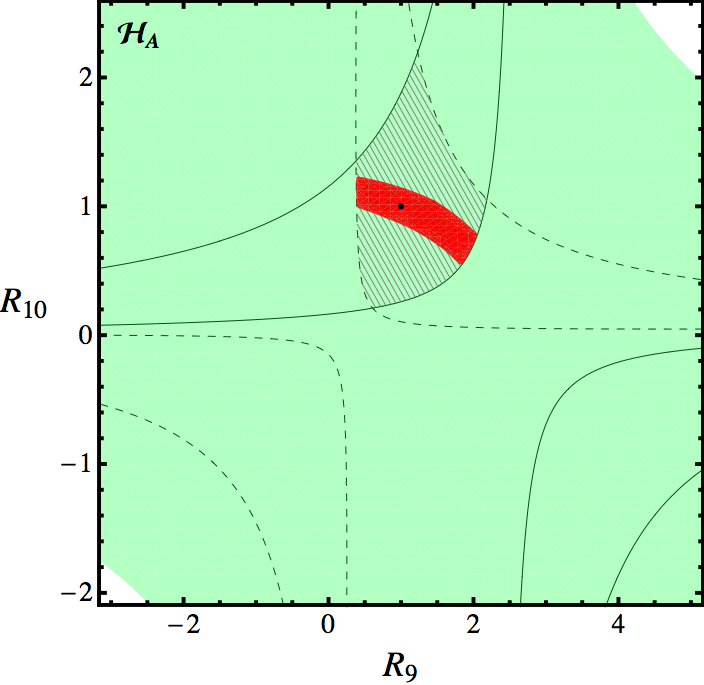} \\
\includegraphics[width=0.47 \linewidth]{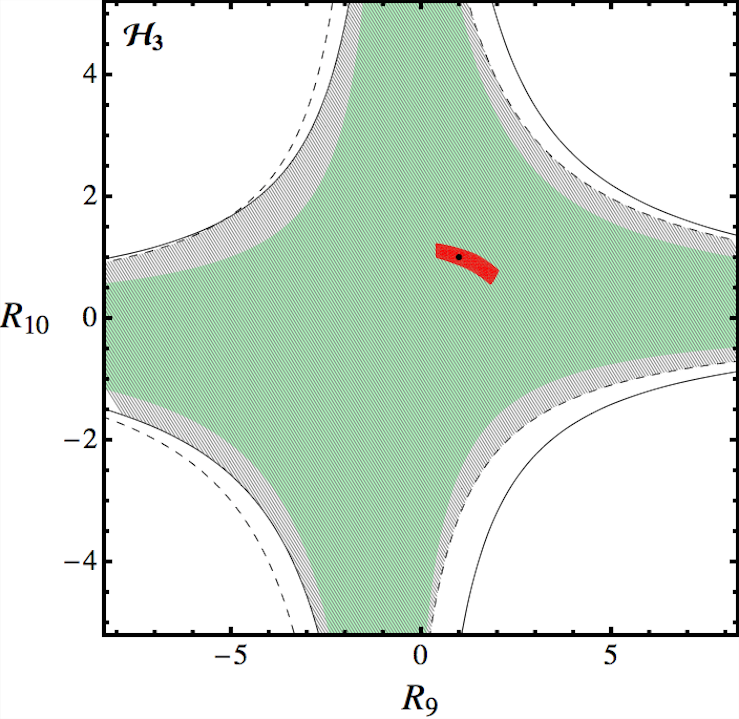}\hspace*{15pt}
\includegraphics[width=0.47 \linewidth]{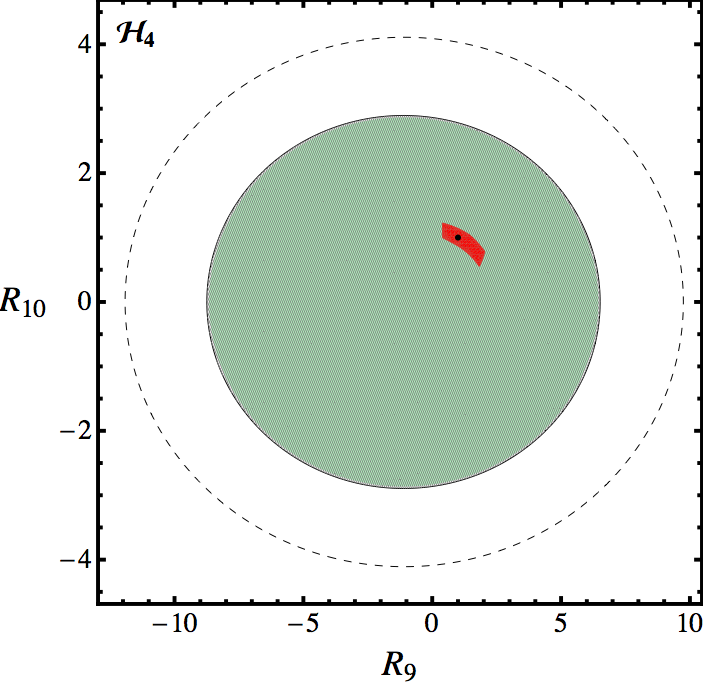}
\caption{Constraints on $[R_9,R_{10}]$ that we expect at 95\% C.L.\ from Belle~II measurements of ${\cal H}_A$ (upper plot), ${\cal H}_3$ (lower left plot) and ${\cal H}_4$ (lower right plot) in the low-$q^2$ region with $50\; {\rm ab}^{-1}$ of integrated luminosity. See figure~\ref{fig:Ri-br-future} for further details.
\label{fig:Ri-hah3h4-future}}
\end{center}
\end{figure}

In order to asses expected uncertainties on observables corresponding to the weights given in eq.~(\ref{weights}) we start from the double differential rate given in eq.~(\ref{diffwidth}) and use the expressions for the $H_I$ in eqs.~(\ref{eq:HT-nlo})--(\ref{eq:HA-nlo}) and use some reference value for the Wilson coefficients. Next we fix the normalization ${\cal L} \;\sigma_{\rm prod}/\Gamma_{\rm tot}$ in such a way to reproduce the $\sim 25\%$ statistical uncertainty that BaBar obtains with an integrated luminosity ${\cal L}_{\rm current} = 0.4242~{\rm ab}^{-1}$~\cite{Lees:2013nxa}. Finally we rescale the normalization by the factor ${\cal L}_{\rm future}/{\cal L}_{\rm current}$ where ${\cal L}_{\rm future} = 50 \; {\rm ab}^{-1}$ is the Belle~II expected final integrated luminosity. 

This procedure produces acceptable error estimates for ${\cal H}_T$, ${\cal H}_L$ and ${\cal H}_A$, while fails for ${\cal H}_3$ and ${\cal H}_4$. The reason is that the integral in eq.~(\ref{oexp}) vanishes when integrated the simple NLO formula given in eq.~(\ref{diffwidth}) against the weights $W_{3,4}$. We bypass this problem by extracting ${\cal O}_{\rm exp}$ from the exact results presented in section~\ref{sec:results} and using eq.~(\ref{errw2}) to calculate the error (in fact the weights $(W_{3,4})^2$ {\it do not} annihilate the NLO differential width). 

Following the discussion summarized in section 5 of ref.~\cite{Hurth:2013ssa}, we add a flat 2\% systematics to the projected statistical errors obtained with the squared weight method and obtain the low-$q^2$ uncertainties collected in table~\ref{table:future-errors}. The expected uncertainty on the high-$q^2$ branching ratio is taken directly from ref.~\cite{Hurth:2013ssa}; in fact, near the end-point of the spectrum our method fails to take into account the improvement in the signal-to-background ratio.

In figures~\ref{fig:Ri-br-future}, \ref{fig:Ri-hthl-future} and \ref{fig:Ri-hah3h4-future} we show the expected impact of Belle~II 
measurements on the various observables we consider in the $[R_9,R_{10}]$ plane. Each contour is drawn at 95\% C.L.\ by combining linearly theoretical and experimental uncertainties. In the scenario we consider the strongest bounds on the Wilson coefficients are driven by measurements of the low-$q^2$ branching ratio and of ${\cal H}_A$ and ${\cal H}_T$ in the two bins. The latter statement is driven by the assumption that the future experimental central values will coincide with the respective SM expectations. If deviations are seen, all observables become crucial to pin down the structure of new physics.


\section{On the connection between theory and experiments}
\label{sec:PHOTOS}

\subsection{Various experimental settings}
Here we discuss how to compare integrated low and high--$q^2$ observables, calculated with the inclusion of QED corrections, to quantities measured by BaBar, Belle
and also by  the future Belle~II experiment. As we explain below, we find that our results can be directly compared to integrated observables measured at BaBar, Belle, and Belle~II with the exception of the di-electron case at BaBar. In the latter case we have to increase our predictions for the integrated branching ratio in the low (high) $q^2$ region by 1.65\% (6.8\%), see eqs.~(\ref{val1}) and (\ref{val2}).

\begin{figure}
\begin{center}
\includegraphics[scale=0.27]{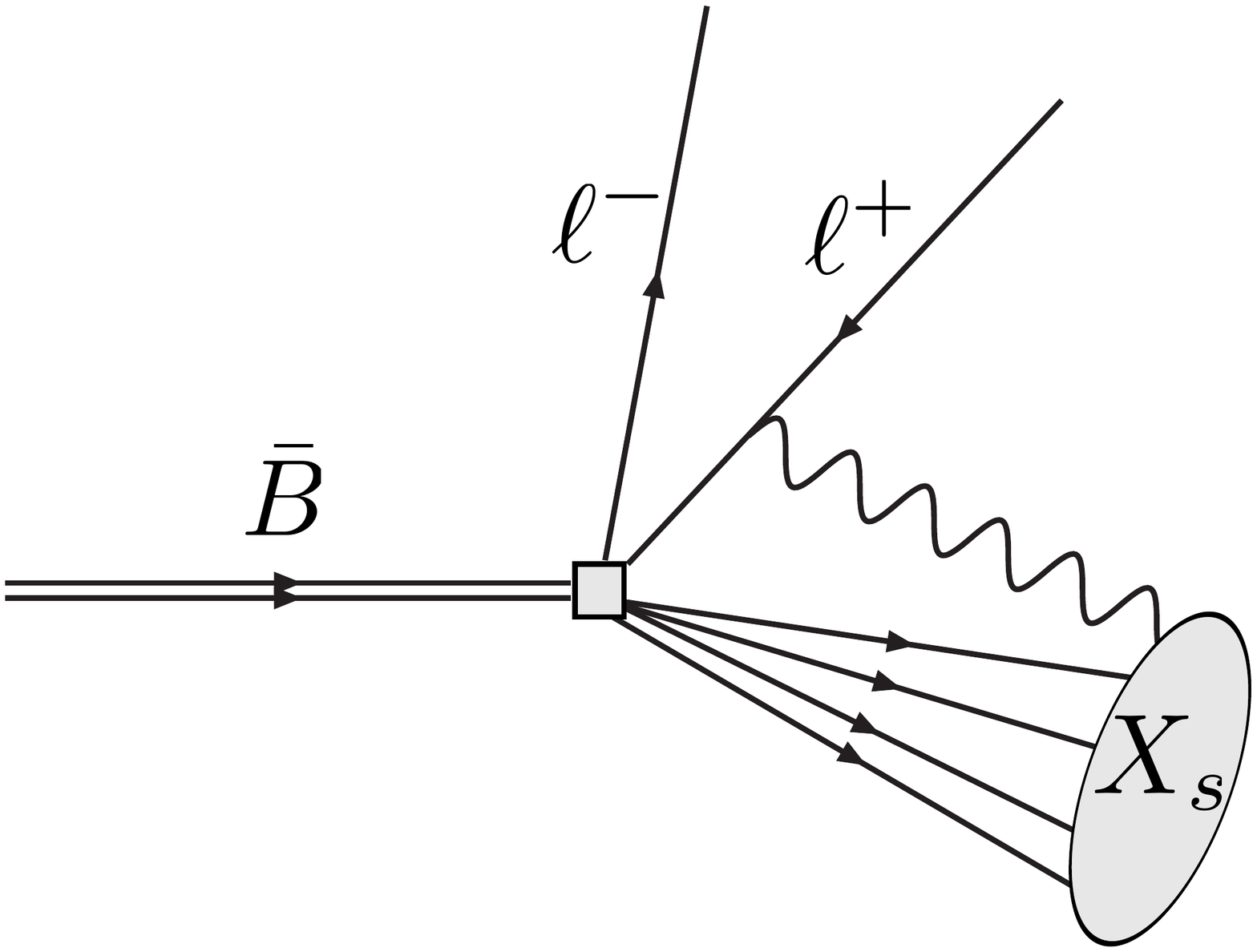}
\quad\quad\quad
\includegraphics[scale=0.33]{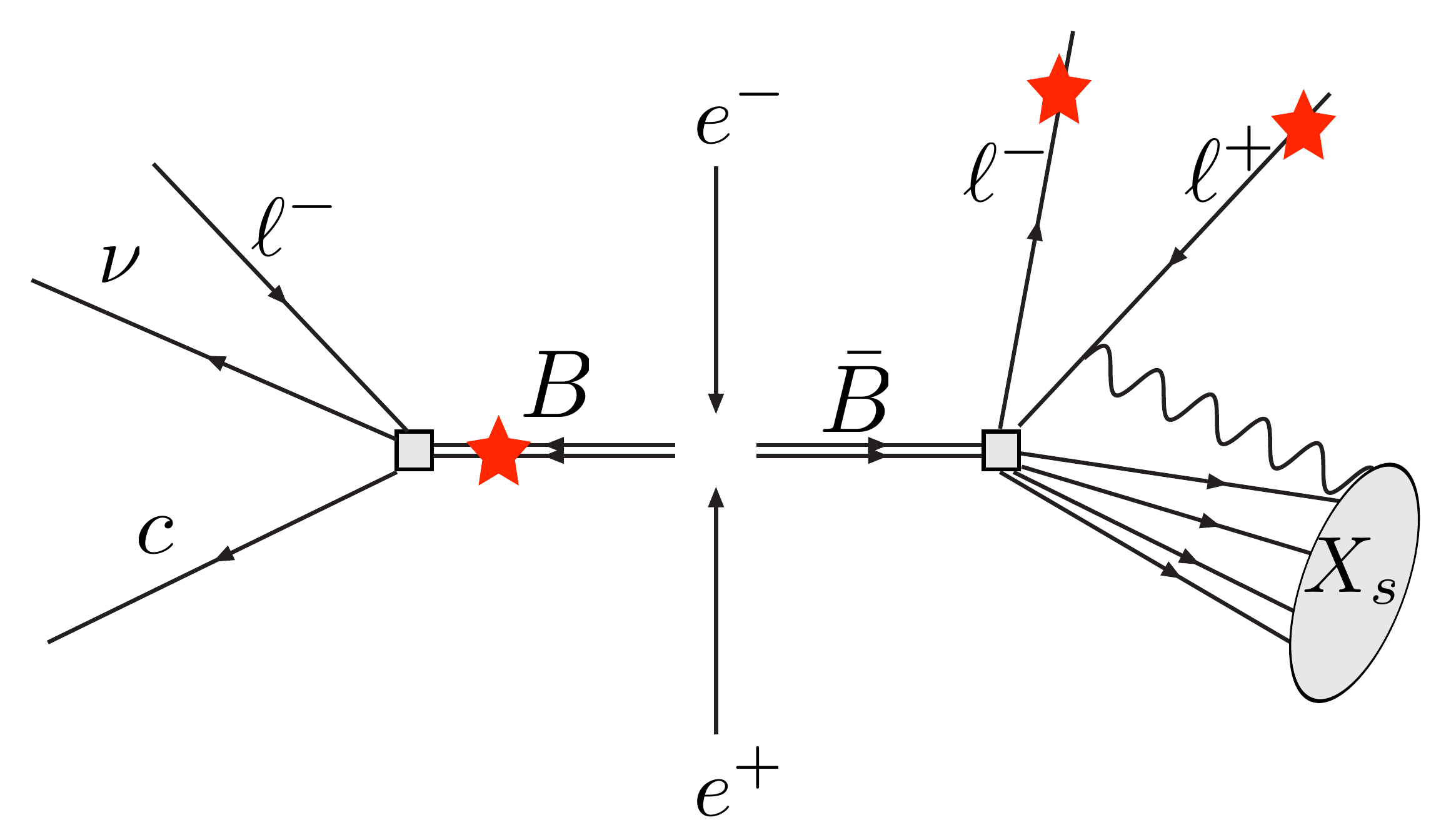}
\caption{Pictorial descriptions of the theoretical definition (left) of $\bar B\to X_s\ell^+\ell^-$ and of the experimental recoil technique. \label{fig:bsll-exp}}
\end{center}
\end{figure}

From the theoretical standpoint the $X_s$ system, in the inclusive $X_s \ell^+\ell^-$ final state, 
contains all the electromagnetic radiation produced in the hard interaction, see the diagram on the left in figure~\ref{fig:bsll-exp}. From the experimental point of view there are two distinct techniques to measure the inclusive $B\to X_s \ell^+\ell^-$ rate: the recoil and sum-over-exclusive methods. In the recoil technique, whose luminosity requirement makes it viable only at super flavor factories, one of the $B$ mesons produced in the $e^+ e^-$ hard interaction is tagged using a semileptonic or hadronic decay and the final state is identified by the two leptons only, see the diagram on the right in figure~\ref{fig:bsll-exp}. In the sum-over-exclusive method, the recoling heavy meson is not looked at and the decaying $B$ is fully reconstructed in final states with a $K^{(*)}$ and up to four pions. The fully inclusive rate is then reconstructed using JETSET~\cite{Sjostrand:1993yb}. 

The comparison between the measured branching ratio (BR) and the results of our inclusive calculations depends critically on the definition of $q^2$. If no photons are included in the definition of the di-lepton invariant mass (i.e.\ $q^2 \equiv (p_{\ell^+} + p_{\ell^-})^2$) our results can be used directly in the comparison with experiments. This is the case for the di-muon channel at both experiments~\cite{privatebelle,privatebabar} and for the di-electron channel at Belle~\cite{privatebelle}.  This will  be exactly the case in a fully inclusive analysis using the recoil technique at Belle~II. However, 
at BaBar photons that belong to a $B\to X_s e^+e^-$ event and that are emitted in a cone of 35 mrad angular opening around either final state electron are included in the calculation of the $q^2$~\cite{privatebabar}. 

In order to calculate the shift that the latter  $q^2$ definition has on to the inclusive theory prediction we generate inclusive $B\to X_s \ell^+\ell^-$ events using EVTGEN~\cite{Lange:2001uf}, hadronize them with JETSET and include electromagnetic radiation with PHOTOS~\cite{Barberio:1990ms,Barberio:1993qi}. Following the BaBar and Belle procedure we build a fully inclusive sample in the whole $q^2$ and $m_{X_s}$ phase space by fully inclusive events (parton level supplemented by a Fermi Motion Model~\cite{Ali:1996bm}) for $m_{X_s} > 1.1 \; {\rm GeV}$ with exclusive $B\to K^{(*)} \ell^+\ell^-$ events (to describe the low $m_{X_s}$ region). Using this large event sample we were able to calculate the impact of including photons emitted in a 35 mrad cone around either electron in the $q^2$ calculation. We find:
\bea
\frac{
\left[{\cal B}_{ee}^{\rm low} \right]_{q = p_{e^+} + p_{e^-} + p_{\gamma_{\rm coll}}}
}{
\left[{\cal B}_{ee}^{\rm low} \right]_{q = p_{e^+} + p_{e^-}}
} -1  &=&  1.65  \%  \label{val1} \\
\frac{
\left[{\cal B}_{ee}^{\rm high} \right]_{q = p_{e^+} + p_{e^-} + p_{\gamma_{\rm coll}}}
}{
\left[{\cal B}_{ee}^{\rm high} \right]_{q = p_{e^+} + p_{e^-}}
} -1 &=&  6.8  \% \; . \label{val2}
\eea
where the suffixes $q = p_{e^+} + p_{e^-}$ and $q = p_{e^+} + p_{e^-} + p_{\gamma_{\rm coll}}$ refer to quantities we calculate and observables measured at BaBar, respectively.

\subsection{Validation}

The results presented in the previous subsection depend crucially on the reliability of using PHOTOS to model photon radiation in $b\to s\ell^+\ell^-$ decays. In this subsection we perform several checks to validate this approach; in particular we show that PHOTOS can be used to reproduce (to a good enough extent) the effects of QED radiation that we calculate analytically. 

\begin{figure}
\begin{center}
\includegraphics[scale=0.375, angle=90]{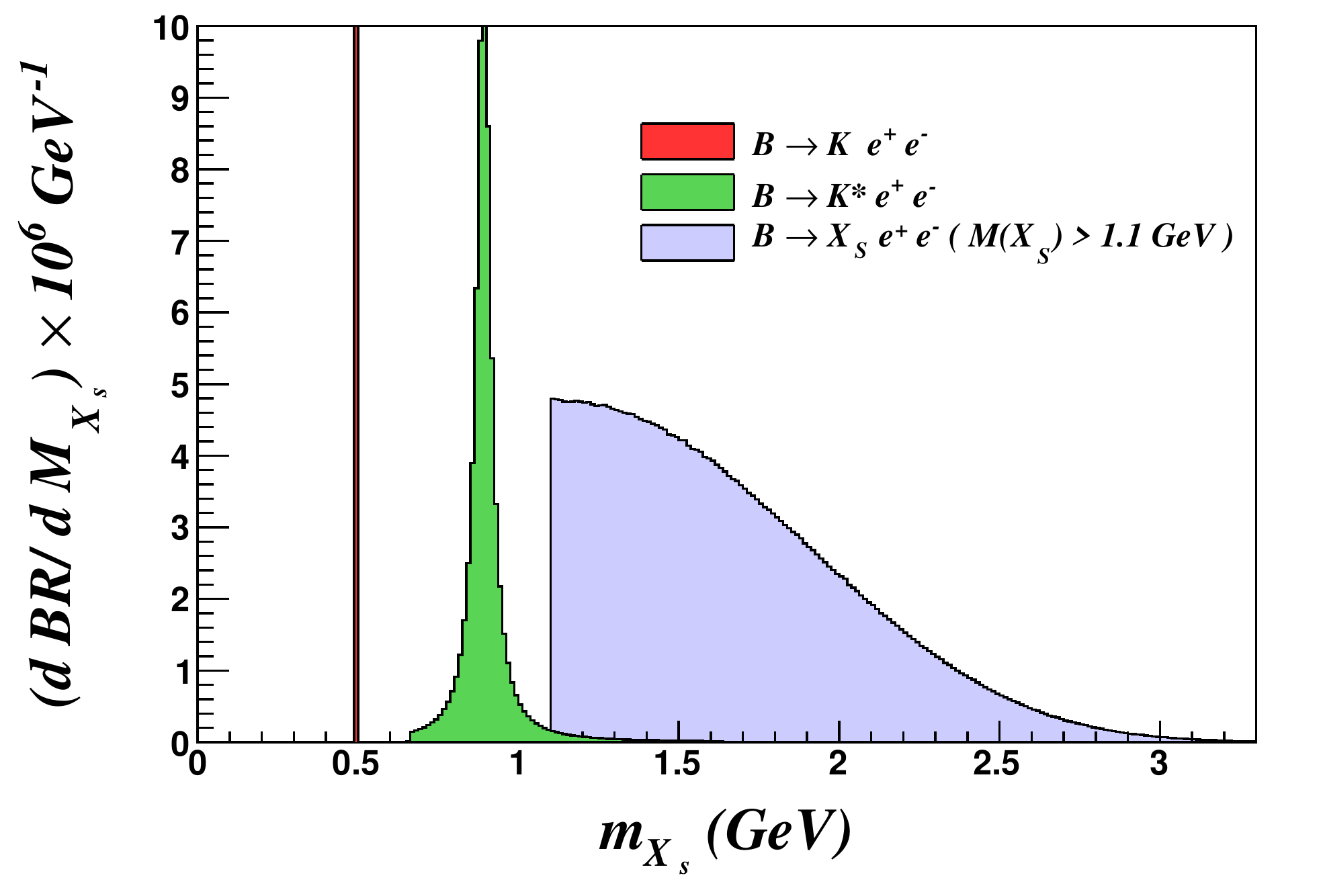}
\hfill
\includegraphics[scale=0.375, angle=90]{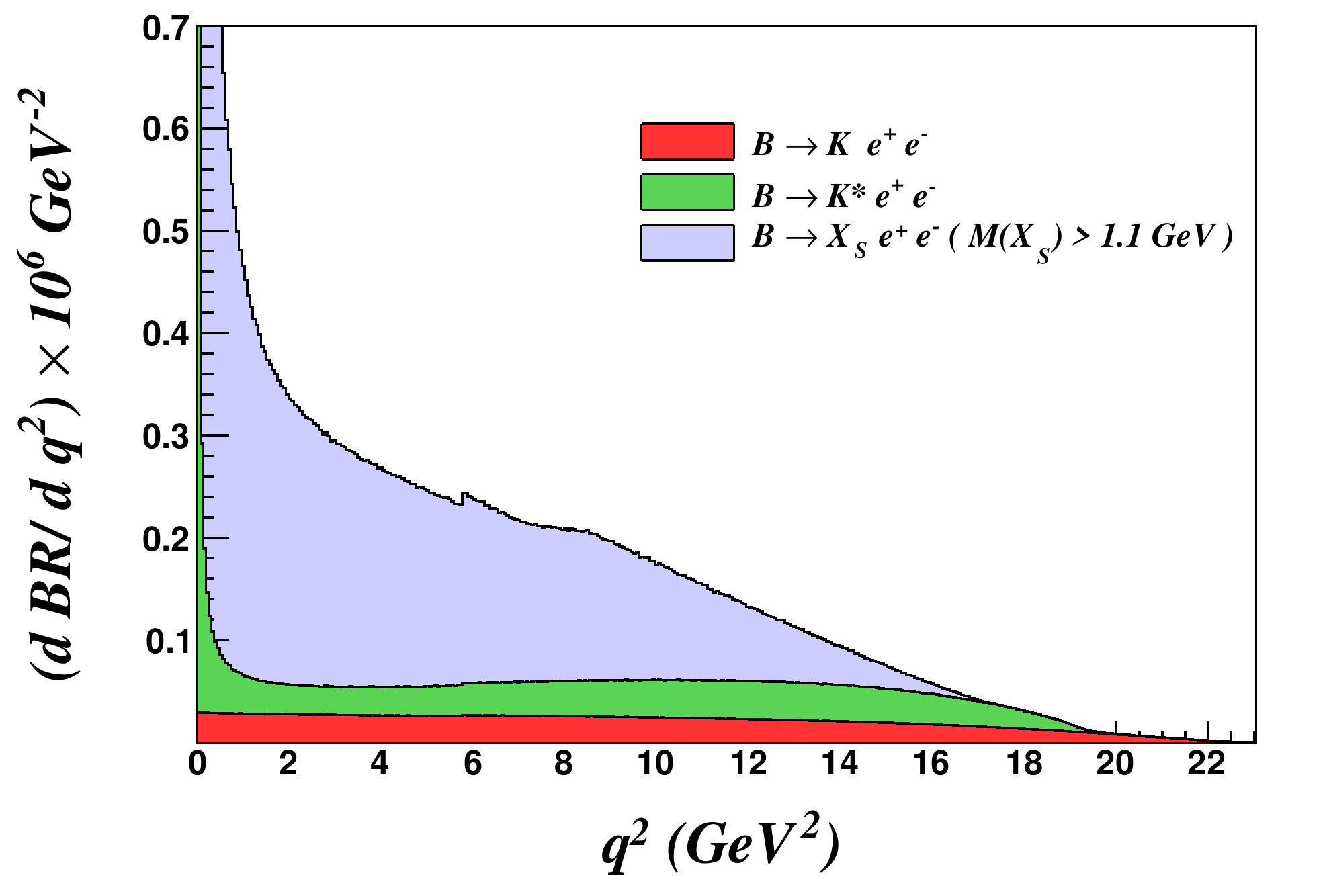}
\caption{$m_{X_s}$ and $q^2$ spectra that we obtain in a $B\to X_s\ell^+\ell^-$ sample generated combining the exclusive $B\to K^{(*)} \ell^+\ell^-$ modes with a pure inclusive calculation for $m_{X_s} > 1.1 \; {\rm GeV}$. \label{fig:bsll-spectra}}
\end{center}
\end{figure}

As discussed above, we generate inclusive $B\to X_s \ell^+\ell^-$ events using EVTGEN, hadronize them with JETSET and include electromagnetic radiation with PHOTOS. In order to obtain a fully inclusive event set we combine $K$, $K^*$ and $X_s (m_{X_s} > 1.1 \; {\rm GeV})$ samples. The $m_{X_s}$ and $q^2$ spectra that we obtain are presented in figure~\ref{fig:bsll-spectra}. The relative weights of the $K$ and $K^*$ samples with respect to the inclusive ($m_{X_s} > 1.1 \; {\rm GeV}$) one have to be provided externally. The actual weights we adopt are extracted from experimental results for the exclusive and inclusive modes and their precise values do not impact much the shape of the $q^2$ spectrum. In fact, as we can see in the plot on the right of figure~\ref{fig:bsll-spectra} only the very high di-lepton invariant mass region, $q^2 > 17 \; {\rm GeV}^2$, is affected. 
\begin{figure}
\begin{center}
\includegraphics[scale=0.375]{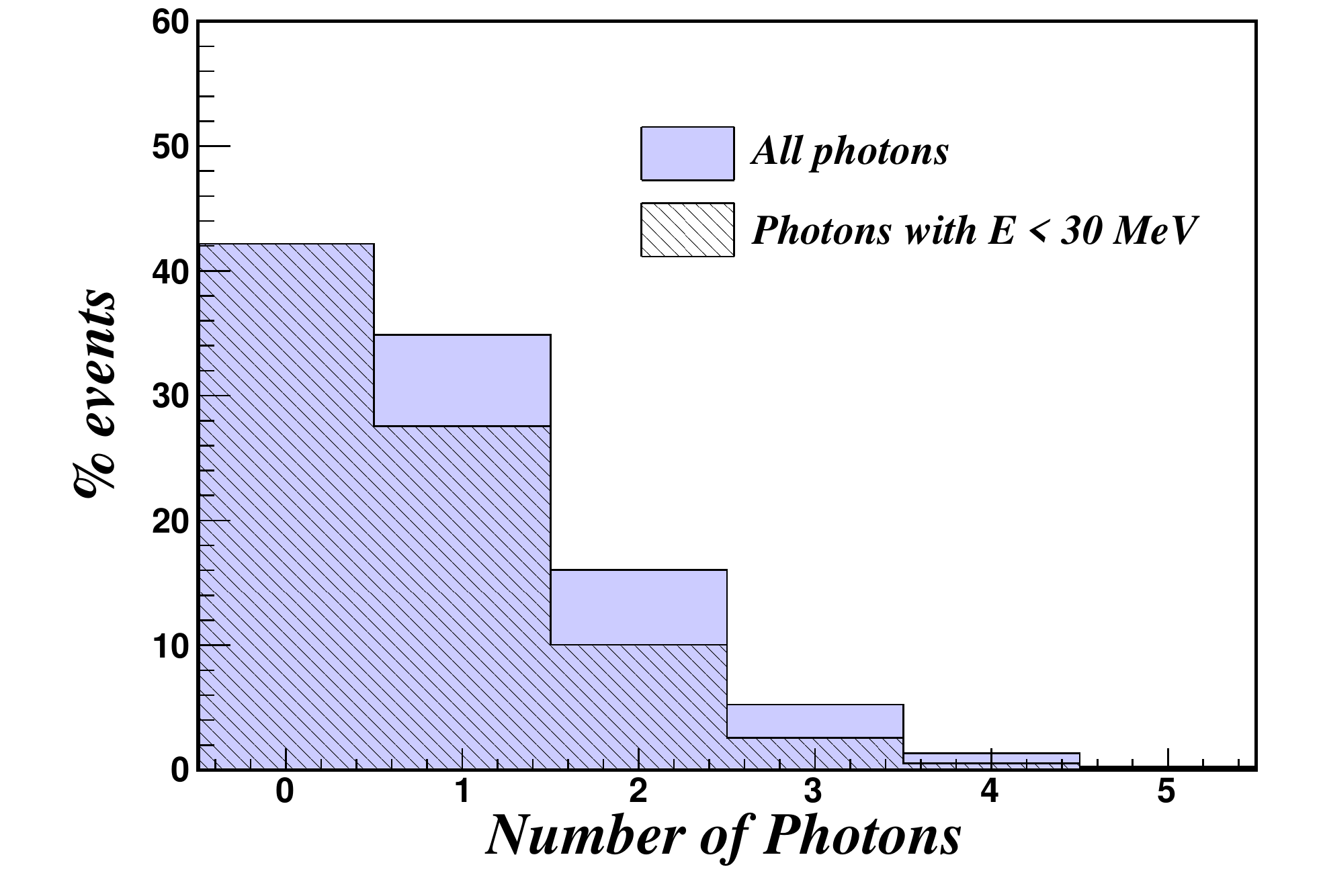}
\hfill
\includegraphics[scale=0.375, angle=0]{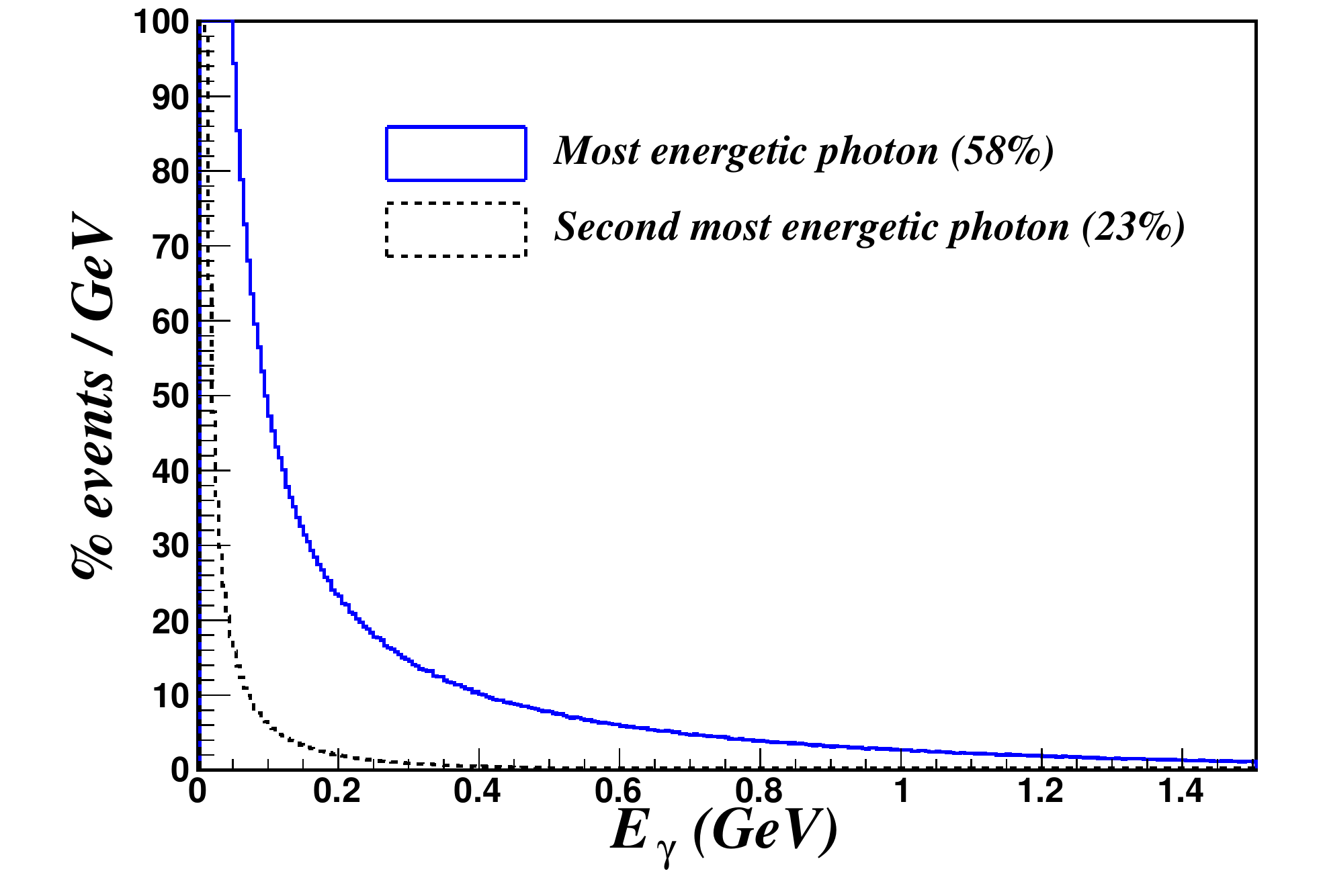}
\caption{Left: Distribution of events with $n_\gamma \leq 5$. Right: distribution of the most energetic and second most energetic photons. \label{fig:photons}}
\end{center}
\end{figure}

A point that is important to mention is that PHOTOS generates events with large photon multiplicity while analytic calculations are confined to a single photon emission. Obviously the vast majority of photons emitted are soft and/or collinear to the final state leptons; moreover, only relatively high energy collinear photons can impact the shape of the $q^2$ spectrum. 

In the left panel of figure~\ref{fig:photons} we show the photon multiplicity we observe in the generated events. The shaded area corresponds to events for which the most energetic photon has $E_\gamma < 30 \; {\rm MeV}$ and that, at the experimental level, are identified as purely hadronic $B\to X_s\ell^+\ell^-$. As expected there is a very large multiplicity of soft photons. We find that only 17\% of all events (this is the integral of the purple unshaded region) correspond to final states with at least one photon with energy larger than 30 MeV. These photons are resolved experimentally and need to be included in the hadronic ($X_s$) or leptonic ($\ell^+\ell^-$) system least the event is rejected (cf.\ also the last paragraph of this subsection).

In the right panel of figure~\ref{fig:photons} we show the distribution of the most and second most energetic photon. The integral of the upper (lower) curve over a photon energy range $[E_{\gamma 1},E_{\gamma 2}]$ yields the percentage of events in which the most (second--most) energetic photon has energy in that interval. The fraction of events with at least one (two) photons is 58\% (23\%), is given by the integral of these curves and can also be easily read off from the left panel of figure~\ref{fig:photons}. Since the impact of including certain collinear photons in the definition of the $q^2$ is more pronounced for more energetic photons, we see that these effects are completely described by a single photon emission: the analytic calculation of QED radiation is, therefore, completely adequate to discuss this phenomenon.

\begin{figure}
\begin{center}
\includegraphics[width=0.99\linewidth]{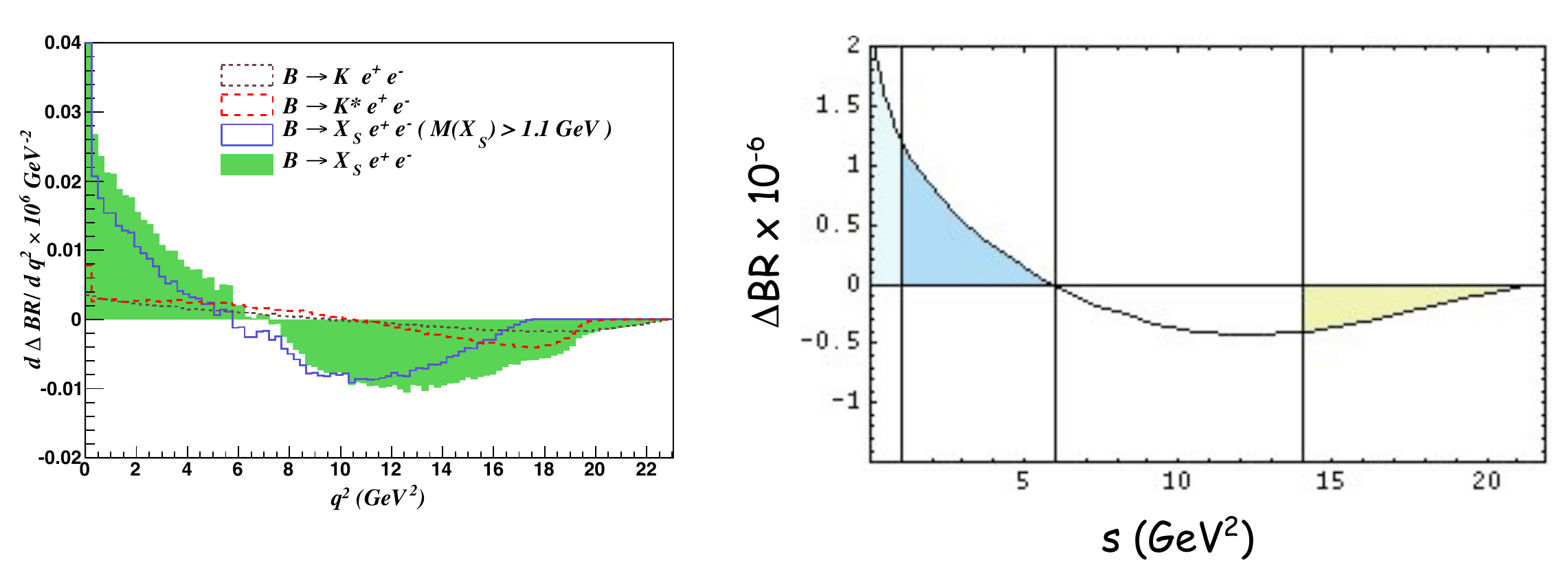}
\caption{Effect of the inclusion of electromagnetic radiation calculated using EVTGEN + PHOTOS (left) and using analytical methods (right). \label{fig:qed-correction}}
\end{center}
\end{figure}

Finally, in order to verify whether PHOTOS correctly models photon radiation in this decay, we need to compare $q^2$ spectra calculated with and without the inclusion of QED radiation. Therefore, we generated a second set of events in which we switched PHOTOS off. The result of this analysis is presented in figure~\ref{fig:qed-correction}. In the left and right panels we show the Monte Carlo study and the result of our analytical calculation, respectively. Numerically, the relative shifts that we obtain for the branching ratio in the low and high--$q^2$ regions are (in round brackets we present the analytical results):
\bea
\delta {\rm BR} (B\to X_s \mu^+\mu^-) &=
\begin{cases}
+1.5\% (+2.0\%) & {\rm low} \; q^2 \cr
-4.4\% (-6.8\%) & {\rm high} \; q^2\cr
\end{cases} \\
\delta {\rm BR} (B\to X_s e^+e^-) &=  
\begin{cases}
+3.6\% (+5.2\%) & {\rm low} \; q^2 \cr
-12.9\% (-17.6\%) & {\rm high} \; q^2\; .\cr
\end{cases} 
\eea
Given the differences in the techniques used, the agreement is remarkable. We conclude that the PHOTOS description of electromagnetic radiation is sufficiently close to the exact calculation to be used to reliably calculate the shifts we presented in eqs.~(\ref{val1}) and (\ref{val2}).

Before concluding this subsection, we would like to stress that validating the use of PHOTOS is important in its own right because experiments use it to estimate the impact of missing photons on their efficiencies. Legitimate $B\to X_s\ell^+\ell^-$ events might be rejected because of two possible reasons. First, if a large number of soft photons ($E_\gamma < 30 \; {\rm MeV}$ and $20 \; {\rm MeV}$ for BaBar and Belle, respectively) is present, they might push the event out of the $m_{ES}$\footnote{Belle names this quantity $m_{bc}$.} and $\Delta E$ acceptance windows (see, for instance, refs.~\cite{Aubert:2004it, Iwasaki:2005sy} for a definition of these kinematical quantities). Second, if a photon with energy larger than $30 \;(20) \; {\rm MeV}$ is not identified, most likely the event is discarded because the total momentum fails to reconstruct a decaying $B$ meson. The latter effect can be quite substantial because, as we discussed above, about 17\% (18\%) of all $B\to X_s \ell^+\ell^-$ events have at least one photon with energy larger than 30 MeV (20 MeV). The fraction of events that is lost to these two mechanisms is taken into account, in the calculation of the efficiencies, using PHOTOS.

\subsection{Monte Carlo estimate of QED corrections to ${\cal H}_T$ and ${\cal H}_L$}
\label{sec:MontecarloHTHL}
\begin{figure}
\begin{center}
\includegraphics[scale=0.37]{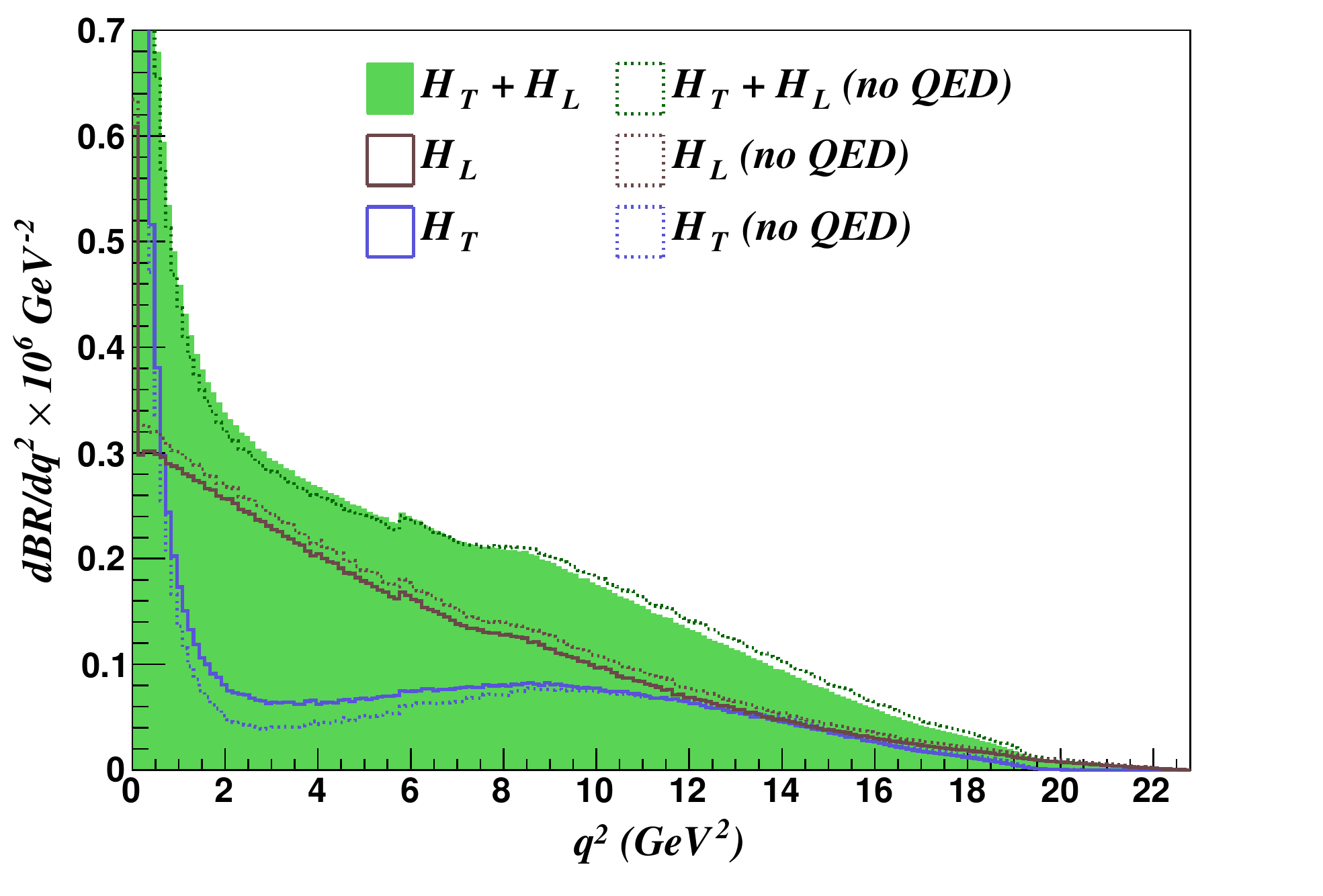}
\includegraphics[scale=0.37]{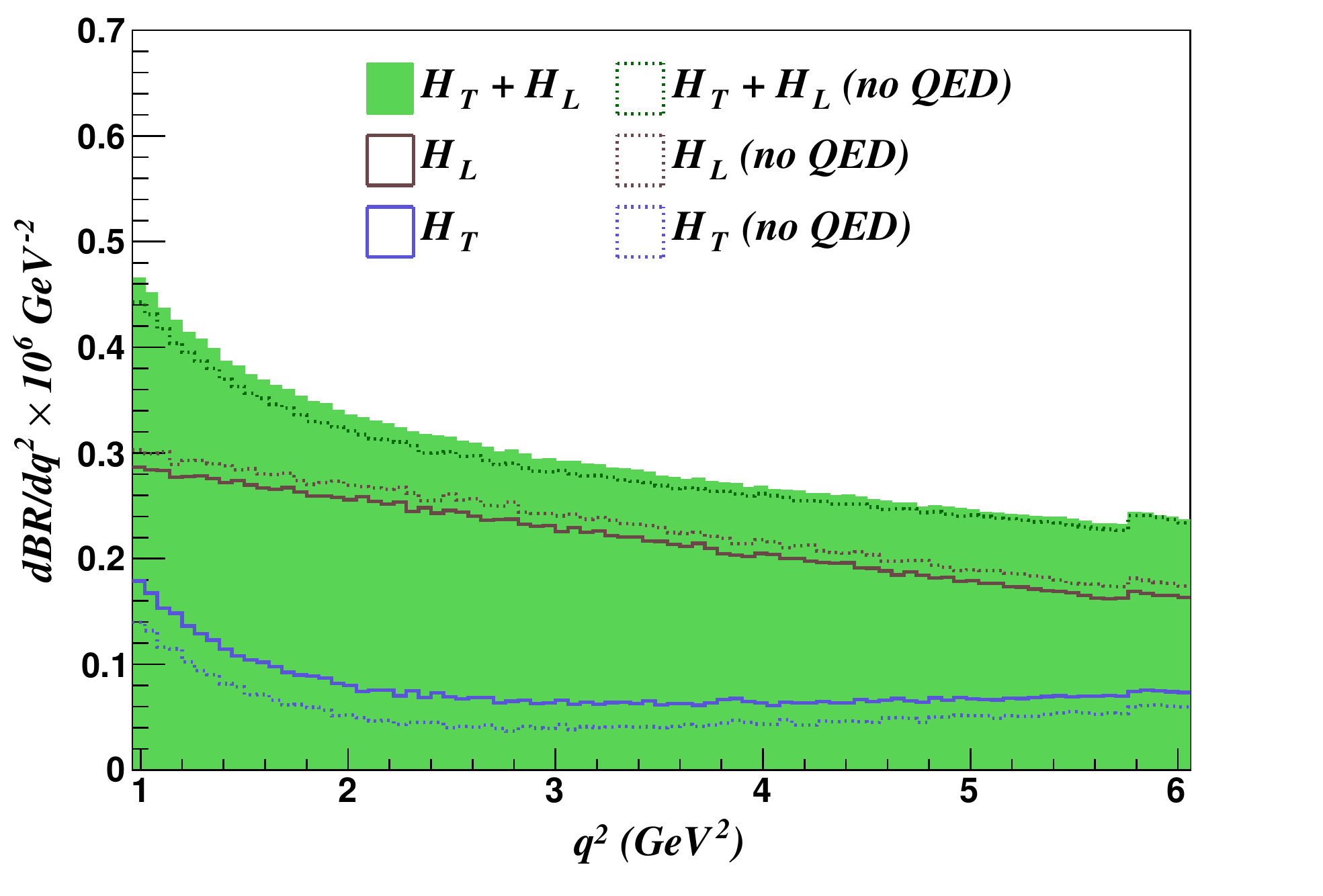}
\caption{$q^2$-dependence of ${\cal H}_T$, ${\cal H}_L$ and branching ratio (${\cal H}_T+{\cal H}_L$) that we extract from a $B\to X_s\ell^+\ell^-$ sample generated combining the exclusive $B\to K^{(*)} \ell^+\ell^-$ modes with a pure inclusive calculation for $m_{X_s} > 1.1 \; {\rm GeV}$. The dotted lines are obtained by switching off QED radiation. \label{fig:bsll-spectrum}}
\end{center}
\end{figure}

The results presented in section~\ref{sec:HT} indicate that the relative size of QED corrections to ${\cal H}_T$ are about an order of magnitude larger than the corresponding corrections to ${\cal H}_L$ and to the branching ratio. In this section we show that this result is actually reproduced in our Monte Carlo study. As a first step we plot in figure~\ref{fig:bsll-spectrum} the $q^2$ spectra for ${\cal H}_T$, ${\cal H}_L$ and the branching ratio with (solid lines) and without (dotted lines) the inclusion of QED radiation. 

\begin{figure}
\begin{center}
\includegraphics[scale=0.45]{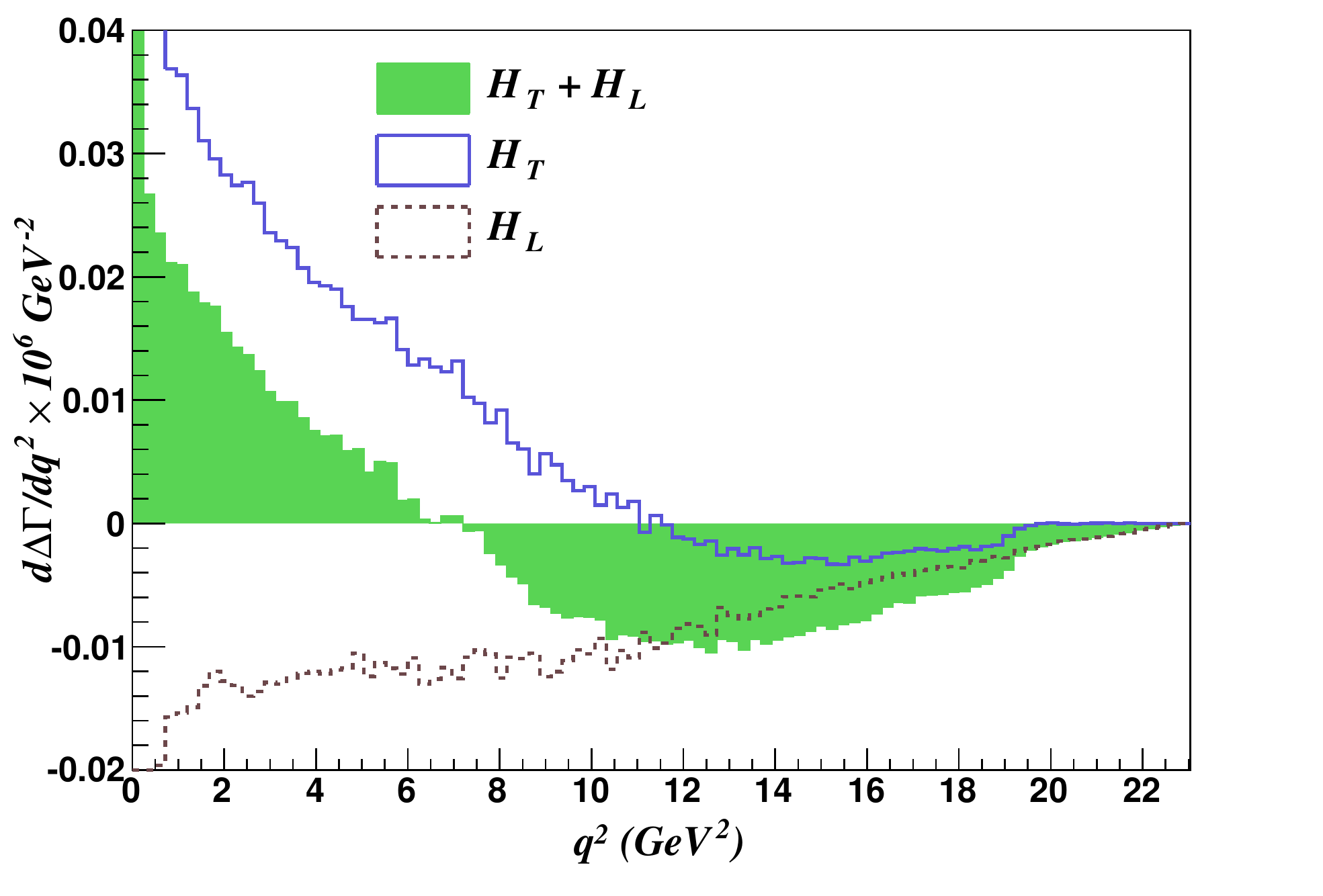}
\caption{Differential $q^2$ distributions of the QED corrections to ${\cal H}_T$, ${\cal H}_L$ and branching ratio (${\cal H}_T+{\cal H}_L$) that we obtain in a $B\to X_s\ell^+\ell^-$ sample generated using EVTGEN and PHOTOS and combining the exclusive $B\to K^{(*)} \ell^+\ell^-$ modes with a pure inclusive calculation for $m_{X_s} > 1.1 \; {\rm GeV}$. \label{fig:bsll-QEDcorrectionTL}}
\end{center}
\end{figure}

Note that the absolute size of QED effects on ${\cal H}_T$, ${\cal H}_L$ and ${\cal H}_T+{\cal H}_L$ is very similar and natural in size; in particular, a small positive net contribution to the integrated branching ratio in the low-$q^2$ region is the sum of a small negative shift on ${\cal H}_L$ and a slightly larger positive shift on ${\cal H}_T$. We plot the actual QED corrections to the three observables in figure~\ref{fig:bsll-QEDcorrectionTL}.


From inspection of the left plot in figure~\ref{fig:bsll-spectrum} we see that, in the low-$q^2$ region ${\cal H}_T$ is much smaller than ${\cal H}_L$. We can understand the origin of this effect by looking at the ratio ${\cal H}_T/{\cal H}_L$ at leading order:
\begin{align}
\frac{{\cal H}_T}{{\cal H}_L} &= 2 \hat s \frac{ C_{10}^2 + \left( C_9 +\frac{2 C_7}{\hat s}\right)^2}{C_{10}^2 + (C_9 + 2 C_7)^2} \; .
\end{align}
The suppression comes from the small $2 \hat s \lesssim 1 $ factor and from the accidental strong cancellation between $C_9$ and $2 C_7/\hat s$ at low $\hat s$ (in fact, the combination $C_9 + 2 C_7/\hat s$ vanishes for $\hat s \sim 0.15$). In the Standard Model $C_7$ is negative; if its sign was reversed we would obtain $C_9 + 2 C_7/\hat s > C_9 + 2 C_7$ and  the integrated ${\cal H}_T$ and ${\cal H}_L$ observables at low-$q^2$ would assume very similar values.

In table~\ref{table:QED} we present the results we obtain by integrating the Monte Carlo generated $b\to s\ell\ell$ histograms. For each bin ($[s_1,s_2]$) and for each  observable $O$ (${\cal H}_T+ {\cal H}_L$, ${\cal H}_T$ and ${\cal H}_L$) we show the total integrated observable ($\int_{s_1}^{s_2} O/\int_1^6 ({\cal H}_T+{\cal H}_L)$), the total integrated QED effect ($\int_{s_1}^{s_2} \Delta O/\int_1^6 ({\cal H}_T+{\cal H}_L)$) and the relative size of the QED correction ($\int_{s_1}^{s_2} \Delta O/\int_{s_1}^{s_2} O$). We see that the absolute size of QED corrections is very similar amongst the three observables (with the effect on ${\cal H}_T$ being only slightly larger) and that the suppression of ${\cal H}_T$ with respect to ${\cal H}_L$ is responsible for very large relative effects in the $30$-$50$\% range. 

Finally we must point out that the numerical estimates presented in table~\ref{table:QED} are affected by sizable uncertainties that are hard to quantify and that only the analytical results presented in table~\ref{table:QEDexact} should be utilized. The Monte Carlo study was nevertheless extremely valuable to build confidence in our study.

\begin{table}
\begin{center}
\begin{tabular}{|c|lll|lll|lll|}
\hline
& \multicolumn{3}{c|}{$q^2 \in [1,6] \; {\rm GeV}^2$} & \multicolumn{3}{c|}{$q^2 \in [1,3.5] \; {\rm GeV}^2$} & \multicolumn{3}{c|}{$q^2 \in [3.5,6] \; {\rm GeV}^2$} \cr
& $\frac{O_{[1,6]}}{{\cal B}_{[1,6]}}$ & $ \frac{\Delta O_{[1,6]}}{{\cal B}_{[1,6]}}$ & $\frac{\Delta O_{[1,6]}}{O_{[1,6]}}$ 
& $\frac{O_{[1,3.5]}}{{\cal B}_{[1,6]}}$ & $ \frac{\Delta O_{[1,3.5]}}{{\cal B}_{[1,6]}}$ & $\frac{\Delta O_{\rm [1,3.5]}}{O_{\rm [1,3.5]}}$
& $\frac{O_{[3.5,6]}}{{\cal B}_{[1,6]}}$ & $ \frac{\Delta O_{[3.5,6]}}{{\cal B}_{[1,6]}}$ & $\frac{\Delta O_{\rm [3.5,6]}}{O_{\rm [3.5,6]}}$ \cr \hline
$\cal B$ & 100 & 3.5 & 3.5 & 56.5 & 2.5 & 4.5 & 43.5 & 1.0 & 2.5\cr
${\cal H}_T$ & 19.0 & 8.0 & 43.0 & 10.0 & 5.0 & 48.5 & 8.5 & 3.0 & 36.0 \cr
${\cal H}_L$ & 81.0 & -4.5 & -5.5 & 46.5 & -2.5 & -5.0 & 35.0 & -2.0 & -6.0\cr
\hline
\end{tabular}
\caption{Relative size of QED effects at low-$q^2$ that we extract from our Monte Carlo $b\to se^+e^-$ sample (All entries are given in percent). For each of the three bins the two columns are the integrated observable and its QED correction normalized to the total low-$q^2$ branching ratio ($\int_{s_1}^{s_2} O/\int_1^6 {\cal B}$ and $\int_{s_1}^{s_2} \Delta O/\int_1^6 {\cal B}$). The third column is the relative size of the QED correction ($\int_{s_1}^{s_2} \Delta O/\int_{s_1}^{s_2} O$). \label{table:QED}}
\end{center}
\end{table}


\section{Conclusion}
\label{sec:conclusions}

The inclusive decay $\bar  B \to X_s \ell^+\ell^-$ is one of the most important modes in the indirect search for new physics via quark flavour observables. It is theoretically 
 clean, while the exclusive mode is affected by unknown power corrections. Thus, besides allowing for a nontrivial check of the recent LHCb data on the exclusive mode, it contains complementary information both in Standard Model predictions and in pinning down new physics. It is therefore a precious channel to be measured at Belle~II, and might be accessible even at LHCb. 
 
In the present article we perform a complete angular analysis of the inclusive decay $\bsll$ by taking into account all perturbative and power corrections that are available to date. We confirm the findings of ref.~\cite{Lee:2006gs} that a separation of the double differential decay width into three observables $H_{T,A,L}(q^2)$, as well as subdivision of the low-$q^2$ region into two bins (see also~\cite{Huber:2007vv}), provides significantly more information than the branching ratio or forward-backward asymmetry in the entire low-$q^2$ region alone.

We compute logarithmically enhanced QED corrections to these observables and find that they do not obey the simple second-order polynomial in $z=\cos(\theta)$ exhibited by the double differential decay width in the absence of QED corrections. We therefore propose to project out $H_{T,A,L}(q^2)$ using weight functions, and argue that the Legendre polynomials $P_n(z)$ are the optimal choice for the latter. Besides reproducing $H_T(q^2)$ and $H_L(q^2)$ in the absence of QED radiation, they allow to construct observables $H_{3,4}(q^2)$ (eq.~(\ref{eq:projectionHI})) that vanish if only QCD corrections are taken into account, and are therefore particular sensitive to QED effects. In view of the benefits of the Legendre weight functions we {\emph{urgently}} recommend the experiments to use the weights (\ref{eq:projectionHI}) to extract single-differential distributions, and to refrain from attempting polynomial fits to the data.

The absolute values of the QED effects that we compute are natural in size. However, due to the phase-space and Wilson coefficient suppression of $H_T(q^2)$ the relative size of the QED corrections is large in this observable. We argue carefully that this does clearly not indicate a breakdown of perturbation theory. On the contrary, we can benefit from the fact that QED corrections lift the smallness of $H_T(q^2)$ to a certain extent, which makes it an observable that is particular sensitive to QED radiation.

To supplement our calculation we carry out a dedicated Monte Carlo study, whose main purpose is three-fold. First, we investigate how the electromagnetic logarithms are treated correctly in the presence of angular and energy cuts. We find that our analytical predictions can be directly applied, with the exception of the electron channel at BaBar, where our numbers have to be modified according to eqs.~(\ref{val1}) and~(\ref{val2}). Second, the size of the QED corrections, in particular their large relative size in $H_T(q^2)$, are confirmed by the Monte Carlo (cf.\ tables~\ref{table:QEDexact} and~\ref{table:QED}). Last but not least, it consitutes also a validation of PHOTOS, which is used by experiments to estimate QED effects in the calculation of efficiencies.

We update the Standard Model predictions for all angular observables integrated over two bins in the low-$q^2$ region. The branching ratio and the observable ${\cal R}(s_0)$ are also evaluated in the high-$q^2$ region. Moreover, we provide our prediction for the zero crossing of the forward-backward asymmetry (or, equivalently, ${\cal H}_A$). The parametric and perturbative uncertainties are in general in the $5-15$\% range, exceptions are ${\cal H}_A[1,6]$ and the high-$q^2$ branching ratio, where the relative errors are much larger. In the former case the reason is the zero crossing of ${\cal H}_A$ which entails a cancellation between the central values of the two bins in the low-$q^2$ region. In the latter case we suffer from poorly known hadronic parameters in the $1/m_b^{2,3}$ power-corrections, a drawback that is circumvented in the ratio ${\cal R}(s_0)$, which normalizes the $\bsll$ rate to the inclusive $\bar B^0 \to X_u \ell\nu$ rate with the same cut in $q^2$~\cite{Ligeti:2007sn}.

We also study the sensitivity of the $\bsll$ decay to new physics in a model-independent way. We give all observables in terms of ratios $R_{7,8,9,10}$ of high-scale Wilson coefficients, which we assume to be altered by the new interactions. We also study correlations between different observables, bins and channels in the $R_9-R_{10}$ plane, and extrapolate to the final Belle~II data set of 50~ab$^{-1}$. We find that ${\cal H}_T$ and ${\cal H}_A$ give the tightest constraints. On the other hand, if deviations from the Standard Model are seen, all observables become crucial to pin down the structure of new physics. 

 In view of the recent measurement by LHCb~\cite{Aaij:2014ora} which reports a value for $R_K = {\rm BR}(B^+ \to K^+ \mu^+ \mu^-) / {\rm BR}(B^+ \to K^+ e^+ e^-)$ in the low-$q^2$ region that is significantly different from unity, one might wonder whether this sign of lepton non-universality could be traced back to logarithmically enhanced QED corrections. LHCb uses the PHOTOS Monte Carlo to eliminate the impact of collinear photon emissions from the final state electrons. Therefore, the corrections calculated in this paper do not seem to apply to the ratio $R_K$. Given that the agreement between PHOTOS and our analytical calculations is not perfect (see e.g.\ tables~\ref{table:QEDexact} and \ref{table:QED}), it would be advisable to correct for photon radiation using data-driven methods that do not rely on PHOTOS.


\section*{Acknowledgments}
We would like to thank Javier Virto for useful discussions, and Kevin Flood, Chris Schilling and Owen Long for logistic and technical support that allowed the Monte Carlo study presented in section~\ref{sec:PHOTOS}. We are indebted to Mathias Brucherseifer, Fabrizio Caola, and Kirill Melnikov for providing us with the data for the two-loop functions $\omega^{(2)}_{ij,I}(\s)$, based on their studies~\cite{Brucherseifer:2013iv,Brucherseifer:2013cu}. T.~Huber acknowledges support from Deutsche Forschungsgemeinschaft within research unit FOR 1873 (QFET). T.~Hurth thanks the CERN theory group for its hospitality during his regular visits to CERN. 
  All authors are grateful to  the Mainz Institute for Theoretical Physics (MITP) for its hospitality at the Capri-Institute in May 2014,  where part of this work was done.


\appendix
\section{QED and QCD functions} \label{app:functions}

\subsection{QED functions for the double differential rate}

Here we list the polynomials that appear in the functions $\xi^{\rm{(em)}}_{ij}(s,z)$ of the log-enhanced QED corrections to the double differential rate in eq.~(\ref{eq:QEDdoublediff}).

\allowdisplaybreaks{
\bea
p_1(s,z) &=& 2 s^2 \left(z^4+6 z^2+1\right)+s \left(11 z^4-8 z^2-3\right)+\left(z^2-1\right)^2 \; , \nnb \\
p_2(s,z) &=& 4 s^3\left(z^2+1\right)+3 s^2 \left(z^2-1\right)-4 s \left(z^2-1\right)-9 z^2-7 \; , \nnb \\
p_3(s,z) &=& s^3 \left(z^2-1\right)^3+s^2 \left(z^2-1\right)^2 \left(19 z^2+5\right)\nnb \\
         && +s \left(6 z^6+37 z^4-36 z^2-7\right) + 5 z^4+24 z^2+3 \; , \nnb \\
p_4(s,z) &=& s^3 \left(z^8-4z^6+2 z^4-28 z^2-3\right)-3 s^2 \left(z^8-4 z^6+8 z^4-4 z^2-1\right) \nnb \\
         && +4 s \left(z^6-5 z^4+3 z^2+1\right)-2 \left(5 z^4+24 z^2+3\right)\; , \nnb \\
p_5(s,z) &=& s^4 \left(13 z^8-56 z^6+210 z^4-112 z^2-55\right) \nnb \\
         && +s^3 \left(-15 z^8+31 z^6-127 z^4+149 z^2+154\right) \nnb \\
	 && +3 s^2 \left(5 z^8-9 z^6+55 z^4-31 z^2-84\right) \nnb \\
	 && +s\left(-13 z^8+65 z^6-285 z^4+355 z^2+262\right) \nnb \\
	 && -13 z^6+37 z^4-299 z^2-109 \; , \nnb \\
p_6(s,z) &=& s \left(z^2-1\right)-z^2-1 \; , \nnb \\
p_7(s,z) &=& s^2 \left(43 z^4+106 z^2+27\right)+24 s \left(2z^4-z^2-1\right)+3 \left(z^2-1\right)^2 \; , \nnb \\
p_8(s,z) &=& s^2 \left(-z^{10}+3 z^8+32 z^6+364 z^4+289 z^2+17\right) \nnb \\
         &&  +s \left(3 z^{10}-19 z^8+106 z^6+102 z^4-173 z^2-19\right) \nnb \\
	 &&  +2 \left(-z^8+7 z^6-9 z^4+z^2+2\right)\; , \nnb \\
p_9(s,z) &=& 2 s^4 \left(17 z^6+183 z^4+143 z^2+9\right) +s^3 \left(77z^8+922 z^6-92 z^4-842 z^2-65\right) \nnb \\
	 && +s^2 \left(z^2-1\right)^2 \left(46z^6+889 z^4+1030 z^2+87\right) \nnb \\
	 && +s \left(z^2-1\right)^3 \left(256z^4+483 z^2+51\right) +\left(z^2-1\right)^4 \left(74z^2+11\right) \; , \nnb \\
p_{10}(s,z) &=&-s^5 \left(13z^8-66 z^6+1288 z^4+2706 z^2+283\right) \nnb \\
            && +s^4 \left(-26z^{10}+173 z^8-2504 z^6-2098 z^4+7690 z^2+989\right) \nnb \\
	    && +s^3 \left(-13z^{12}+122 z^{10}-1190 z^8+830 z^6+8809 z^4-7288 z^2-1270\right) \nnb \\
	    && +s^2 \left(z^2-1\right)^2 \left(15 z^8-18 z^6+397 z^4+3716 z^2+706\right) \nnb \\
	    && -s\left(z^2-1\right)^3 \left(15 z^6+19 z^4-403z^2-143\right) +\left(z^2-1\right)^4 \left(13 z^4-22 z^2+1\right) \; , \nnb \\
p_{11}(s,z) &=& s^2 \left(5 z^2+3\right)+z^2-1 \; , \nnb \\
p_{12}(s,z) &=& s^2\left(z^6-6 z^4-9 z^2-2\right)-s \left(z^2-1\right)^3+z^4-1 \; , \nnb \\
p_{13}(s,z) &=& s^3 \left(z^4+22z^2+9\right)+s^2 \left(z^6+11 z^4-33 z^2-11\right) \nnb \\
            && -s \left(2 z^6+17 z^4-24 z^2+5\right)+\left(z^2-1\right)^2\left(z^2+7\right)\; , \nnb \\
p_{14}(s,z) &=& s^3 \left(3 z^4+12z^2+1\right)+s^2 \left(4 z^6+15 z^4-18 z^2-1\right) \nnb \\
            && +s\left(z^2-1\right)^2 \left(7 z^2-1\right)-\left(z^2-1\right)^3\; , \nnb \\
p_{15}(s,z) &=& s\left(z^2-1\right)+z^2+1\; , \nnb \\
p_{16}(s,z) &=& s^3 \left(5 z^4+24 z^2+3\right)+s^2 \left(6 z^6+37 z^4-36 z^2-7\right) \nnb \\
            &&+s\left(z^2-1\right)^2 \left(19 z^2+5\right)+\left(z^2-1\right)^3\; , \nnb \\
p_{17}(s,z) &=& s^2 \left(z^6-3 z^4+39 z^2+27\right) -2 s \left(z^6-3 z^4-5 z^2+7\right)+z^6-3 z^4+7 z^2-5\; , \nnb \\
p_{18}(s,z) &=& s^2 \left(3 z^4-18 z^2-49\right)-2 s^{3/2} \left(z^4-10z^2+9\right) +3 s \left(z^6-6 z^4-11z^2+16\right) \nnb \\
            &&-2 \sqrt{s} \left(z^2-5\right) \left(z^2-1\right)^2+\left(z^2-1\right)^2 \left(3z^2-7\right)\; , \nnb \\
p_{19}(s,z) &=& s^2 \left(37 z^4+86 z^2+21\right)+16 s \left(2 z^4-z^2-1\right)+\left(z^2-1\right)^2\; , \nnb \\
p_{20}(s,z) &=& s^2 \left(z^8-4 z^6+154z^4+340 z^2+85\right)-2 s \left(z^8-4 z^6-58 z^4+28z^2+33\right) \nnb \\
            && +\left(z^2-1\right)^2 \left(z^4-2 z^2+5\right)\; , \nnb \\
p_{21}(s,z) &=& s^3 \left(3 z^6-37 z^4-359 z^2-183\right)-2 s^{5/2} \left(z^6-35 z^4-5z^2+39\right) \nnb \\
	    && +s^2 \left(6z^8-77 z^6-613 z^4+305 z^2+379\right)-4 s^{3/2} \left(z^2-1\right)^2\left(z^4-27 z^2-34\right) \nnb \\	       
	    && +s \left(z^2-1\right)^2 \left(3 z^6-31z^4-323 z^2-229\right)-2 \sqrt{s} \left(z^2-1\right)^3\left(z^4-14 z^2-29\right) \nnb \\
	    && +\left(z^2-1\right)^3 \left(3 z^4-10z^2-33\right) \; , \nnb \\
p_{22}(s,z) &=& 2 s^4 \left(15 z^6+153 z^4+113 z^2+7\right)
               +s^3 \left(69z^8+754 z^6-132 z^4-642 z^2-49\right) \nnb \\
	    && +s^2 \left(z^2-1\right)^2\left(42 z^6+717 z^4+742 z^2+63\right)
	       +5 s \left(z^2-1\right)^3\left(40 z^4+63 z^2+7\right) \nnb \\
	    && +\left(z^2-1\right)^4 \left(38z^2+7\right)\; .
\eea
}

\subsection{Functions for the QCD corrections to the ${\cal H}_I$}

The one-loop QCD functions~\cite{Asatrian:2002va,Lee:2006gs} can be computed analytically,
\begin{align}
\dps\omega^{(1)}_{77,T}(\s) & = -\frac{8}{3} \log \left(\frac{\mu_b}{m_b}\right)-\frac{\left(\sqrt{\s}+1\right)^2
   \left(\s^{3/2}-10 \s+13 \sqrt{\s}-8\right) \text{Li}_2(1-\s)}{6 (\s-1)^2}\nnb \\
   &+\frac{2 \sqrt{\s} \left(\s^2-6
   \s-3\right) \text{Li}_2\left(1-\sqrt{\s}\right)}{3 (\s-1)^2}-\frac{\pi ^2 \left(3 \s^{3/2}+22 \s+23
   \sqrt{\s}+16\right) \left(\sqrt{\s}-1\right)^2}{36 (\s-1)^2}\nnb \\
   &+\frac{5 \s^3-54 \s^2+57 \s-8}{18 (\s-1)^2}-\log
   (1-\s)+\frac{\s (5 \s+1) \log (\s)}{3 (\s-1)^2}+\frac{2}{3} \log (1-\s) \log(\s) \, , \nnb \\
&\nnb \\
\dps\omega^{(1)}_{79,T}(\s) & = \dps -\frac{4}{3} \log \left(\frac{\mu_b}{m_b}\right)-\frac{2 \sqrt{\s} (\s+3)\text{Li}_2\left(1-\sqrt{\s}\right)}{3 (\s-1)^2}-\frac{\pi ^2 \left(16 \s+29
   \sqrt{\s}+19\right) \left(\sqrt{\s}-1\right)^2}{36 (\s-1)^2}\nnb \\
   &+\frac{\s^2-6 \s+5}{6 (\s-1)^2}+\frac{\left(\sqrt{\s}+1\right)^2 \left(8 \s-15
   \sqrt{\s}+9\right) \text{Li}_2(1-\s)}{6 (\s-1)^2}\nnb \\
   &-\frac{(5 \s+1) \log (1-\s)}{6 \s}+\frac{\s (3
   \s+1) \log (\s)}{6 (\s-1)^2}+\frac{2}{3} \log(1-\s) \log(\s) \, , \nnb \\
&\nnb \\
\dps\omega^{(1)}_{99,T}(\s) & = \dps \frac{\left(\sqrt{\s}+1\right)^2 \left(8 \s^{3/2}-15 \s+4 \sqrt{\s}-5\right) \text{Li}_2(1-\s)}{6 (\s-1)^2
   \sqrt{\s}}-\frac{2 \left(\s^2-12 \s-5\right) \text{Li}_2\left(1-\sqrt{\s}\right)}{3 (\s-1)^2
   \sqrt{\s}} \nnb \\
   &-\frac{\pi ^2 \left(16 \s^{3/2}+29 \s+4 \sqrt{\s}+15\right) \left(\sqrt{\s}-1\right)^2}{36
   (\s-1)^2 \sqrt{\s}}+\frac{\left(2 \s^2-7 \s-5\right) \log (\s)}{3
   (\s-1)^2}\nnb \\
   &+\frac{\s^2+18 \s-19}{6 (\s-1)^2}-\frac{(2 \s+1) \log (1-\s)}{3 \s}+\frac{2}{3} \log(1-\s)\log(\s)\, ,\nnb \\
&\nnb \\
\dps\omega^{(1)}_{710,A}(\s) & = -\frac{4}{3} \log \left(\frac{\mu_b}{m_b}\right)+\frac{2 \left(4 \s^2-13 \s-1\right)
   \text{Li}_2\left(1-\sqrt{\s}\right)}{3 (\s-1)^2}-\frac{\left(2 \s^2-9 \s-3\right) \text{Li}_2(1-\s)}{3
   (\s-1)^2}\nnb \\
   &-\frac{\left(3 \s^2-16 \s+13\right) \log \left(1-\sqrt{\s}\right)}{3 (\s-1)^2}+\frac{\left(4
   \s^2-13 \s-1\right) \log \left(1-\sqrt{\s}\right) \log (\s)}{3 (\s-1)^2}\nnb\\
   &-\frac{\left(2 \s^2-9 \s-3\right)
   \log (1-\s) \log (\s)}{3 (\s-1)^2}+\frac{\left(\s^3-23 \s^2+23 \s-1\right) \log (1-\s)}{6 (\s-1)^2
   \s}\nnb\\
   &+\frac{\left(\s-20 \sqrt{\s}+5\right) \left(\sqrt{\s}-1\right)^2}{6 (\s-1)^2}-\frac{\pi ^2}{3}\, , \nnb \\
&\nnb \\
\dps\omega^{(1)}_{910,A}(\s) & = -\frac{2 \left(\s^2-3 \s-1\right) \text{Li}_2(1-\s)}{3 (\s-1)^2}-\frac{4 (5-2 \s) \s \text{Li}_2\left(1-\sqrt{\s}\right)}{3 (\s-1)^2}-\frac{\left(4
   \sqrt{\s}-3\right) \left(\sqrt{\s}-1\right)^2}{3 (\s-1)^2}\nnb \\
&-\frac{2 \left(2 \s^2-7 \s+5\right) \log
   \left(1-\sqrt{\s}\right)}{3 (\s-1)^2}-\frac{2 \left(\s^2-3 \s-1\right) \log (1-\s) \log (\s)}{3
   (\s-1)^2}\nnb \\
   &+\frac{\left(2 \s^3-11 \s^2+10 \s-1\right) \log (1-\s)}{3 (\s-1)^2 \s}+\frac{2 \s (2 \s-5) \log \left(1-\sqrt{\s}\right)
   \log (\s)}{3 (\s-1)^2}-\frac{\pi ^2}{3} \, , \nnb \\
&\nnb \\
\dps\omega^{(1)}_{77,L}(\s) & = -\frac{8}{3} \log \left(\frac{\mu_b}{m_b}\right)+\frac{\left(\sqrt{\s}+1\right)^2 \left(4\s^{3/2}-7 \s+2 \sqrt{\s}-3\right) \text{Li}_2(1-\s)}{3 (\s-1)^2 \sqrt{\s}}-\frac{9 \s^2-38 \s+29}{6 (\s-1)^2} \nnb \\
&-\frac{4 \left(\s^2-6 \s-3\right)\text{Li}_2\left(1-\sqrt{\s}\right)}{3 (\s-1)^2 \sqrt{\s}}-\frac{\pi ^2 \left(8 \s^{3/2}+13 \s+2 \sqrt{\s}+9\right) \left(\sqrt{\s}-1\right)^2}{18 (\s-1)^2 \sqrt{\s}}\nnb \\
&-\frac{\left(\s^3-3 \s+2\right) \log
   (1-\s)}{3 (\s-1)^2 \s}+\frac{2 \left(\s^2-3 \s-3\right) \log(\s)}{3
   (\s-1)^2}+\frac{2}{3} \log (1-\s) \log (\s)\, , \nnb \\
&\nnb \\
\dps\omega^{(1)}_{79,L}(\s) & = -\frac{4}{3} \log \left(\frac{\mu_b}{m_b}\right)+\frac{4 \sqrt{\s} (\s+3)
   \text{Li}_2\left(1-\sqrt{\s}\right)}{3 (\s-1)^2}+\frac{\left(\sqrt{\s}+1\right)^2 \left(4 \s-9
   \sqrt{\s}+3\right) \text{Li}_2(1-\s)}{3 (\s-1)^2}\nnb \\
   &+\frac{7 \s^2-2 \s-5}{6 (\s-1)^2}-\frac{\pi ^2 \left(8 \s+19
   \sqrt{\s}+5\right) \left(\sqrt{\s}-1\right)^2}{18 (\s-1)^2}-\frac{(2 \s+1) \log (1-\s)}{3 \s}\nnb \\
   &+\frac{(\s-7) \s \log (\s)}{3 (\s-1)^2}+\frac{2}{3} \log (1-\s) \log(\s) \, , \nnb \\
&\nnb \\
\dps\omega^{(1)}_{99,L}(\s) & = -\frac{\left(\sqrt{\s}+1\right)^2 \left(\s^{3/2}-8 \s+3 \sqrt{\s}-4\right) \text{Li}_2(1-\s)}{3
   (\s-1)^2}+\frac{4 \sqrt{\s} \left(\s^2-12 \s-5\right) \text{Li}_2\left(1-\sqrt{\s}\right)}{3
   (\s-1)^2}\nnb \\
   &-\frac{\pi ^2 \left(3 \s^{3/2}+20 \s+\sqrt{\s}+8\right) \left(\sqrt{\s}-1\right)^2}{18
   (\s-1)^2}+\frac{4 \s^3-51 \s^2+42 \s+5}{6 (\s-1)^2}-\log (1-\s)\nnb \\
   &+\frac{8 \s (2 \s+1) \log (\s)}{3(\s-1)^2}+\frac{2}{3} \log (1-\s) \log (\s)\, .
\end{align}
The two-loop QCD functions~\cite{Brucherseifer:2013iv,Brucherseifer:2013cu} are obtained from least-squares fits and are also valid for all $q^2$. The necessary data was kindly provided by the authors of~\cite{Brucherseifer:2013iv,Brucherseifer:2013cu}.
\begin{align}
\dps\omega^{(2)}_{99,T}(\s) & = \beta_0^{(5)} \log\left(\frac{\mu_b}{m_b}\right) \, \omega^{(1)}_{99,T}(\s) +
54.919 (1-\s)^4-136.374 (1-\s)^3\nnb \\
& +119.344 (1-\s)^2-15.6175 (1-\s)-31.1706 \, ,\nnb \\
&\nnb \\
\dps\omega^{(2)}_{910,A}(\s) & =\beta_0^{(5)} \log\left(\frac{\mu_b}{m_b}\right) \, \omega^{(1)}_{910,A}(\s) +
74.3717 (1-\s)^4-183.885 (1-\s)^3\nnb \\
&+158.739 (1-\s)^2-29.0124 (1-\s)-30.8056 \, ,\nnb \\
&\nnb \\
\dps\omega^{(2)}_{99,L}(\s) & =\beta_0^{(5)} \log\left(\frac{\mu_b}{m_b}\right) \, \omega^{(1)}_{99,L}(\s) 
-5.95974 (1-s)^3+11.7493 (1-s)^2\nnb \\
&+12.2293 (1-s)-38.6457 \, .
\end{align}
They are given for $n_h=2$ and $n_l=3$. $\beta_0^{(5)} = 23/3$ denotes the one-loop QCD $\beta$-function for five active flavours.

\subsection{Functions for the QED corrections to the ${\cal H}_I$}

The following functions are again obtained by least-squares fits. They are valid in the low-$q^2$ region ($1 \; {\rm GeV}^2 < q^2 < 6 \; {\rm GeV}^2$) only.

\allowdisplaybreaks{
\begin{align}
\dps \omega^{(\rm{em})}_{77,T}(\s)&= \dps \ln\left(\frac{m_b^2}{m_\ell^2}\right) \, \frac{1.54986-1703.72 \, \s^5+1653.38 \, \s^4-683.608 \, \s^3+179.279 \, \s^2-35.5047 \, \s}{8 (1-\s)^2} \, , \nnb \\
&\nnb\\
\dps \omega^{(\rm{em})}_{77,L}(\s)&= \dps \ln\left(\frac{m_b^2}{m_\ell^2}\right) \, \frac{9.73761 +647.747 \, \s^4-642.637 \, \s^3+276.839 \, \s^2-68.3562 \, \s-\frac{1.6755}{\s}}{4 (1-\s)^2} \, , \nnb \\
&\nnb\\
\dps \omega^{(\rm{em})}_{99,T}(\s)&= \dps \ln\left(\frac{m_b^2}{m_\ell^2}\right) \, \frac{2.2596 +157.984 \, \s^4-141.281 \, \s^3+52.8914 \, \s^2-13.5377 \, \s+\frac{0.0284049}{\s}}{2\s (1-\s)^2} \, , \nnb \\
&\nnb\\
\dps \omega^{(\rm{em})}_{99,L}(\s)&= \dps \ln\left(\frac{m_b^2}{m_\ell^2}\right) \, \frac{-0.768521 -80.8068 \, \s^4+70.0821 \, \s^3-21.2787 \, \s^2+2.9335\, \s-\frac{0.0180809}{\s}}{(1-\s)^2} \, , \nnb \\
&\nnb\\
\dps \omega^{(\rm{em})}_{79,T}(\s)&= \dps \ln\left(\frac{m_b^2}{m_\ell^2}\right) \, \frac{19.063 +2158.03 \s^4-2062.92 \s^3+830.53 \s^2-186.12 \s+\frac{0.324236}{\s}}{8 (1-\s)^2} \, , \nnb \\
&\nnb\\
\dps \omega^{(\rm{em})}_{79,L}(\s)&= \dps \ln\left(\frac{m_b^2}{m_\ell^2}\right) \, \frac{-6.03641 -896.643 \s^4+807.349 \s^3-278.559 \s^2+47.6636 \s-\frac{0.190701}{\s}}{4 (1-\s)^2} \, , \nnb \\
&\nnb\\
\dps \omega^{(\rm{em})}_{27,T}(\s)&= \dps \ln\left(\frac{m_b^2}{m_\ell^2}\right) \, \left[ \frac{21.5291 +3044.94 \s^4-2563.05 \s^3+874.074 \s^2-175.874 \s+\frac{0.121398}{\s}}{8 (1-\s)^2}\right. \nnb \\
&\dps + \left. i \, \frac{2.49475 +598.376 \s^4-456.831 \s^3+117.683 \s^2-9.90525 \s-\frac{0.0116501}{\s}}{8 (1-\s)^2} \right] \nnb \\
& \dps + \frac{8}{9} \, \omega^{(\rm{em})}_{79,T}(\s) \, \ln\left(\frac{\mu_b}{5\rm{GeV}}\right) \, ,\nnb \\
&\nnb\\
\dps \omega^{(\rm{em})}_{27,L}(\s)&= \dps \ln\left(\frac{m_b^2}{m_\ell^2}\right) \, \left[ \frac{-8.01684 -1121.13 \s^4+882.711 \s^3-280.866 \s^2+54.1943 \s-\frac{0.128988}{\s}}{4 (1-\s)^2}\right. \nnb \\
&\dps + \left. i \, \frac{-2.14058 -588.771 \s^4+483.997 \s^3-124.579 \s^2+12.3282 \s+\frac{0.0145059}{\s}}{4 (1-\s)^2} \right] \nnb \\
& \dps + \frac{8}{9} \, \omega^{(\rm{em})}_{79,L}(\s) \, \ln\left(\frac{\mu_b}{5\rm{GeV}}\right) \, ,\nnb \\
&\nnb\\
\dps \omega^{(\rm{em})}_{29,T}(\s)&= \dps \ln\left(\frac{m_b^2}{m_\ell^2}\right) \, \left[ 
\frac{4.54727 +330.182 \s^4-258.194 \s^3+79.8713 \s^2-19.6855
   \s+\frac{0.0371348}{\s}}{2\s (1-\s)^2}
\right. \nnb \\
&\dps + \left. i \, \frac{73.9149 \s^4-61.1338 \s^3+14.6517 \s^2-0.102331 \s+0.710037}{2\s (1-\s)^2} \right] \nnb \\
& \dps + \frac{16}{9} \, \omega^{(\rm{em})}_{99,T}(\s) \, \ln\left(\frac{\mu_b}{5\rm{GeV}}\right) \, ,\nnb \\
&\nnb\\
\dps \omega^{(\rm{em})}_{29,L}(\s)&= \dps \ln\left(\frac{m_b^2}{m_\ell^2}\right) \, \left[ 
\frac{-2.27221 -298.369 \s^4+224.662 \s^3-65.1375 \s^2+11.5686
   \s-\frac{0.0233098}{\s}}{(1-\s)^2}\right. \nnb \\
&\dps + \left. i \, \frac{-0.666157 -120.303 \s^4+109.315 \s^3-28.2734 \s^2+2.44527
   \s+\frac{0.00279781}{\s}}{(1-\s)^2} \right] \nnb \\
& \dps + \frac{16}{9} \, \omega^{(\rm{em})}_{99,L}(\s) \, \ln\left(\frac{\mu_b}{5\rm{GeV}}\right) \, ,\nnb \\
&\nnb\\
\dps \omega^{(\rm{em})}_{22,T}(\s)&= \dps \ln\left(\frac{m_b^2}{m_\ell^2}\right) \, \left[ 
\frac{2.84257 +269.974 \s^4-194.443 \s^3+48.4535 \s^2-8.24929\s+\frac{0.0111118}{\s}}{2\s (1-\s)^2}\right. \nnb\\
&\dps + \left. \ln\left(\frac{\mu_b}{5\rm{GeV}}\right) \! \frac{4(4.54727 +330.182 \s^4-258.194 \s^3+79.8713 \s^2-19.6855\s+\frac{0.0371348}{\s})}{9\s (1-\s)^2} \right] \nnb \\
& \dps + \frac{64}{81} \, \omega^{(\rm{em})}_{99,T}(\s) \, \ln^2\left(\frac{\mu_b}{5\rm{GeV}}\right) \, ,\nnb \\
&\nnb\\
\dps \omega^{(\rm{em})}_{22,L}(\s)&= \dps \ln\left(\frac{m_b^2}{m_\ell^2}\right) \, \left[ 
\frac{-1.71832 -234.11 \s^4+162.126 \s^3-37.2361 \s^2+6.29949\s-\frac{0.00810233}{\s}}{(1-\s)^2}\right. \nnb \\
&\dps + \left. \ln\left(\frac{\mu_b}{5\rm{GeV}}\right) \! \frac{8(224.662 \s^3-2.27221 -298.369 \s^4-65.1375 \s^2+11.5686\s-\frac{0.0233098}{\s})}{9(1-\s)^2} \right] \nnb \\
& \dps + \frac{64}{81} \, \omega^{(\rm{em})}_{99,L}(\s) \, \ln^2\left(\frac{\mu_b}{5\rm{GeV}}\right) \, ,\nnb \\
&\nnb\\
\dps \omega^{\rm (em)}_{710,A}(\s) & =  \dps
\ln \left(\frac{m_b^2}{m_\ell^2}\right)\,\left[ \frac{7 - 16\,\sqrt{\s} + 9\,\s}{
 4\,\left( 1 - \s \right)} + \ln (1 - \sqrt{\s}) + \frac{1+3 \,\s}{1-\s} \, \ln \!\left(\frac{1 + \sqrt{\s}}{2}\right)
- \frac{\s \,\ln \s}{\left( 1 - \s \right)} \right] \;,\nnb\\
&\nnb\\
\dps \omega^{\rm (em)}_{910,A}(\s) & =  \dps
\ln \left(\frac{m_b^2}{m_\ell^2}\right)\!\left[\ln (1 - \sqrt{\s}) -\frac{5 - 16\,\sqrt{\s} + 11\,\s}{
 4\,\left( 1 - \s \right)} + \frac{1-5 \,\s}{1-\s} \, \ln \!\left(\frac{1 + \sqrt{\s}}{2}\right)
- \frac{(1-3 \,\s) \ln \s}{\left( 1 - \s \right)} \right] , \nnb\\
&\nnb\\
\dps\omega^{\rm (em)}_{210,A}(\s) & = \dps
\ln \left(\frac{m_b^2}{m_\ell^2}\right)\,\left[ \frac{-351.322 \s^4+378.173 \s^3-160.158 \s^2+24.2096 \s+0.305176}{24\s(1-\s)^2}\right. \nnb\\
&\left. +i \, \frac{7.98625 +238.507 \, (\s-a) -766.869\, (\s-a)^2}{24\s(1-\s)^2}\, (\s-a)^2\, \theta(\s-a)\right] \nnb \\
&+ \frac{8}{9} \, \omega^{\rm (em)}_{910,A}(\s)\,\ln\!\left(\frac{\mu_b}{5\gev}\right)\, ,
\end{align}
}
with $a=(4 m_c^2/m_b^2)^2$.

The respective high-$q^2$ functions for the branching ratio that are obtained by a least-squares fit (for fixed values of $m_b$ and $m_c$) read
\bea
\omega_{29}^{\rm (em)}(\s) & = & 
\ln \left(\frac{m_b^2}{m_\ell^2}\right)\,\left[\frac{\Sigma_4(\s)+ i \,\Sigma_4^I(\s)}{8 (1-\s)^2 (1+2\s)}\right] + \frac{16}{9} \,
\omega_{1010}^{\rm (em)}(\s)\,\ln\!\left(\frac{\mu_b}{5\gev}\right)\;,\\
\omega_{22}^{\rm (em)}(\s) & = & 
\ln \left(\frac{m_b^2}{m_\ell^2}\right)\,\left[\frac{\Sigma_5(\s)}{8 (1-\s)^2 (1+2\s)} +  
\, \frac{\Sigma_4(\s)}{9 (1-\s)^2 (1+2\s)}\ln\!\left(\frac{\mu_b}{5\gev}\right)\right] \nnb\\ & & \nnb \\
    &&+ \, \frac{64}{81} \; \omega_{1010}^{\rm (em)}(\s)\, \ln^2\!\left(\frac{\mu_b}{5\gev}\right)
\;,\\ & & \nnb\\
\omega_{27}^{\rm (em)}(\s) & = & 
\ln \left(\frac{m_b^2}{m_\ell^2}\right)\,\left[\frac{\Sigma_6(\s)+ i \, \Sigma_6^I(\s)}{96 (1-\s)^2}\right] + \frac{8}{9} \,
\omega_{79}^{\rm (em)}(\s) \, \ln\!\left(\frac{\mu_b}{5\gev}\right)\; .
\eea
The functions $\Sigma_i$ are polynomials in $\delta = 1-\s$ and are valid for $\s>0.65$.
\bea
\Sigma_4(\s) &=&  -153.673 \, \delta^2 + 498.823 \, \delta^3 - 1146.74 \, \delta^4 + 1138.81 \, \delta^5 \; , \nnb \\
\Sigma_4^I(\s) &=& - 255.712 \, \delta^2 + 1139.10 \, \delta^3 - 2414.21 \, \delta^4 + 2379.91 \, \delta^5 \; ,\nnb\\%
\Sigma_5(\s) &=& - 220.101 \, \delta^2 + 875.703 \, \delta^3 - 1920.56 \, \delta^4 + 1822.07 \, \delta^5 \; ,\nnb\\ %
\Sigma_6(\s) &=& - 310.113 \, \delta^2 + 834.253 \, \delta^3 - 2181.94 \, \delta^4 + 2133.78 \, \delta^5\; ,\nnb\\ %
\Sigma_6^I(\s) &=& - 518.180 \, \delta^2 + 2047.18 \, \delta^3 - 4470.04 \, \delta^4 + 4827.74 \, \delta^5\; . \label{eq:sigma6}
\eea  


\section{New Physics formulas}
\label{app:NPformulae}

\allowdisplaybreaks{
\begin{align}
%
%
\dps {\cal H}_T[1,3.5]_{ee} =& \Big[
0.0162226 \,{\cal I}(R_7 R_8^\ast)+0.00186782 \,{\cal I}(R_7 R_9^\ast)+0.00985919 \,{\cal I}(R_8 R_9^\ast) \nnb \\
&-0.000201564 \,{\cal I}(R_8 R_{10}^\ast)+0.0465868 \,{\cal I}(R_7)-0.00822885 \,{\cal I}(R_8) \nnb \\
&-0.0187815 \,{\cal I}(R_9)+0.000379966 \,{\cal I}(R_{10})+0.393156 \,{\cal R}(R_7) \nnb \\
&+0.0400072 \,{\cal R}(R_8)+0.0531851 \,{\cal R}(R_9)-0.0385002 \,{\cal R}(R_{10}) \nnb \\
&+0.0458427 \,{\cal R}(R_7 R_8^\ast)-0.369964 \,{\cal R}(R_7 R_9^\ast)+0.00570607 \,{\cal R}(R_7 R_{10}^\ast) \nnb \\
&-0.0369498 \,{\cal R}(R_8 R_9^\ast)+0.000616422 \,{\cal R}(R_8 R_{10}^\ast)-0.00978058 \,{\cal R}(R_9 R_{10}^\ast) \nnb \\
&+0.204994 \,|R_7|^2+0.00230146 \,|R_8|^2+0.244813 \,|R_9|^2 \nnb \\
&+1.74294 \,|R_{10}|^2+0.632156 \, \Big] \, \times \, 10^{-7} \, , \\[0.5em]
%
%
\dps {\cal H}_T[3.5,6]_{ee} =& \Big[
0.00519889 \,{\cal I}(R_7 R_8^\ast)+0.00141211 \,{\cal I}(R_7 R_9^\ast)+0.00745377 \,{\cal I}(R_8 R_9^\ast) \nnb \\
&-0.000152386 \,{\cal I}(R_8 R_{10}^\ast)+0.0151043 \,{\cal I}(R_7)+0.00358335 \,{\cal I}(R_8) \nnb \\
&-0.0100672 \,{\cal I}(R_9)+0.000148662 \,{\cal I}(R_{10})-0.138516 \,{\cal R}(R_7) \nnb \\
&-0.0131665 \,{\cal R}(R_8)+0.375959 \,{\cal R}(R_9)-0.074623 \,{\cal R}(R_{10}) \nnb \\
&+0.0143568 \,{\cal R}(R_7 R_8^\ast)-0.254325 \,{\cal R}(R_7 R_9^\ast)+0.00431139 \,{\cal R}(R_7 R_{10}^\ast) \nnb \\
&-0.0260943 \,{\cal R}(R_8 R_9^\ast)+0.000467687 \,{\cal R}(R_8 R_{10}^\ast)-0.0157259 \,{\cal R}(R_9 R_{10}^\ast) \nnb \\
&+0.0631028 \,|R_7|^2+0.000727107 \,|R_8|^2+0.273706 \,|R_9|^2 \nnb \\
&+1.96638 \,|R_{10}|^2+0.257773 \, \Big] \, \times \, 10^{-7} \, , \\[0.5em]
%
%
\dps {\cal H}_T[1,6]_{ee} =& \Big[
0.0214215 \,{\cal I}(R_7 R_8^\ast)+0.00327993 \,{\cal I}(R_7 R_9^\ast)+0.017313 \,{\cal I}(R_8 R_9^\ast) \nnb \\
&-0.000353949 \,{\cal I}(R_8 R_{10}^\ast)+0.0616911 \,{\cal I}(R_7)-0.0046455 \,{\cal I}(R_8) \nnb \\
&-0.0288487 \,{\cal I}(R_9)+0.000528628 \,{\cal I}(R_{10})+0.25464 \,{\cal R}(R_7) \nnb \\
&+0.0268407 \,{\cal R}(R_8)+0.429144 \,{\cal R}(R_9)-0.113123 \,{\cal R}(R_{10}) \nnb \\
&+0.0601994 \,{\cal R}(R_7 R_8^\ast)-0.624289 \,{\cal R}(R_7 R_9^\ast)+0.0100175 \,{\cal R}(R_7 R_{10}^\ast) \nnb \\
&-0.0630441 \,{\cal R}(R_8 R_9^\ast)+0.00108411 \,{\cal R}(R_8 R_{10}^\ast)-0.0255065 \,{\cal R}(R_9 R_{10}^\ast) \nnb \\
&+0.268097 \,|R_7|^2+0.00302857 \,|R_8|^2+0.518519 \,|R_9|^2 \nnb \\
&+3.70932 \,|R_{10}|^2+0.889929 \, \Big] \, \times \, 10^{-7} \, , \\[0.5em]
%
%
\dps {\cal H}_T[1,3.5]_{\mu\mu} =& \Big[
0.0162226 \,{\cal I}(R_7 R_8^\ast)+0.00186782 \,{\cal I}(R_7 R_9^\ast)+0.00985919 \,{\cal I}(R_8 R_9^\ast) \nnb \\
&-0.000201564 \,{\cal I}(R_8 R_{10}^\ast)+0.0478295 \,{\cal I}(R_7)-0.00813434 \,{\cal I}(R_8) \nnb \\
&-0.0247652 \,{\cal I}(R_9)+0.000379966 \,{\cal I}(R_{10})+0.459563 \,{\cal R}(R_7) \nnb \\
&+0.0451794 \,{\cal R}(R_8)-0.155638 \,{\cal R}(R_9)-0.0385002 \,{\cal R}(R_{10}) \nnb \\
&+0.0460521 \,{\cal R}(R_7 R_8^\ast)-0.337431 \,{\cal R}(R_7 R_9^\ast)+0.00570607 \,{\cal R}(R_7 R_{10}^\ast) \nnb \\
&-0.0344757 \,{\cal R}(R_8 R_9^\ast)+0.000616422 \,{\cal R}(R_8 R_{10}^\ast)-0.00978058 \,{\cal R}(R_9 R_{10}^\ast) \nnb \\
&+0.206371 \,|R_7|^2+0.00230943 \,|R_8|^2+0.179467 \,|R_9|^2 \nnb \\
&+1.28881 \,|R_{10}|^2+0.436438 \, \Big] \, \times \, 10^{-7} \, , \\[0.5em]
%
%
\dps {\cal H}_T[3.5,6]_{\mu\mu} =& \Big[
0.00519889 \,{\cal I}(R_7 R_8^\ast)+0.00141211 \,{\cal I}(R_7 R_9^\ast)+0.00745377 \,{\cal I}(R_8 R_9^\ast) \nnb \\
&-0.000152386 \,{\cal I}(R_8 R_{10}^\ast)+0.0165184 \,{\cal I}(R_7)+0.00369089 \,{\cal I}(R_8) \nnb \\
&-0.0169196 \,{\cal I}(R_9)+0.000148662 \,{\cal I}(R_{10})-0.112376 \,{\cal R}(R_7) \nnb \\
&-0.0111424 \,{\cal R}(R_8)+0.249027 \,{\cal R}(R_9)-0.074623 \,{\cal R}(R_{10}) \nnb \\
&+0.0146547 \,{\cal R}(R_7 R_8^\ast)-0.244671 \,{\cal R}(R_7 R_9^\ast)+0.00431139 \,{\cal R}(R_7 R_{10}^\ast) \nnb \\
&-0.0253601 \,{\cal R}(R_8 R_9^\ast)+0.000467687 \,{\cal R}(R_8 R_{10}^\ast)-0.0157259 \,{\cal R}(R_9 R_{10}^\ast) \nnb \\
&+0.0650616 \,|R_7|^2+0.000738436 \,|R_8|^2+0.239011 \,|R_9|^2 \nnb \\
&+1.72527 \,|R_{10}|^2+0.123204 \, \Big] \, \times \, 10^{-7} \, , \\[0.5em]
%
%
\dps {\cal H}_T[1,6]_{\mu\mu} =& \Big[
0.0214215 \,{\cal I}(R_7 R_8^\ast)+0.00327993 \,{\cal I}(R_7 R_9^\ast)+0.017313 \,{\cal I}(R_8 R_9^\ast) \nnb \\
&-0.000353949 \,{\cal I}(R_8 R_{10}^\ast)+0.0643479 \,{\cal I}(R_7)-0.00444346 \,{\cal I}(R_8) \nnb \\
&-0.0416848 \,{\cal I}(R_9)+0.000528628 \,{\cal I}(R_{10})+0.347186 \,{\cal R}(R_7) \nnb \\
&+0.034037 \,{\cal R}(R_8)+0.0933889 \,{\cal R}(R_9)-0.113123 \,{\cal R}(R_{10}) \nnb \\
&+0.0607068 \,{\cal R}(R_7 R_8^\ast)-0.582101 \,{\cal R}(R_7 R_9^\ast)+0.0100175 \,{\cal R}(R_7 R_{10}^\ast) \nnb \\
&-0.0598358 \,{\cal R}(R_8 R_9^\ast)+0.00108411 \,{\cal R}(R_8 R_{10}^\ast)-0.0255065 \,{\cal R}(R_9 R_{10}^\ast) \nnb \\
&+0.271433 \,|R_7|^2+0.00304786 \,|R_8|^2+0.418478 \,|R_9|^2 \nnb \\
&+3.01408 \,|R_{10}|^2+0.559642 \, \Big] \, \times \, 10^{-7} \, , \\[0.5em]
%
%
%
\dps {\cal H}_A[1,3.5]_{ee} =& \Big[
-0.0000761415 \,{\cal I}(R_8 R_9^\ast)+0.0259112 \,{\cal I}(R_8 R_{10}^\ast)+0.0031943 \,{\cal I}(R_9 R_{10}^\ast) \nnb \\
&-0.000083788 \,{\cal I}(R_8)+0.00025712 \,{\cal I}(R_9)-0.112552 \,{\cal I}(R_{10}) \nnb \\
&+0.0230277 \,{\cal R}(R_7)+0.00181543 \,{\cal R}(R_8)-0.0133235 \,{\cal R}(R_9) \nnb \\
&-0.826626 \,{\cal R}(R_{10})+0.00214715 \,{\cal R}(R_7 R_9^\ast)-0.849154 \,{\cal R}(R_7 R_{10}^\ast) \nnb \\
&+0.000222401 \,{\cal R}(R_8 R_9^\ast)-0.0847389 \,{\cal R}(R_8 R_{10}^\ast)+0.722934 \,{\cal R}(R_9 R_{10}^\ast) \nnb \\
&-0.00174093 \,|R_9|^2-0.0120987 \,|R_{10}|^2+0.0121072 \, \Big] \, \times \, 10^{-7} \, , \\[0.5em]
%
%
\dps {\cal H}_A[3.5,6]_{ee} =& \Big[
-0.000057133 \,{\cal I}(R_8 R_9^\ast)+0.0194427 \,{\cal I}(R_8 R_{10}^\ast)+0.00509883 \,{\cal I}(R_9 R_{10}^\ast) \nnb \\
&-0.000062727 \,{\cal I}(R_8)+0.000151953 \,{\cal I}(R_9)-0.0912157 \,{\cal I}(R_{10}) \nnb \\
&+0.0172872 \,{\cal R}(R_7)+0.00136744 \,{\cal R}(R_8)-0.0259495 \,{\cal R}(R_9) \nnb \\
&+0.356293 \,{\cal R}(R_{10})+0.00160379 \,{\cal R}(R_7 R_9^\ast)-0.605103 \,{\cal R}(R_7 R_{10}^\ast) \nnb \\
&+0.000169807 \,{\cal R}(R_8 R_9^\ast)-0.0623319 \,{\cal R}(R_8 R_{10}^\ast)+1.08406 \,{\cal R}(R_9 R_{10}^\ast) \nnb \\
&-0.0027675 \,|R_9|^2-0.0192329 \,|R_{10}|^2-0.0115297 \, \Big] \, \times \, 10^{-7} \, , \\[0.5em]
%
%
\dps {\cal H}_A[1,6]_{ee} =& \Big[
-0.000133274 \,{\cal I}(R_8 R_9^\ast)+0.0453539 \,{\cal I}(R_8 R_{10}^\ast)+0.00829314 \,{\cal I}(R_9 R_{10}^\ast) \nnb \\
&-0.000146515 \,{\cal I}(R_8)+0.000409073 \,{\cal I}(R_9)-0.203767 \,{\cal I}(R_{10}) \nnb \\
&+0.0403149 \,{\cal R}(R_7)+0.00318287 \,{\cal R}(R_8)-0.0392731 \,{\cal R}(R_9) \nnb \\
&-0.470333 \,{\cal R}(R_{10})+0.00375094 \,{\cal R}(R_7 R_9^\ast)-1.45426 \,{\cal R}(R_7 R_{10}^\ast) \nnb \\
&+0.000392209 \,{\cal R}(R_8 R_9^\ast)-0.147071 \,{\cal R}(R_8 R_{10}^\ast)+1.80699 \,{\cal R}(R_9 R_{10}^\ast) \nnb \\
&-0.00450843 \,|R_9|^2-0.0313316 \,|R_{10}|^2+0.000577448 \, \Big] \, \times \, 10^{-7} \, , \\[0.5em]
%
%
\dps {\cal H}_A[1,3.5]_{\mu\mu} =& \Big[
-0.0000761415 \,{\cal I}(R_8 R_9^\ast)+0.0259112 \,{\cal I}(R_8 R_{10}^\ast)+0.0031943 \,{\cal I}(R_9 R_{10}^\ast) \nnb \\
&-0.000083788 \,{\cal I}(R_8)+0.00025712 \,{\cal I}(R_9)-0.112552 \,{\cal I}(R_{10}) \nnb \\
&+0.0230277 \,{\cal R}(R_7)+0.00181543 \,{\cal R}(R_8)-0.0133235 \,{\cal R}(R_9) \nnb \\
&-0.875607 \,{\cal R}(R_{10})+0.00214715 \,{\cal R}(R_7 R_9^\ast)-0.845327 \,{\cal R}(R_7 R_{10}^\ast) \nnb \\
&+0.000222401 \,{\cal R}(R_8 R_9^\ast)-0.0844478 \,{\cal R}(R_8 R_{10}^\ast)+0.694542 \,{\cal R}(R_9 R_{10}^\ast) \nnb \\
&-0.00174093 \,|R_9|^2-0.0120987 \,|R_{10}|^2+0.0131242 \, \Big] \, \times \, 10^{-7} \, , \\[0.5em]
%
%
\dps {\cal H}_A[3.5,6]_{\mu\mu} =& \Big[
-0.000057133 \,{\cal I}(R_8 R_9^\ast)+0.0194427 \,{\cal I}(R_8 R_{10}^\ast)+0.00509883 \,{\cal I}(R_9 R_{10}^\ast) \nnb \\
&-0.000062727 \,{\cal I}(R_8)+0.000151953 \,{\cal I}(R_9)-0.091289 \,{\cal I}(R_{10}) \nnb \\
&+0.0172872 \,{\cal R}(R_7)+0.00136744 \,{\cal R}(R_8)-0.0259495 \,{\cal R}(R_9) \nnb \\
&+0.318008 \,{\cal R}(R_{10})+0.00160379 \,{\cal R}(R_7 R_9^\ast)-0.619516 \,{\cal R}(R_7 R_{10}^\ast) \nnb \\
&+0.000169807 \,{\cal R}(R_8 R_9^\ast)-0.063428 \,{\cal R}(R_8 R_{10}^\ast)+1.07786 \,{\cal R}(R_9 R_{10}^\ast) \nnb \\
&-0.0027675 \,|R_9|^2-0.0192329 \,|R_{10}|^2-0.0113078 \, \Big] \, \times \, 10^{-7} \, , \\[0.5em]
%
%
\dps {\cal H}_A[1,6]_{\mu\mu} =& \Big[
-0.000133274 \,{\cal I}(R_8 R_9^\ast)+0.0453539 \,{\cal I}(R_8 R_{10}^\ast)+0.00829314 \,{\cal I}(R_9 R_{10}^\ast) \nnb \\
&-0.000146515 \,{\cal I}(R_8)+0.000409073 \,{\cal I}(R_9)-0.203841 \,{\cal I}(R_{10}) \nnb \\
&+0.0403149 \,{\cal R}(R_7)+0.00318287 \,{\cal R}(R_8)-0.0392731 \,{\cal R}(R_9) \nnb \\
&-0.557599 \,{\cal R}(R_{10})+0.00375094 \,{\cal R}(R_7 R_9^\ast)-1.46484 \,{\cal R}(R_7 R_{10}^\ast) \nnb \\
&+0.000392209 \,{\cal R}(R_8 R_9^\ast)-0.147876 \,{\cal R}(R_8 R_{10}^\ast)+1.77241 \,{\cal R}(R_9 R_{10}^\ast) \nnb \\
&-0.00450843 \,|R_9|^2-0.0313316 \,|R_{10}|^2+0.00181642 \, \Big] \, \times \, 10^{-7} \, , \\[0.5em]
%
%
%
\dps {\cal H}_3[1,3.5]_{ee} =& \Big[
0.0264036 \,{\cal I}(R_{10}) + 3.07156 \,{\cal R}(R_{10}) - 1.74043 \,{\cal R}(R_7 R_{10}^\ast) \nnb \\
&-  0.132357 \,{\cal R}(R_8 R_{10}^\ast)  + 2.94364 \,{\cal R}(R_9 R_{10}^\ast)  - 0.105444   \, 
\Big] \, \times \, 10^{-9} \, , \\[0.5em]
%
%
\dps {\cal H}_3[3.5,6]_{ee} =& \Big[
0.132813 \,{\cal I}(R_{10}) + 3.51904 \,{\cal R}(R_{10}) - 0.913353 \,{\cal R}(R_7 R_{10}^\ast) \nnb \\
& -  0.0694587 \,{\cal R}(R_8 R_{10}^\ast)  + 2.4359 \,{\cal R}(R_9 R_{10}^\ast)  - 0.0872558   
 \, \Big] \, \times \, 10^{-9} \, , \\[0.5em]
%
%
\dps {\cal H}_3[1,6]_{ee} =& \Big[
0.159216 \,{\cal I}(R_{10}) + 6.5906 \,{\cal R}(R_{10}) - 2.65379 \,{\cal R}(R_7 R_{10}^\ast) \nnb \\
& -  0.201815 \,{\cal R}(R_8 R_{10}^\ast)  + 5.37954 \,{\cal R}(R_9 R_{10}^\ast)  - 0.192699   
 \, \Big] \, \times \, 10^{-9} \, , \\[0.5em]
%
%
\dps {\cal H}_3[1,3.5]_{\mu\mu} =& \Big[
0.010976 \,{\cal I}(R_{10}) + 1.27946 \,{\cal R}(R_{10}) - 0.723502 \,{\cal R}(R_7 R_{10}^\ast) \nnb \\
& -  0.0550209 \,{\cal R}(R_8 R_{10}^\ast)  + 1.22368 \,{\cal R}(R_9 R_{10}^\ast)  - 0.0438331   
 \, \Big] \, \times \, 10^{-9} \, , \\[0.5em]
%
%
\dps {\cal H}_3[3.5,6]_{\mu\mu} =& \Big[
0.0552105 \,{\cal I}(R_{10}) + 1.46503 \,{\cal R}(R_{10}) - 0.379682 \,{\cal R}(R_7 R_{10}^\ast)  \nnb \\- 
& 0.0288741 \,{\cal R}(R_8 R_{10}^\ast)  + 1.01261 \,{\cal R}(R_9 R_{10}^\ast)  - 0.0362724   
 \, \Big] \, \times \, 10^{-9} \, , \\[0.5em]
%
%
\dps {\cal H}_3[1,6]_{\mu\mu} =& \Big[
0.0661865 \,{\cal I}(R_{10}) + 2.74449 \,{\cal R}(R_{10}) - 1.10318 \,{\cal R}(R_7 R_{10}^\ast) \nnb \\
& -  0.083895 \,{\cal R}(R_8 R_{10}^\ast)  + 2.23628 \,{\cal R}(R_9 R_{10}^\ast)  - 0.0801055
 \, \Big] \, \times \, 10^{-9} \, , \\[0.5em]
%
%
%
\dps {\cal H}_4[1,3.5]_{ee} =& \Big[
-0.0412679 \,{\cal I}(R_{7}) - 0.00313835 \,{\cal I}(R_{8}) + 0.200198 \,{\cal I}(R_{9}) \nnb \\
&-  0.430034 \,{\cal R}(R_{7}) - 0.034058 \,{\cal R}(R_{8}) + 1.46516 \,{\cal R}(R_{9}) \nnb \\
&+  0.0135748 \,{\cal R}(R_7 R_{8}^\ast) - 0.361104 \,{\cal R}(R_7 R_{9}^\ast) - 0.0274613 \,{\cal R}(R_8 R_{9}^\ast) \nnb \\
&+  0.482688 \,|R_9|^2 + 0.0892516 \,|R_7|^2 + 0.00051617 \,|R_8|^2 \nnb \\
&+ 3.35446 \,|R_{10}|^2 + 1.6742   
 \, \Big] \, \times \, 10^{-9} \, , \\[0.5em]
%
%
\dps {\cal H}_4[3.5,6]_{ee} =& \Big[
-0.0257056 \,{\cal I}(R_{7}) - 0.00195486 \,{\cal I}(R_{8}) + 0.127314 \,{\cal I}(R_{9}) \nnb \\
&-  0.17595 \,{\cal R}(R_{7}) - 0.0138586 \,{\cal R}(R_{8}) + 0.528054 \,{\cal R}(R_{9})\nnb \\
& +  0.00348411 \,{\cal R}(R_7 R_{8}^\ast) - 0.127392 \,{\cal R}(R_7 R_{9}^\ast) -  0.00968792 \,{\cal R}(R_8 R_{9}^\ast) \nnb \\
&+ 0.179914 \,|R_9|^2 +  0.0229073 \,|R_7|^2 + 0.00013248 \,|R_8|^2\nnb \\
& +  1.25032 \,|R_{10}|^2 + 0.529364   
 \, \Big] \, \times \, 10^{-9} \, , \\[0.5em]
%
%
\dps {\cal H}_4[1,6]_{ee} =& \Big[
-0.0669735 \,{\cal I}(R_{7}) - 0.0050932 \,{\cal I}(R_{8}) + 0.327512 \,{\cal I}(R_{9}) \nnb \\
&-  0.605984 \,{\cal R}(R_{7}) - 0.0479166 \,{\cal R}(R_{8}) + 1.99322 \,{\cal R}(R_{9}) \nnb \\
&+  0.0170589 \,{\cal R}(R_7 R_{8}^\ast) - 0.488496 \,{\cal R}(R_7 R_{9}^\ast) - 0.0371492 \,{\cal R}(R_8 R_{9}^\ast) \nnb \\
&+  0.662601 \,|R_9|^2 + 0.112159 \,|R_7|^2 + 0.00064865 \,|R_8|^2 \nnb \\
&+ 4.60478 \,|R_{10}|^2 + 2.20357    \, \Big] \, \times \, 10^{-9} \, , \\[0.5em]
%
%
\dps {\cal H}_4[1,3.5]_{\mu\mu} =& \Big[
-0.0171551 \,{\cal I}(R_{7}) - 0.00130462 \,{\cal I}(R_{8}) + 0.0832226 \,{\cal I}(R_{9}) \nnb \\
&-  0.179086 \,{\cal R}(R_{7}) - 0.0141823 \,{\cal R}(R_{8}) + 0.609926 \,{\cal R}(R_{9}) \nnb \\
&+  0.00564308 \,{\cal R}(R_7 R_{8}^\ast) - 0.150112 \,{\cal R}(R_7 R_{9}^\ast) - 0.0114157 \,{\cal R}(R_8 R_{9}^\ast) \nnb \\
&+  0.200654 \,|R_9|^2 + 0.0371021 \,|R_7|^2 + 0.000214573 \,|R_8|^2 \nnb \\
&+ 1.39446 \,|R_{10}|^2 + 0.697498   \, \Big] \, \times \, 10^{-9} \, , \\[0.5em]
%
%
\dps {\cal H}_4[3.5,6]_{\mu\mu} =& \Big[
-0.0106858 \,{\cal I}(R_{7}) - 0.000812638 \,{\cal I}(R_{8}) + 0.0529245 \,{\cal I}(R_{9}) \nnb \\
&- 0.0732557 \,{\cal R}(R_{7}) - 0.00576964 \,{\cal R}(R_{8}) + 0.219832 \,{\cal R}(R_{9}) \nnb \\
&+  0.00144835 \,{\cal R}(R_7 R_{8}^\ast) - 0.0529571 \,{\cal R}(R_7 R_{9}^\ast) - 0.00402729 \,{\cal R}(R_8 R_{9}^\ast) \nnb \\
&+ 0.0747905 \,|R_9|^2 + 0.00952261 \,|R_{7}|^2 + 0.0000550722 \,|R_{8}|^2 \nnb \\
&+ 0.51976 \,|R_{10}|^2 + 0.22061
 \, \Big] \, \times \, 10^{-9} \, , \\[0.5em]
%
%
\dps {\cal H}_4[1,6]_{\mu\mu} =& \Big[
-0.027841 \,{\cal I}(R_{7}) - 0.00211725 \,{\cal I}(R_{8}) + 0.136147 \,{\cal I}(R_{9}) \nnb \\
&-  0.252341 \,{\cal R}(R_{7}) - 0.0199519 \,{\cal R}(R_{8}) + 0.829758 \,{\cal R}(R_{9}) \nnb \\
&+ 0.00709143 \,{\cal R}(R_7 R_{8}^\ast) - 0.203069 \,{\cal R}(R_7 R_{9}^\ast) - 0.015443 \,{\cal R}(R_8 R_{9}^\ast) \nnb \\
&+ 0.275445 \,|R_9|^2 + 0.0466247 \,|R_{7}|^2 +  0.000269645 \,|R_{8}|^2 \nnb \\
&+ 1.91422 \,|R_{10}|^2 + 0.918108
\, \Big] \, \times \, 10^{-9} \, , \\[0.5em]
%
%
%
\dps {\cal H}_L[1,3.5]_{ee} =& \Big[
0.000741931 \,{\cal I}(R_7 R_8^\ast)+0.000952641 \,{\cal I}(R_7 R_9^\ast)+0.0050284 \,{\cal I}(R_8 R_9^\ast) \nnb \\
&-0.000102803 \,{\cal I}(R_8 R_{10}^\ast)+0.00124959 \,{\cal I}(R_7)+0.00594309 \,{\cal I}(R_8) \nnb \\
&+0.00735758 \,{\cal I}(R_9)-0.00113202 \,{\cal I}(R_{10})-0.194866 \,{\cal R}(R_7) \nnb \\
&-0.0251935 \,{\cal R}(R_8)+1.42501 \,{\cal R}(R_9)-0.25154 \,{\cal R}(R_{10}) \nnb \\
&+0.00213751 \,{\cal R}(R_7 R_8^\ast)-0.136283 \,{\cal R}(R_7 R_9^\ast)+0.00300802 \,{\cal R}(R_7 R_{10}^\ast) \nnb \\
&-0.0187453 \,{\cal R}(R_8 R_9^\ast)+0.000402421 \,{\cal R}(R_8 R_{10}^\ast)-0.0462841 \,{\cal R}(R_9 R_{10}^\ast) \nnb \\
&+0.00589466 \,|R_7|^2+0.000128527 \,|R_8|^2+0.575967 \,|R_9|^2 \nnb \\
&+4.20578 \,|R_{10}|^2+0.806915 \, \Big] \, \times \, 10^{-7} \, , \\[0.5em]
%
%
\dps {\cal H}_L[3.5,6]_{ee} =& \Big[
0.000562052 \,{\cal I}(R_7 R_8^\ast)+0.000724099 \,{\cal I}(R_7 R_9^\ast)+0.00382208 \,{\cal I}(R_8 R_9^\ast) \nnb \\
&-0.0000781401 \,{\cal I}(R_8 R_{10}^\ast)+0.00223749 \,{\cal I}(R_7)+0.0047901 \,{\cal I}(R_8) \nnb \\
&-0.00211229 \,{\cal I}(R_9)-0.000740423 \,{\cal I}(R_{10})-0.161117 \,{\cal R}(R_7) \nnb \\
&-0.0192094 \,{\cal R}(R_8)+1.14892 \,{\cal R}(R_9)-0.193345 \,{\cal R}(R_{10}) \nnb \\
&+0.0017624 \,{\cal R}(R_7 R_8^\ast)-0.107501 \,{\cal R}(R_7 R_9^\ast)+0.00228636 \,{\cal R}(R_7 R_{10}^\ast) \nnb \\
&-0.0136079 \,{\cal R}(R_8 R_9^\ast)+0.000286423 \,{\cal R}(R_8 R_{10}^\ast)-0.0355109 \,{\cal R}(R_9 R_{10}^\ast) \nnb \\
&+0.00631092 \,|R_7|^2+0.0000975709 \,|R_8|^2+0.439598 \,|R_9|^2 \nnb \\
&+3.20293 \,|R_{10}|^2+0.701014 \, \Big] \, \times \, 10^{-7} \, , \\[0.5em]
%
%
\dps {\cal H}_L[1,6]_{ee} =& \Big[
0.00130398 \,{\cal I}(R_7 R_8^\ast)+0.00167674 \,{\cal I}(R_7 R_9^\ast)+0.00885049 \,{\cal I}(R_8 R_9^\ast) \nnb \\
&-0.000180943 \,{\cal I}(R_8 R_{10}^\ast)+0.00348707 \,{\cal I}(R_7)+0.0107332 \,{\cal I}(R_8) \nnb \\
&+0.00524529 \,{\cal I}(R_9)-0.00187244 \,{\cal I}(R_{10})-0.355982 \,{\cal R}(R_7) \nnb \\
&-0.0444029 \,{\cal R}(R_8)+2.57393 \,{\cal R}(R_9)-0.444885 \,{\cal R}(R_{10}) \nnb \\
&+0.00389991 \,{\cal R}(R_7 R_8^\ast)-0.243784 \,{\cal R}(R_7 R_9^\ast)+0.00529438 \,{\cal R}(R_7 R_{10}^\ast) \nnb \\
&-0.0323532 \,{\cal R}(R_8 R_9^\ast)+0.000688843 \,{\cal R}(R_8 R_{10}^\ast)-0.081795 \,{\cal R}(R_9 R_{10}^\ast) \nnb \\
&+0.0122056 \,|R_7|^2+0.000226098 \,|R_8|^2+1.01556 \,|R_9|^2 \nnb \\
&+7.40871 \,|R_{10}|^2+1.50793 \, \Big] \, \times \, 10^{-7} \, , \\[0.5em]
%
%
\dps {\cal H}_L[1,3.5]_{\mu\mu} =& \Big[
0.000741931 \,{\cal I}(R_7 R_8^\ast)+0.000952641 \,{\cal I}(R_7 R_9^\ast)+0.0050284 \,{\cal I}(R_8 R_9^\ast) \nnb \\
&-0.000102803 \,{\cal I}(R_8 R_{10}^\ast)+0.000345511 \,{\cal I}(R_7)+0.00587433 \,{\cal I}(R_8) \nnb \\
&+0.0117155 \,{\cal I}(R_9)-0.00113202 \,{\cal I}(R_{10})-0.217245 \,{\cal R}(R_7) \nnb \\
&-0.0269584 \,{\cal R}(R_8)+1.53068 \,{\cal R}(R_9)-0.25154 \,{\cal R}(R_{10}) \nnb \\
&+0.00260573 \,{\cal R}(R_7 R_8^\ast)-0.153057 \,{\cal R}(R_7 R_9^\ast)+0.00300802 \,{\cal R}(R_7 R_{10}^\ast) \nnb \\
&-0.0200209 \,{\cal R}(R_8 R_9^\ast)+0.000402421 \,{\cal R}(R_8 R_{10}^\ast)-0.0462841 \,{\cal R}(R_9 R_{10}^\ast) \nnb \\
&+0.00897313 \,|R_7|^2+0.000146331 \,|R_8|^2+0.609248 \,|R_9|^2 \nnb \\
&+4.43707 \,|R_{10}|^2+0.914888 \, \Big] \, \times \, 10^{-7} \, , \\[0.5em]
%
%
\dps {\cal H}_L[3.5,6]_{\mu\mu} =& \Big[
0.000562052 \,{\cal I}(R_7 R_8^\ast)+0.000724099 \,{\cal I}(R_7 R_9^\ast)+0.00382208 \,{\cal I}(R_8 R_9^\ast) \nnb \\
&-0.0000781401 \,{\cal I}(R_8 R_{10}^\ast)+0.00132426 \,{\cal I}(R_7)+0.00472065 \,{\cal I}(R_8) \nnb \\
&+0.00235817 \,{\cal I}(R_9)-0.000740423 \,{\cal I}(R_{10})-0.177392 \,{\cal R}(R_7) \nnb \\
&-0.0204867 \,{\cal R}(R_8)+1.23911 \,{\cal R}(R_9)-0.193345 \,{\cal R}(R_{10}) \nnb \\
&+0.00194094 \,{\cal R}(R_7 R_8^\ast)-0.118058 \,{\cal R}(R_7 R_9^\ast)+0.00228636 \,{\cal R}(R_7 R_{10}^\ast) \nnb \\
&-0.0144108 \,{\cal R}(R_8 R_9^\ast)+0.000286423 \,{\cal R}(R_8 R_{10}^\ast)-0.0355109 \,{\cal R}(R_9 R_{10}^\ast) \nnb \\
&+0.00748476 \,|R_7|^2+0.00010436 \,|R_8|^2+0.466941 \,|R_9|^2 \nnb \\
&+3.39295 \,|R_{10}|^2+0.791074 \, \Big] \, \times \, 10^{-7} \, , \\[0.5em]
%
%
\dps {\cal H}_L[1,6]_{\mu\mu} =& \Big[
0.00130398 \,{\cal I}(R_7 R_8^\ast)+0.00167674 \,{\cal I}(R_7 R_9^\ast)+0.00885049 \,{\cal I}(R_8 R_9^\ast) \nnb \\
&-0.000180943 \,{\cal I}(R_8 R_{10}^\ast)+0.00166977 \,{\cal I}(R_7)+0.010595 \,{\cal I}(R_8) \nnb \\
&+0.0140737 \,{\cal I}(R_9)-0.00187244 \,{\cal I}(R_{10})-0.394638 \,{\cal R}(R_7) \nnb \\
&-0.0474451 \,{\cal R}(R_8)+2.76979 \,{\cal R}(R_9)-0.444885 \,{\cal R}(R_{10}) \nnb \\
&+0.00454667 \,{\cal R}(R_7 R_8^\ast)-0.271115 \,{\cal R}(R_7 R_9^\ast)+0.00529438 \,{\cal R}(R_7 R_{10}^\ast) \nnb \\
&-0.0344317 \,{\cal R}(R_8 R_9^\ast)+0.000688843 \,{\cal R}(R_8 R_{10}^\ast)-0.081795 \,{\cal R}(R_9 R_{10}^\ast) \nnb \\
&+0.0164579 \,|R_7|^2+0.000250691 \,|R_8|^2+1.07619 \,|R_9|^2 \nnb \\
&+7.83003 \,|R_{10}|^2+1.70596 \, \Big] \, \times \, 10^{-7} \, , \\[0.5em]
%
%
%
\dps {\cal B}[1,3.5]_{ee} =& \Big[
0.0169646 \,{\cal I}(R_7 R_8^\ast)+0.00282046 \,{\cal I}(R_7 R_9^\ast)+0.0148876 \,{\cal I}(R_8 R_9^\ast) \nnb \\
&-0.000304367 \,{\cal I}(R_8 R_{10}^\ast)+0.0347138 \,{\cal I}(R_7)-0.00283044 \,{\cal I}(R_8) \nnb \\
&+0.000660238 \,{\cal I}(R_9)-0.00100106 \,{\cal I}(R_{10})+0.189792 \,{\cal R}(R_7) \nnb \\
&+0.0139496 \,{\cal R}(R_8)+1.46271 \,{\cal R}(R_9)-0.290285 \,{\cal R}(R_{10}) \nnb \\
&+0.0507378 \,{\cal R}(R_7 R_8^\ast)-0.506251 \,{\cal R}(R_7 R_9^\ast)+0.00871409 \,{\cal R}(R_7 R_{10}^\ast) \nnb \\
&-0.0584716 \,{\cal R}(R_8 R_9^\ast)+0.00107643 \,{\cal R}(R_8 R_{10}^\ast)-0.0560647 \,{\cal R}(R_9 R_{10}^\ast) \nnb \\
&+0.210889 \,|R_7|^2+0.0028916 \,|R_8|^2+0.813297 \,|R_9|^2 \nnb \\
&+5.94874 \,|R_{10}|^2+1.46402 \, \Big] \, \times \, 10^{-7} \, , \\[0.5em]
%
%
\dps {\cal B}[3.5,6]_{ee} =& \Big[
0.00576094 \,{\cal I}(R_7 R_8^\ast)+0.00213621 \,{\cal I}(R_7 R_9^\ast)+0.0112758 \,{\cal I}(R_8 R_9^\ast) \nnb \\
&-0.000230526 \,{\cal I}(R_8 R_{10}^\ast)+0.0117001 \,{\cal I}(R_7)+0.00792519 \,{\cal I}(R_8) \nnb \\
&-0.000973809 \,{\cal I}(R_9)-0.000822616 \,{\cal I}(R_{10})-0.304197 \,{\cal R}(R_7) \nnb \\
&-0.0338418 \,{\cal R}(R_8)+1.538 \,{\cal R}(R_9)-0.268205 \,{\cal R}(R_{10}) \nnb \\
&+0.0166482 \,{\cal R}(R_7 R_8^\ast)-0.361825 \,{\cal R}(R_7 R_9^\ast)+0.00659775 \,{\cal R}(R_7 R_{10}^\ast) \nnb \\
&-0.0407383 \,{\cal R}(R_8 R_9^\ast)+0.000775603 \,{\cal R}(R_8 R_{10}^\ast)-0.0512368 \,{\cal R}(R_9 R_{10}^\ast) \nnb \\
&+0.0694138 \,|R_7|^2+0.000881518 \,|R_8|^2+0.714084 \,|R_9|^2 \nnb \\
&+5.16931 \,|R_{10}|^2+0.985134 \, \Big] \, \times \, 10^{-7} \, , \\[0.5em]
%
%
\dps {\cal B}[1,6]_{ee} =& \Big[
0.0227255 \,{\cal I}(R_7 R_8^\ast)+0.00495667 \,{\cal I}(R_7 R_9^\ast)+0.0261634 \,{\cal I}(R_8 R_9^\ast) \nnb \\
&-0.000534893 \,{\cal I}(R_8 R_{10}^\ast)+0.0464139 \,{\cal I}(R_7)+0.00509475 \,{\cal I}(R_8) \nnb \\
&-0.000313571 \,{\cal I}(R_9)-0.00182368 \,{\cal I}(R_{10})-0.114406 \,{\cal R}(R_7) \nnb \\
&-0.0198921 \,{\cal R}(R_8)+3.00071 \,{\cal R}(R_9)-0.55849 \,{\cal R}(R_{10}) \nnb \\
&+0.067386 \,{\cal R}(R_7 R_8^\ast)-0.868076 \,{\cal R}(R_7 R_9^\ast)+0.0153118 \,{\cal R}(R_7 R_{10}^\ast) \nnb \\
&-0.0992099 \,{\cal R}(R_8 R_9^\ast)+0.00185203 \,{\cal R}(R_8 R_{10}^\ast)-0.107301 \,{\cal R}(R_9 R_{10}^\ast) \nnb \\
&+0.280302 \,|R_7|^2+0.00377311 \,|R_8|^2+1.52738 \,|R_9|^2 \nnb \\
&+11.1181 \,|R_{10}|^2+2.44915 \, \Big] \, \times \, 10^{-7} \, , \\[0.5em]
%
%
\dps {\cal B}[1,3.5]_{\mu\mu} =& \Big[
0.0169646 \,{\cal I}(R_7 R_8^\ast)+0.00282046 \,{\cal I}(R_7 R_9^\ast)+0.0148876 \,{\cal I}(R_8 R_9^\ast) \nnb \\
&-0.000304367 \,{\cal I}(R_8 R_{10}^\ast)+0.0350544 \,{\cal I}(R_7)-0.00280454 \,{\cal I}(R_8) \nnb \\
&-0.000975567 \,{\cal I}(R_9)-0.00100106 \,{\cal I}(R_{10})+0.233832 \,{\cal R}(R_7) \nnb \\
&+0.017358 \,{\cal R}(R_8)+1.35952 \,{\cal R}(R_9)-0.290285 \,{\cal R}(R_{10}) \nnb \\
&+0.0514155 \,{\cal R}(R_7 R_8^\ast)-0.490489 \,{\cal R}(R_7 R_9^\ast)+0.00871409 \,{\cal R}(R_7 R_{10}^\ast) \nnb \\
&-0.0572729 \,{\cal R}(R_8 R_9^\ast)+0.00107643 \,{\cal R}(R_8 R_{10}^\ast)-0.0560647 \,{\cal R}(R_9 R_{10}^\ast) \nnb \\
&+0.215344 \,|R_7|^2+0.00291736 \,|R_8|^2+0.78123 \,|R_9|^2 \nnb \\
&+5.7259 \,|R_{10}|^2+1.3762 \, \Big] \, \times \, 10^{-7} \, , \\[0.5em]
%
%
\dps {\cal B}[3.5,6]_{\mu\mu} =& \Big[
0.00576094 \,{\cal I}(R_7 R_8^\ast)+0.00213621 \,{\cal I}(R_7 R_9^\ast)+0.0112758 \,{\cal I}(R_8 R_9^\ast) \nnb \\
&-0.000230526 \,{\cal I}(R_8 R_{10}^\ast)+0.0122024 \,{\cal I}(R_7)+0.00796339 \,{\cal I}(R_8) \nnb \\
&-0.00336638 \,{\cal I}(R_9)-0.000822616 \,{\cal I}(R_{10})-0.29433 \,{\cal R}(R_7) \nnb \\
&-0.0330948 \,{\cal R}(R_8)+1.50123 \,{\cal R}(R_9)-0.268205 \,{\cal R}(R_{10}) \nnb \\
&+0.0171247 \,{\cal R}(R_7 R_8^\ast)-0.362728 \,{\cal R}(R_7 R_9^\ast)+0.00659775 \,{\cal R}(R_7 R_{10}^\ast) \nnb \\
&-0.040807 \,{\cal R}(R_8 R_9^\ast)+0.000775603 \,{\cal R}(R_8 R_{10}^\ast)-0.0512368 \,{\cal R}(R_9 R_{10}^\ast) \nnb \\
&+0.0725464 \,|R_7|^2+0.000899635 \,|R_8|^2+0.706733 \,|R_9|^2 \nnb \\
&+5.11822 \,|R_{10}|^2+0.940534 \, \Big] \, \times \, 10^{-7} \, , \\[0.5em]
%
%
\dps {\cal B}[1,6]_{\mu\mu} =& \Big[
0.0227255 \,{\cal I}(R_7 R_8^\ast)+0.00495667 \,{\cal I}(R_7 R_9^\ast)+0.0261634 \,{\cal I}(R_8 R_9^\ast) \nnb \\
&-0.000534893 \,{\cal I}(R_8 R_{10}^\ast)+0.0472568 \,{\cal I}(R_7)+0.00515885 \,{\cal I}(R_8) \nnb \\
&-0.00434195 \,{\cal I}(R_9)-0.00182368 \,{\cal I}(R_{10})-0.0604983 \,{\cal R}(R_7) \nnb \\
&-0.0157368 \,{\cal R}(R_8)+2.86075 \,{\cal R}(R_9)-0.55849 \,{\cal R}(R_{10}) \nnb \\
&+0.0685402 \,{\cal R}(R_7 R_8^\ast)-0.853217 \,{\cal R}(R_7 R_9^\ast)+0.0153118 \,{\cal R}(R_7 R_{10}^\ast) \nnb \\
&-0.09808 \,{\cal R}(R_8 R_9^\ast)+0.00185203 \,{\cal R}(R_8 R_{10}^\ast)-0.107301 \,{\cal R}(R_9 R_{10}^\ast) \nnb \\
&+0.287891 \,|R_7|^2+0.003817 \,|R_8|^2+1.48796 \,|R_9|^2 \nnb \\
&+10.8441 \,|R_{10}|^2+2.31673 \, \Big] \, \times \, 10^{-7} \, , \\[0.5em]
%
%
%
\dps {\cal B}[>14.4]_{ee} =& \Big[
0.000264356 \,{\cal I}(R_7 R_8^\ast)+0.000401975 \,{\cal I}(R_7 R_9^\ast)+0.00161219 \,{\cal I}(R_8 R_9^\ast) \nnb \\
&-0.0000328066 \,{\cal I}(R_8 R_{10}^\ast)-0.0158129 \,{\cal I}(R_7)+0.000478008 \,{\cal I}(R_8) \nnb \\
&+0.125395 \,{\cal I}(R_9)-0.00293188 \,{\cal I}(R_{10})-0.0723471 \,{\cal R}(R_7) \nnb \\
&-0.00827793 \,{\cal R}(R_8)+0.511715 \,{\cal R}(R_9)-0.0806142 \,{\cal R}(R_{10}) \nnb \\
&+0.000709678 \,{\cal R}(R_7 R_8^\ast)-0.0516424 \,{\cal R}(R_7 R_9^\ast)+0.00111614 \,{\cal R}(R_7 R_{10}^\ast) \nnb \\
&-0.00651216 \,{\cal R}(R_8 R_9^\ast)+0.000119004 \,{\cal R}(R_8 R_{10}^\ast)-0.0168936 \,{\cal R}(R_9 R_{10}^\ast) \nnb \\
&+0.00287361 \,|R_7|^2+0.0000373632 \,|R_8|^2+0.211548 \,|R_9|^2 \nnb \\
&+1.50748 \,|R_{10}|^2+0.200589 \, \Big] \, \times \, 10^{-7} \, , \\[0.5em]
%
%
\dps {\cal B}[>14.4]_{\mu\mu} =& \Big[
0.000264356 \,{\cal I}(R_7 R_8^\ast)+0.000401975 \,{\cal I}(R_7 R_9^\ast)+0.00161219 \,{\cal I}(R_8 R_9^\ast) \nnb \\
&-0.0000328066 \,{\cal I}(R_8 R_{10}^\ast)-0.0175987 \,{\cal I}(R_7)+0.000342205 \,{\cal I}(R_8) \nnb \\
&+0.134924 \,{\cal I}(R_9)-0.00293188 \,{\cal I}(R_{10})-0.0871863 \,{\cal R}(R_7) \nnb \\
&-0.00943852 \,{\cal R}(R_8)+0.594393 \,{\cal R}(R_9)-0.0806142 \,{\cal R}(R_{10}) \nnb \\
&+0.000835527 \,{\cal R}(R_7 R_8^\ast)-0.0601984 \,{\cal R}(R_7 R_9^\ast)+0.00111614 \,{\cal R}(R_7 R_{10}^\ast) \nnb \\
&-0.00716282 \,{\cal R}(R_8 R_9^\ast)+0.000119004 \,{\cal R}(R_8 R_{10}^\ast)-0.0168936 \,{\cal R}(R_9 R_{10}^\ast) \nnb \\
&+0.00370104 \,|R_7|^2+0.0000421485 \,|R_8|^2 \nnb \\
&+0.234333 \,|R_9|^2+1.66583 \,|R_{10}|^2+0.292268 \, \Big] \, \times \, 10^{-7} \, , \\[0.5em]
%
%
%
\dps {\cal R}(14.4)_{ee} =& \Big[
0.000352294 \,{\cal I}(R_7 R_8^\ast)+0.000544926 \,{\cal I}(R_7 R_9^\ast)+0.00213997 \,{\cal I}(R_8 R_9^\ast) \nnb \\
&-0.0000442492 \,{\cal I}(R_8 R_{10}^\ast)-0.0160419 \,{\cal I}(R_7)+0.000523537 \,{\cal I}(R_8) \nnb \\
&+0.130938 \,{\cal I}(R_9)-0.00323922 \,{\cal I}(R_{10})-0.0669411 \,{\cal R}(R_7) \nnb \\
&-0.00821459 \,{\cal R}(R_8)+0.458105 \,{\cal R}(R_9)-0.0958901 \,{\cal R}(R_{10}) \nnb \\
&+0.000807558 \,{\cal R}(R_7 R_8^\ast)-0.054864 \,{\cal R}(R_7 R_9^\ast)+0.00123432 \,{\cal R}(R_7 R_{10}^\ast) \nnb \\
&-0.00734198 \,{\cal R}(R_8 R_9^\ast)+0.000139543 \,{\cal R}(R_8 R_{10}^\ast)-0.0189772 \,{\cal R}(R_9 R_{10}^\ast) \nnb \\
&+0.00293717 \,|R_7|^2+0.0000444449 \,|R_8|^2+0.228597 \,|R_9|^2 \nnb \\
&+1.6322 \,|R_{10}|^2+0.174573 \, \Big] \, \times \, 10^{-3} \, , \\[0.5em]
%
%
\dps {\cal R}(14.4)_{\mu\mu} =& \Big[
0.000352294 \,{\cal I}(R_7 R_8^\ast)+0.000544926 \,{\cal I}(R_7 R_9^\ast)+0.00213997 \,{\cal I}(R_8 R_9^\ast) \nnb \\
&-0.0000442492 \,{\cal I}(R_8 R_{10}^\ast)-0.0181914 \,{\cal I}(R_7)+0.000360068 \,{\cal I}(R_8) \nnb \\
&+0.142407 \,{\cal I}(R_9)-0.00323922 \,{\cal I}(R_{10})-0.0825154 \,{\cal R}(R_7) \nnb \\
&-0.00943762 \,{\cal R}(R_8)+0.54544 \,{\cal R}(R_9)-0.0958901 \,{\cal R}(R_{10}) \nnb \\
&+0.000959044 \,{\cal R}(R_7 R_8^\ast)-0.0651631 \,{\cal R}(R_7 R_9^\ast)+0.00123432 \,{\cal R}(R_7 R_{10}^\ast) \nnb \\
&-0.0081252 \,{\cal R}(R_8 R_9^\ast)+0.000139543 \,{\cal R}(R_8 R_{10}^\ast)-0.0189772 \,{\cal R}(R_9 R_{10}^\ast) \nnb \\
&+0.00393316 \,|R_7|^2+0.000050205 \,|R_8|^2+0.256024 \,|R_9|^2 \nnb \\
&+1.82281 \,|R_{10}|^2+0.266662 \, \Big] \, \times \, 10^{-3} \, .
\end{align}
}

\setlength {\baselineskip}{0.2in}

\end{document}